\documentclass[11pt,a4paper]{article}
\usepackage{jheppub}

\usepackage[dvipsnames]{xcolor}
\usepackage{bm} 
\usepackage[inline]{enumitem}

\usepackage{graphicx}
\usepackage{amsmath,amssymb}
\usepackage{soul}
\usepackage{textcomp}
\usepackage{subfigure}
\usepackage{float}
\usepackage[toc,page]{appendix}
\usepackage{comment}
\usepackage[normalem]{ulem}

\newcommand{\T}{t}
\newcommand{\Tt}{\tilde{t}}
\newcommand{\U}{u}
\newcommand{\Ut}{\tilde{u}}
\newcommand{\g}{{g}}
\newcommand{\gt}{{\tilde{g}}}
\newcommand{\gb}{G}
\newcommand{\A}{a}
\newcommand{\B}{b}
\newcommand{\C}{c}
\newcommand{\At}{\tilde{a}}
\newcommand{\Bt}{\tilde{b}}
\newcommand{\Ct}{\tilde{c}}
\newcommand{\x}{\chi}
\newcommand{\xt}{\tilde{\chi}}

\newcommand{\tc}{\hat{\tau}}

\newcommand{\tildetauzero}{\check{{\tau}}_0}
\newcommand{\tildekappa}{\check{{\kappa}}}
\newcommand{\fullaction}{\mathbf{{S}}}

\newcommand{\e}{\varepsilon}
\newcommand{\et}{\tilde{\varepsilon}}
\newcommand{\p}{P}
\newcommand{\pt}{\tilde{P}}
\newcommand{\tp}{\tau_\pi}
\newcommand{\ttp}{\tilde{\tau}_\pi}

\newcommand{\de}{\delta \varepsilon}
\newcommand{\tde}{\delta \tilde{\varepsilon}}
\newcommand{\dph}{\delta \phi}
\newcommand{\dpht}{\delta \tilde{\phi}}
\newcommand{\ctp}{C_{\tau}}
\newcommand{\ctpt}{\tilde{C}_{\tau}}
\newcommand{\cet}{C_{\eta}}
\newcommand{\cett}{\tilde{C}_{\eta}}
\newcommand{\gtwo}{{g_2}}

\newcommand{\taum}{\tau_i}
\newcommand{\gammap}{\gamma^\prime}

\newcommand{\coef}{m_1}
\newcommand{\tcoef}{m_2}
\newcommand{\coefp}{n_1}
\newcommand{\tcoefp}{n_2}

\newcommand{\mn}{{\mu\nu}}
\newcommand{\ab}{{\alpha\beta}}

 \newcommand{\AS}[1]{{\color{black}{#1}}}
\newcommand{\AM}[1]{{\color{black}{#1}}}
 \newcommand{\AR}[1]{{\color{black}{#1}}}

\newcommand{\AMnew}[1]{{\color{black}{#1}}}
 \newcommand{\ASnew}[1]{{\color{black}{#1}}}
 \newcommand{\ARnew}[1]{{\color{black}{#1}}}

\author[a,b]{Toshali Mitra,}
\author[c,d]{Sukrut Mondkar,}
\author[e,c]{Ayan Mukhopadhyay,}
\author[f]{Anton Rebhan,}
\author[f,g]{and Alexander Soloviev}

\affiliation[a]{The Institute of Mathematical Sciences, Chennai 600113, India}
\affiliation[b]{Homi Bhabha National Institute, Training School Complex, Anushakti Nagar, Mumbai 400094,
India}
\affiliation[c]{Center for Strings, Gravitation and Cosmology, Indian Institute of Technology Madras,
Chennai 600036, India}
\affiliation[d]{Harish-Chandra Research Institute, Chhatnag Road, Jhunsi, Prayagraj (Allahabad) 211019, India}
\affiliation[e]{Instituto de F\'{\i}sica,
Pontificia Universidad Cat\'{o}lica de Valpara\'{\i}so,
Avenida Universidad 330, Valpara\'{\i}so, Chile}
\affiliation[f]{Institut f{\"u}r Theoretische Physik, Technische Universit{\"a}t Wien,
Wiedner Hauptstr. 8-10, A-1040 Vienna, Austria}
\affiliation[g]{Faculty of Mathematics and Physics, University of Ljubljana, Jadranska ulica 19, SI-1000 Ljubljana, Slovenia}

\emailAdd{toshalim@imsc.res.in}
\emailAdd{sukrut@physics.iitm.ac.in}
\emailAdd{ayan.mukhopadhyay@pucv.cl}
\emailAdd{anton.rebhan@tuwien.ac.at}
\emailAdd{alexander.soloviev@fmf.uni-lj.si}

\abstract{Hybrid fluid models, consisting of two sectors with more weakly and more strongly self-interacting
degrees of freedom coupled consistently as in the semi-holographic framework, have been shown to exhibit an attractor surface for Bjorken flow. Retaining only the simple viscid fluid descriptions of both sectors, we find that, on the attractor surface, the hydrodynamization times of both subsectors decrease with increasing
total energy density at the respective point of hydrodynamization following a conformal scaling, 
reach their minimum values, and subsequently rise rapidly. The minimum values are obtained when the respective energy densities are of the order of the inverse of the dimensionful inter-system coupling.  
Restricting to attractor curves which can be matched to glasma models at a time set by the saturation scale for both $p$-$p$ and Pb-Pb collisions, we find that the
more weakly coupled sector hydrodynamizes much \textit{later}, and the
{strongly coupled} sector hydrodynamizes \textit{earlier} in $p$-$p$ collisions, since the total energy densities at the respective hydrodynamization times of these sectors fall inside and outside {of} the conformal window. 
This holds true also for phenomenologically relevant solutions that are significantly away from the attractor surface at the time we match to glasma models.
}


\date{\today}

\title{Hydrodynamization in hybrid {Bjorken flow} attractors} 
\begin{document}

\maketitle


\section{Introduction}
\label{intro}

One of the most remarkable discoveries in non-equilibrium dynamics recently has been that quantum many-body systems can \textit{hydrodynamize} far away from equilibrium \cite{Chesler:2009cy,Chesler:2010bi,Heller:2011ju,Chesler:2015bba,Attems:2016tby,Attems:2017zam,Romatschke:2017vte,Florkowski:2017olj}. Hydrodynamization refers to the stage of dynamical evolution in which the expectation values of the energy-momentum tensor and conserved currents start following the constitutive relations of hydrodynamics to a very good approximation.
\AR{In fact}, hydrodynamization can happen in both weakly coupled and strongly coupled systems even when they are far away from equilibrium and have large pressure anisotropies \cite{Keegan:2015avk}. Theoretically, hydrodynamization can be formalized in terms of a hydrodynamic attractor in phase space which is built out of a resummation of the divergent series of the late time effective hydrodynamic expansion such that the time-dependent energy-momentum tensor approaches that of the attractor for any arbitrary initial condition \cite{Heller:2011ju, Heinz:2013th,Heller:2013fn,Heller:2015dha,Heller:2016gbp,Romatschke:2017vte, Romatschke:2017acs,Denicol:2017lxn,Casalderrey-Solana:2017zyh,Denicol:2018pak,Kurkela:2019set,Heller:2020anv,Almaalol:2022ijz,Jaiswal:2022mdk}, see \cite{Soloviev:2021lhs} for a recent review. The hydrodynamic attractor in diverse conformal systems, irrespective of whether they are weakly or strongly coupled (e.g., a kinetic system or a holographic strongly interacting field theory), are quasi-universal in the sense that the evolution is almost identical when the phase space is described by suitable variables. The study of the hydrodynamic attractor has been extensively carried out in the context of the Bjorken flow to learn about the hydrodynamization of the quark-gluon plasma (QGP) produced in heavy-ion collisions
(see e.g.\ \cite{Berges:2020fwq,Busza:2018rrf} for recent reviews),
although such a simplified set-up captures the longitudinal expansion only and not the transverse flow.

To understand the hydrodynamization of the QGP, one needs to follow its evolution from the earliest stages, where a perturbative description in the forms of the glasma effective theory \cite{Gelis:2010nm} and subsequently kinetic theory \cite{Arnold:2002zm,Kurkela:2015qoa} is applicable, to later epochs when the soft gluons form a strongly coupled thermal bath. The semi-holographic framework, briefly reviewed in Sec. \ref{sec:set-up}, allows for simple phenomenological constructions where the evolution of both perturbative (weakly self-interacting) and non-perturbative (strongly self-interacting) degrees of freedom can be described simultaneously in a consistent way \cite{Faulkner:2010tq,Mukhopadhyay:2013dqa,Iancu:2014ava,Mukhopadhyay:2015smb,Banerjee:2017ozx,Kurkela:2018dku,Ecker:2018ucc}. In the simplest model, one retains only two effective metric couplings, and reduces both sectors to viscous fluids each following the M{\"u}ller-Israel-Stewart (MIS) \cite{Muller:1967zza,Israel:1979wp} description with different amounts of viscosities. The effective metric couplings permit the exchange of energy and momentum, while the total energy-momentum tensor, which can be constructed readily from the subsector energy-momentum tensors, is conserved in the physical background (Minkowski) metric. For further simplicity, the fluids describing both sectors are assumed to be conformal. The full system deviates from conformality, however, since the effective metric couplings (denoted by $\gamma$, $\tilde\gamma=r\gamma$) have mass dimension $-4$. We assume them to be given in terms of a saturation scale that sets the scale for the energy densities of the QGP when perturbative descriptions tend to break down.\footnote{At this point, we note that although we are calling the full non-perturbative description 
semi-holographic, we are only using an effective fluid description for the strongly coupled non-perturbative sector with the expectation that our simplification should be sufficient to learn about hydrodynamization. The explicit dual holographic description will be necessary to understand the role of irreversible energy transfer to a dynamical black hole, which is expected to be a slow process determined by the coupling between the two sectors 
\AR{given the} results obtained in \cite{Ecker:2018ucc}.}

The existence of hydrodynamic attractors in such hybrid fluid models has been established in \cite{Mitra:2020mei}. These attractor curves form a two-dimensional surface in four-dimensional phase space such that for any initial condition the evolution in phase space approaches a curve on this attractor surface after some time. At late time, the full system can be described as a single fluid with equation of state and transport coefficients dependent on two constant parameters which label the curves on the attractor surface.
A feature of the attractor surface is an initial evolution that is reminiscent of the so-called bottom-up thermalization scenario of Ref.~\cite{Baier:2000sb}, for
at sufficiently early time, the energy density of the perturbative sector always dominates over the strongly self-interacting non-perturbative sector. 
\AR{This ``bottom-up-like'' feature is a consequence of the effective fluid
description of the semi-holographic setup, which \ARnew{however} can be expected to be a good approximation to a full semi-holographic treatment
only at $\tau\gtrsim \gamma^{1/4}$.} 
\ARnew{Nevertheless, it is} a robust
feature of the resulting hybrid attractor itself, where the 
dominance of the weakly coupled sector at earlier times is determined by an exponential factor with an exponent involving only the viscosities and relaxation times of the two sectors. Apart from the overall energy scale, the ratio of the respective energy densities at \AMnew{a conveniently chosen early \ARnew{reference} time} $\tau_0$ turns out to be a useful parameter for the curves which span the attractor surface, even though $\tau_0$ is outside the
phenomenologically relevant range.

In this paper, we systematically study hydrodynamization in these hybrid hydrodynamic attractors following preliminary observations in \cite{Mitra:2020mei} 
indicating that one can obtain
significant new insights into small vs large system collisions 
(high-multiplicity $p$-$p$ or $A$-$p$ vs. $A$-$A$ heavy ion collisions
\cite{Loizides:2016tew,Schlichting:2016sqo})
by a detailed exploration of the attractor surface. \AR{In the application
to such high-energy collisions, we consider only
times $\tau\gtrsim \gamma^{1/4}$, since a fluid description cannot
be expected to be applicable at arbitrarily small times.} 
\AMnew{In particular, our systematic study of hydrodynamization on the attractor surface \ARnew{suggests} a novel explanation for rapid hydrodynamization of the softer degrees of freedom in small system collisions as summarized below.} 

The general characterization of the hydrodynamization times of the two components on the hybrid attractor surface \AMnew{in our simplified model} is as follows. 
Let us denote the hydrodynamization times of the perturbative and non-perturbative (holographic) sectors as $\tau_{hd}$ and $\widetilde{\tau}_{hd}$, respectively. 
When the energy density at the hydrodynamization times is
in the regime 
$\mathcal{E}(\tau_{hd}), \mathcal{E}(\widetilde{\tau}_{hd}) \AR{\ll} \gamma^{-1},$
we find that the hydrodynamization times of both sectors follow conformal behaviors, i.e. $$\tau_{hd}\approx p\times (4\pi C_\eta)\times\mathcal{E}(\tau_{hd})^{-1/4}$$
and $$\widetilde{\tau}_{hd}\approx \widetilde{p}\times (4\pi \widetilde{C}_\eta)\times\mathcal{E}(\widetilde{\tau}_{hd})^{-1/4}$$ with $p$ and $\widetilde{p}$ being constants. Generically, due to the \AR{bottom-up-like} feature discussed above, the energy density of the perturbative sector is dominant at some reference time {{$\tau_0 \ll \gamma^{1/4}$}}. 
For such cases, we find that 
$$p\approx 0.63,$$ 
which is {approximately the same value as} in usual conformal attractors (which we obtain in the decoupling limit). Therefore, the hydrodynamization time of the perturbative sector in the conformal regime is both {universal} and identical to that of conformal attractors studied earlier in the literature. Furthermore, $\widetilde{p}$ for the strongly self-interacting sector depends only on the ratio of energy densities of the two subsectors at early time. As this ratio of the energy densities of the perturbative to holographic sectors at \ARnew{reference} time (say) $\tau_0 = 0.001\gamma^{1/4}$ is increased from $1:1$ to $1000:1$, the constant $\widetilde{p}$ increases from $0.64$ to $2.1$. The hydrodynamization time of the relatively soft sector is thus sensitive to initial conditions, and can be significantly larger than in usual conformal attractors even in the conformal window. 

Remarkably, for fine-tuned initial conditions on the attractor surface, where the holographic sector has dominant energy at a fixed early \AR{reference} time, the roles in the conformal window are reversed. The $\widetilde{p}$ of the holographic sector is then universally $\approx 0.63$ as in the usual conformal attractors, while $p$ of the perturbative sector is determined by the ratio of the energy densities at the initialization time. We can thus deduce that the hydrodynamization time of the sub-dominant sector increases because it is forced to share its energy with the dominant sector prior to hydrodynamization, while the dominant sector is dynamically driven to behave in a universal manner. 

On the attractor surface, the hydrodynamization times of both subsectors take their minimal values when the respective energy densities are close to $\gamma^{-1}$, i.e. $$\mathcal{E}({\tau}_{hd}),\mathcal{E}(\widetilde{\tau}_{hd})\sim \gamma^{-1}.$$
For larger $\mathcal{E}({\tau}_{hd})$, $\mathcal{E}(\widetilde{\tau}_{hd})$ the hydrodynamization times of both sectors increase rapidly and do not follow any scaling behavior like that in the conformal window. 

For pondering potential phenomenological consequences of these general features of the hybrid attractor, we 
interpret solutions with lower total energy densities at the \AM{matching time} $\taum\sim \gamma^{1/4}$ as corresponding to small systems (high-multiplicity $p$-$p$ or $A$-$p$ collisions) \cite{Loizides:2016tew,Schlichting:2016sqo} as opposed to $A$-$A$ heavy-ion collisions. To quantify our results, we 
identify the energy scale of the intersystem coupling $\gamma^{-1/4}$ with the saturation scale $Q_s^{p\mbox{-}p}$ of $p$-$p$ collisions,
and set various initial conditions such that the total energy density $\propto Q_s^4$ at $\tau \approx Q_s^{-1}$, with $Q_s$ in $A$-$A$ systems chosen from typical values in the IPsat and bCGC models \cite{Kowalski:2007rw}. Generically, we find:
\begin{itemize}
    \item The total energy densities at hydrodynamization time of the perturbative sector ($\mathcal{E}({\tau}_{hd})$) are \textit{within} the conformal window approximately for both $p$-$p$ and Pb-Pb collisions. Since $\mathcal{E}({\tau}_{hd})$ is smaller in $p$-$p$ collisions, naturally the hydrodynamization time of the perturbative sector is significantly \textit{larger} in $p$-$p$ collisions as readily follows from the scaling behavior of $\tau_{hd}$ in the conformal window.
    \item The total energy densities at hydrodynamization time of the holographic sector ($\mathcal{E}(\widetilde{\tau}_{hd})$) are \textit{outside} the conformal window for both $p$-$p$ and Pb-Pb collisions, i.e. $\mathcal{E}(\widetilde{\tau}_{hd})> \gamma^{-1}$. Since $\mathcal{E}({\tau}_{hd})$ is smaller in $p$-$p$ collisions,  the hydrodynamization time of the softer sector is \textit{smaller} in $p$-$p$ collisions than in the case of Pb-Pb collisions. This follows from the growth of $\widetilde{\tau}_{hd}$ with $\mathcal{E}(\widetilde{\tau}_{hd})$.
\end{itemize}
Furthermore, we have established that these results remain valid for generic phenomenologically relevant initial conditions which can be matched to the glasma models at time scales of the order the saturation scale {for both $p$-$p$ and Pb-Pb type collisions.} 
{For a large class of such initial conditions, our model exhibits an attractor of a system with both weakly and strongly coupled components that is driving the systems towards hydrodynamic behavior well before the onset of hydrodynamics.}
Our simple hybrid model therefore provides a new perspective on how collective flow can emerge from the softer components even in small system collisions.

The organization of the paper is as follows. In Sec.~\ref{sec:set-up}, we review our set-up of two fluids undergoing Bjorken flow and coupled mutually via the democratic effective metric coupling \cite{Banerjee:2017ozx,Kurkela:2018dku}. We further discuss the method to discern the attractor surface in this hybrid system. In Sec.~\ref{sec:early}, we provide more analytic details of early time behavior on the attractor surface, before turning our attention to the late-time hydrodynamic behavior. In both cases, we expand the discussion beyond our earlier results in \cite{Mitra:2020mei}. Finally in Sec.~\ref{sec:hydrodynamization}, we systematically study the hydrodynamization on the attractor surface and connect to the phenomenology of $p$-$p$ and Pb-Pb collisions. 

\section{Set-up}\label{sec:set-up}

\subsection{The hybrid action}

A minimalist set-up required to obtain our hybrid fluid model involves only two leading \textit{democratic effective metric couplings} \cite{Banerjee:2017ozx,Kurkela:2018dku}. The action formalism for the full theory
describing the perturbative and non-perturbative degrees of freedom coupled via the mutual democratic effective metric coupling is \cite{Kurkela:2018dku}
\begin{align}\label{Eq:FullAction}
\fullaction[\psi,\tilde{\psi},\gb_\mn]&=S[\psi, \g_\mn]+\tilde{S}[\tilde\psi,\gt_\mn]
+S_{\rm int}[\g_\mn, \gt_\mn, \gb_\mn],\nonumber\\
S_{\rm int}&=\frac{1}{2\gamma}\int d^d x\sqrt{-\gb}
(g_{\mu\alpha}-\gb_{\mu\alpha})\gb^{\ab}(\gt_{\nu\beta}-\gb_{\nu\beta})\gb^\mn\nonumber\\
&+\frac{1}{2\gamma}\frac{\gamma^\prime}{d \gamma^\prime-\gamma}
\int d^d x\sqrt{-\gb}(g_\mn \gb^\mn-d)(\gt_\ab \gb^\ab-d).
\end{align}
Above $S$ and $\tilde{S}$ denote the Wilsonian effective actions for the perturbative and non-perturbative sectors respectively, with respective degrees of freedom $\psi$ and $\tilde{\psi}$, and living in respective effective metrics $\g_\mn$ and $\gt_\mn$, {while the physical metric of the full theory is $G_\mn$ (to be set to Minkowski metric eventually)}. Both of these sectors inhabit the same topological space, and they are superficially invisible to each other as they experience different background metrics, but those are determined by the respective opposite system. If $\tilde{S}$ is strongly self-interacting and admits a holographic description, then it is the on-shell action of an extra-dimensional gravitational theory whose boundary metric should be identified with $\gt$. In what follows, we will not need any explicit description of both $S$ and $\tilde{S}$. 

Since both $S$ and $\tilde{S}$ are individually diffeomorphism-invariant in the respective background metrics $\g$ and $\gt$, we obtain the two Ward identities
\begin{align}\label{ward}
    \nabla_\mu \T^\mn=0\quad \text{and}\quad \tilde{\nabla}_\mu \Tt^\mn=0,
\end{align}
where the energy momentum tensors are
\begin{align}
    t^\mn=\frac{2}{\sqrt{-g}}\frac{\delta S}{\delta g_\mn} 
    \quad \text{and}\quad
        \Tt^\mn=\frac{2}{\sqrt{-\gt}}\frac{\delta \tilde{S}}{\delta \gt_\mn},
\end{align}
and $\nabla$ and $\tilde{\nabla}$ are built out of $\g$ and $\gt$, respectively. 

The interaction term $S_{int}$ in \eqref{Eq:FullAction} relates these two different background metrics in terms of the actual physical background metric $\gb$ ultralocally, as follows. We readily see that the two effective metrics appear in the full action as auxiliary fields, and extremizing the action with respect to them leads to imposing
\begin{equation}\label{Eq:coup-var}
    \frac{\delta S}{\delta \g_\mn} =0, \quad \frac{\delta \tilde{S}}{\delta \gt_\mn} =0.
\end{equation}
The above yield the following \textit{coupling equations} 
\begin{align}\label{eq-coup}
\g_\mn&=\gb_\mn+\frac{\sqrt{-\gt}}{\sqrt{-\gb}}\left[\gamma \,\Tt^\ab \gb_{\alpha\mu}\gb_{\beta\nu}+\gammap\,\Tt^{\alpha\beta}\gb_{\alpha\beta}\gb_\mn\right],\nonumber\\
\gt_\mn&=\gb_\mn+\frac{\sqrt{-\g}}{\sqrt{-\gb}}\left[\gamma\, \T^\ab
\gb_{\alpha\mu}\gb_{\beta\nu}+\gammap\,\T^{\alpha\beta}\gb_{\alpha\beta}\gb_\mn\right],
\end{align}
where $\gamma$ and $\gammap$ are the two dimensionful couplings with mass dimension $-4$. It is expected that $\gamma ,\gammap \sim \mathcal{O}(\Lambda_{int}^{-4})$, where $\Lambda_{int}$ is a suitable scale separating perturbative and non-perturbative phenomena (and is also larger than the relevant scale of observation). In the context of heavy-ion collisions, $\Lambda_{int}$ can be identified with a saturation scale $Q_s$ determining the initial conditions. It is evident from \eqref{eq-coup} that higher order effective metric couplings will be suppressed by powers of $(T/Q_s)^4$, where $T$ is an effective temperature determining the scales of the stress tensors. 

We can compute the energy momentum tensor of the full system by varying the action \eqref{Eq:FullAction} with respect to the physical background metric $\gb_\mn$, which yields 
\begin{align}
    T^\mu_{\phantom{\mu}\nu}
    &=\frac{1}{2}\left(\frac{\sqrt{-g}}{\sqrt{-\gb}}(t^\mu_{\phantom{\mu}\nu}+t_\nu^{\phantom{\nu}\mu})
    +\frac{\sqrt{-\gt}}{\sqrt{-\gb}}(\tilde{t}^\mu_{\phantom{\mu}\nu}+\tilde{t}_\nu^{\phantom{\nu}\mu})\right)+\Delta K \delta^\mu_\nu,
    \nonumber\\
\Delta K&=-\frac{1}{2}\frac{\sqrt{-g}}{\sqrt{-\gb}}\frac{\sqrt{-\gt}}{\sqrt{-\gb}}\Big[\gamma\, \T^\mn \gb_{\mu\alpha}\Tt^\ab \gb_{\beta\nu}+\gamma^\prime\, t^{\alpha\beta} \gb_{\alpha\beta}\tilde{t}^{\rho\sigma} \gb_{\rho\sigma}\Big].
\label{eq:fullemt}\end{align}
In the first equation above, all indices for subsystem variables are lowered using the respective background metrics, while that of the full system (on the left hand side) is lowered using the physical background metric. It can be explicitly checked that the coupling equations \eqref{eq-coup} and the two Ward identities \eqref{ward} indeed imply the local conservation of the full energy-momentum tensor in the physical background metric, i.e.
\begin{align}
    \nabla_\mu^{(B)}  T^\mu_{\phantom{\mu}\nu}=0.
\end{align}

The mutual effective metric coupling thus leads to a full energy-momentum tensor \eqref{eq:fullemt} which can be obtained ultra-locally from the subsystem energy-momentum tensors alone, and does not require any further microscopic input. This ensures that a coarse-grained hydrodynamic description of the full energy-momentum tensor in terms of those of the subsystems. It implements the hypothetical Wilsonian perspective that the coarse-grained perturbative description should determine the complete non-perturbative description at any scale. Therefore, assuming that $\tilde{S}$ and the inter-system couplings can be derived from $S$, we can state that the coarse-grained descriptions of the two subsystems are related,\footnote{In \cite{Banerjee:2017ozx}, a derivation of the semi-holographic procedure was proposed by noting that the perturbative series has renormalons which forbid Borel resummation, and this can be cured by considering non-perturbative contributions. A physical observable, such as the full energy-momentum tensor, which receives contributions from both sectors, should be a function of the perturbative coupling with a finite radius of convergence, provided that the intersystem couplings are appropriate functions of the perturbative couplings. This was illustrated in a toy bi-holographic set-up.} and the full hydrodynamic description can be constructed in terms of the subsystem hydrodynamic variables. A detailed description will follow soon.

In most set-ups, the full system is solved in a self-consistent iterative procedure \cite{Mukhopadhyay:2015smb,Ecker:2018ucc,Mondkar:2021qsf}.  At the first step of iteration, the effective metrics are identified with the background metric $\gb_\mn$, and the two systems are solved independently with given initial data. In the next step, the effective metrics of the two subsystems are updated via the coupling equations \eqref{eq-coup} using the respective subsystem stress tensors obtained from the previous iteration. The subsystems are solved once again with the \textit{same} initial conditions. The iterations are thus continued with fixed initial conditions until they converge implying that the full energy-momentum tensor is conserved in the physical background metric. 

This iterative procedure was first proposed in the context of heavy-ion collisions with glasma-like initial conditions for the perturbative sector and an empty $AdS$-like initial conditions for the holographic sector \cite{Iancu:2014ava}. Subsequent works \cite{Mukhopadhyay:2015smb,Ecker:2018ucc,Mondkar:2021qsf} have shown that the convergence of the semi-holographic iterative procedure can not only be achieved, but to a very good accuracy over three to four iterations only. For example, the convergence of the iterative procedure has been demonstrated with non-trivial initial conditions for a scalar field in the boundary coupled to the bulk \cite{Ecker:2018ucc,Mondkar:2021qsf}. Note that since both systems are marginally deformed via their effective metrics in each step of the iteration, the iterative procedure is well defined in the full quantum theory, however we will not discuss this further here. 

Here, we will restrict ourselves to a large $N$ approximation in both sub-sectors which ensures factorization of expectation values of multi-trace operators into those of the single trace operators like the energy-momentum tensor, and thus ignore stochastic hydrodynamic phenomena. Furthermore, we will see that the iterative procedure is also not necessary in the hydrodynamic limit -- the coupling equations and the dynamical equations can be evolved simultaneously ensuring conservation of the full energy-momentum tensor in the physical background metric automatically.

\subsection{Hybrid fluid model}

Although the above discussion included an action, it is not necessary to describe the evolution of the system, which is governed by the effective conservation laws \eqref{ward} and the algebraic coupling equations \eqref{eq-coup} in the hydrodynamic limit. The only required inputs are the energy-momentum tensors of each sector, $\T^\mn$ and $\Tt^\mn$, determined via the constitutive relations, and the physical background metric $\gb_\mn$. Since the hydrodynamic limit is found in many theories (we have in mind a strongly coupled holographic theory for one subsystem and a weakly coupled kinetic theory for the other), we can consider the simpler case in which the two subsystems are two different relativistic fluids.

Hence, consider two conformal fluids with energy momentum tensors in their respective metrics, given by
\begin{align}
\T^\mn &=(\e+\p)\U^\mu \U^\nu+\p \g^\mn+\Pi^\mn,\\
\Tt^\mn &=(\et+\pt)\Ut^\mu \Ut^\nu+\pt \gt^\mn+\tilde{\Pi}^\mn,
\end{align}
where $
\T^\mn \g_\mn =-\e +3\p=0 $, $\Tt^\mn \gt_\mn =- \et +3\pt=0$. The energy momentum tensors are conserved in their respective backgrounds \eqref{ward}.
The four-velocities are properly normalized with respect to their metrics, i.e. 
\begin{align}
\U^\mu \g_\mn \U^\nu =-1 \quad \text{and} \quad \Ut^\mu \gt_\mn \Ut^\nu =-1.
\end{align}
The dissipation tensors, $\Pi^\mn$ and $\tilde{\Pi}^\mn$, are taken to be orthogonal to their respective four velocities
\begin{align}
\Pi^\mn \U_\mu=0 \quad \text{and} \quad \tilde{\Pi}^\mn \Ut_\mu=0.
\end{align}
The two fluids are assumed to have dissipative dynamics dictated by M{{\"u}}ller-Israel-Stewart (MIS) theory\footnote{MIS theory with the Weyl covariant version of the convective derivative, as in Eqs. \eqref{mis}, is just a consistent truncation of BRSSS theory where certain non-linear transport coefficients are considered to be vanishing.} \cite{Muller:1967zza,Israel:1979wp, Baier:2007ix},\footnote{The more weakly and the more strongly coupled sectors are also simply assumed to have higher and lower specific viscosity, respectively, as well as correspondingly shorter and longer relaxation time scales. While not pursued here, one could also consider including second-order derivatives of the shear tensor of the strongly coupled sector as in \cite{Noronha:2011fi,Heller:2014wfa} such as to make the dynamics of the latter more similar to the holographic dynamics, which is governed by quasinormal modes. (We thank Michal Heller for this remark.)} 
namely
\begin{align}
(\tp\U^\alpha \nabla_\alpha+1+\frac{3}{2}\nabla_\alpha \U^\alpha)\Pi^\mn&=-\eta \sigma^\mn,\nonumber\\
(\ttp\Ut^\mu \tilde{\nabla}_\mu+1+\frac{3}{2}\tilde{\nabla}_\alpha \Ut^\alpha)\tilde{\Pi}^\mn&=-\tilde{\eta} \tilde{\sigma}^\mn,
\label{mis}
\end{align}
where $\eta$ and $\tilde{\eta}$ are the subsystem viscosities. Note that we have truncated at first order, and since we are working with conformal subsystems, we have set the bulk viscosities to zero. 

It is worthwhile to underline that although the two subsystems are conformal, the complete system \eqref{eq:fullemt} will in general not be conformal, $T^\mn g^{(B)}_\mn\neq 0$. We will see that there is an emergent conformality for large proper time. It is worthwhile to mention that the full system also has emergent conformality at thermal equilibrium at high temperature (with respect to the scale of the intersystem coupling) \cite{Kurkela:2018dku}.

\subsection{Explicit equations for boost-invariant Bjorken flow}\label{sec:explicit}
We now provide the explicit form of the complete set of equations, \eqref{ward}, \eqref{eq-coup} and \eqref{mis}. We will work in the Milne coordinates: $\tau = \sqrt{t^2 - z^2}$, $x$, $y$, and $\zeta = {\rm arctanh} (z/t)$, where $x$ and $y$ are the transverse coordinates and $z$ is the longitudinal coordinate for the expanding system. In these coordinates, the Minkowski metric assumes the form
\begin{align}\label{bjorken-metric}
\gb_\mn&=\text{diag}(-1,1,1,\tau^2).
\end{align}
The above is thus the physical background metric for the full system. The Bjorken flow is boost-invariant, and therefore in the Milne coordinates, all physical variables depend only on the proper time $\tau$. 

Motivated by the form of the background metric, we make the following boost-invariant ansatz for the effective metrics:
\begin{align}
\g_\mn(\tau)=\text{diag}(-\A(\tau)^2,\B(\tau)^2,\B(\tau)^2,\C(\tau)^2),\nonumber\\
\gt_\mn(\tau)=\text{diag}(-\At(\tau)^2,\Bt(\tau)^2,\Bt(\tau)^2,\Ct(\tau)^2).\label{coup}
\end{align}
The four velocities are then
\begin{align}
\U^\mu=\left(\frac{1}{\A(\tau)},0,0,0\right) \quad \text{and} \quad \U^\mu=\left(\frac{1}{\At(\tau)},0,0,0\right).
\end{align}
Here we consider the dissipation tensors, $ \Pi^{\mu\nu} $ and $ \tilde\Pi^{\mu\nu} $, to be symmetric, transverse and traceless i.e., $ \g_{\mu\nu}\Pi^{\mu\nu} = 0 $ and $\gt_{\mu\nu}\tilde\Pi^{\mu\nu} $. These conditions imply that there is only one independent component of $ \Pi^{\mu\nu} $ for each sector, which we call $ \Pi^\eta_{\,\,\eta} = -\phi $ (the longitudinal non-equilibrium pressure) following the convention in \cite{Heller:2015dha}, which means that
\begin{align}
 \Pi^{\mu\nu} &= \text{diag}\left(0,\frac{\phi(\tau)}{2},\frac{\phi(\tau)}{2},\frac{-\phi(\tau)}{\tau^2}\right) ,\\
\tilde{  \Pi}^{\mu\nu} &= \text{diag}\left(0,\frac{\tilde{\phi}(\tau)}{2},\frac{\tilde{\phi}(\tau)}{2},\frac{-\tilde{\phi}(\tau)}{\tau^2}\right).
 \end{align}
Then explicitly, the only non-vanishing components of the conservation equations \eqref{ward} are
\begin{align}\label{exp-ward}
0&=\partial_\tau \e+\frac{4}{3}\e \partial_\tau \log(\B^{2}\C)+\phi \partial_\tau\log(\frac{\B}{\C}),\\ \label{exp-wardt}
0&=\partial_\tau \et+\frac{4}{3}\et \partial_\tau \log(\Bt^{2}\Ct)+\tilde{\phi} \partial_\tau\log(\frac{\Bt}{\Ct}),
\end{align}
the MIS equations are \eqref{mis}
\begin{align}\label{exp-mis}
0&=\tp \partial_\tau \phi+\frac{4}{3}\eta \partial_\tau \log \frac{\B}{\C}+\left[\A+\frac{4}{3}\tp \log (\B^2 \C)\right]\phi,\\
0&=\ttp \partial_\tau \tilde{\phi}+\frac{4}{3}\tilde\eta \partial_\tau \log \frac{\Bt}{\Ct}+\left[\At+\frac{4}{3}\ttp \log (\Bt^2 \Ct)\right]\tilde{\phi}, \label{exp-mist}
\end{align}
and finally, the algebraic coupling equations \eqref{coup} are
\begin{align}\label{coup1}
1-\A^2&=\gamma\frac{\At   \Bt^2 \Ct}{\tau} \Big{[}\frac{  \et}{\At^2}+  r (-\frac{\et}{\At^2}+\frac{\tilde{\phi}}{\Bt^2}+\frac{2 \et}{3 \Bt^2}+\tau ^2 (\frac{\et}{3 \Ct^2}-\frac{\tilde{\phi}}{\Ct^2}))\Big{]},\\ \label{coup2}
\B^2-1&=\gamma\frac{\At \Bt^2 \Ct}{\tau} \Big{[}\frac{  (\tilde{\phi}+\frac{2 \et}{3})}{2 \Bt^2}-  r (-\frac{\et}{\At^2}+\frac{\tilde{\phi}}{\Bt^2}+\frac{2 \et}{3 \Bt^2}+\tau ^2 (\frac{\et}{3 \Ct^2}-\frac{\tilde{\phi}}{\Ct^2}))\Big{]},\\ \label{coup3}
\C^2-\tau ^2&=\gamma\frac{\At \Bt^2 \Ct}{\tau} \Big{[}\frac{  \tau ^4 }{\Ct^2}(\frac{\et}{3}-\tilde{\phi})-  r \tau ^2 (-\frac{\et}{\At^2}+\frac{\tilde{\phi}}{\Bt^2}+\frac{2 \et}{3 \Bt^2}+\tau ^2 (\frac{\et}{3 \Ct^2}-\frac{\tilde{\phi}}{\Ct^2}))\Big{]},\\ \label{coup4}
1-\At^2&=\gamma\frac{\A \B^2 \C}{\tau} \Big{[} \frac{\e}{\A^2}+ r (-\frac{\e}{\A^2}+\frac{\phi}{\B^2}+\frac{2 \e}{3 \B^2}+\tau
   ^2 (\frac{\e}{3 \C^2}-\frac{\phi}{\C^2}))\Big{]},
   \\\label{coup5}
\Bt^2-1&=\gamma\frac{\A \B^2 \C}{\tau} \Big{[}\frac{  (\phi+\frac{2 \e}{3})}{2 \B^2}-  r (-\frac{\e}{\A^2}+\frac{\phi}{\B^2}+\frac{2 \e}{3 \B^2}+\tau ^2 (\frac{\e}{3 \C^2}-\frac{\phi}{\C^2}))\Big{]},
\\ \label{coup6}
\Ct^2-\tau ^2&=\gamma\frac{\A \B^2 \C}{\tau} \Big{[}\frac{  \tau ^4 }{\C^2}(\frac{\e}{3}-\phi)-  r \tau ^2 (-\frac{\epsilon
   (\tau )}{\A^2}+\frac{\phi}{\B^2}+\frac{2 \e}{3 \B^2}+\tau ^2 (\frac{\e}{3 \C^2}-\frac{\phi}{\C^2}))\Big{]},
   \end{align}
where we introduced $r\equiv-\gamma^\prime/\gamma$. We see that in the decoupling limit, namely $\gamma=0$, the right hand side vanishes and the effective metrics reduce to the usual background Milne metric \eqref{bjorken-metric} in which the two fluids evolve independently.

It is clear that the above system actually has four dynamical variables, namely the energy densities and the non-equilibrium pressures $(\e, \et, \phi, \tilde{\phi})$ which follow the first-order equations \eqref{exp-ward}, \eqref{exp-wardt}, \eqref{exp-mis} and \eqref{exp-mist}. The six effective metric variables, $\A, \At, \B,\Bt, \C, \Ct$, can be eliminated by determining them as functions of the four dynamical variables by solving the six algebraic equations  \eqref{coup1}-\eqref{coup6}. The evolution of the full system can thus be readily determined numerically for given initial values of $(\e, \et, \phi, \tilde{\phi})$.

In what follows, we parametrize the transport coefficients in the two conformal subsystems, and the respective MIS relaxation times via dimensionless quantities $C_\eta$, $\tilde{C}_\eta$, $C_\tau$ and $\tilde{C}_\tau$ defined via
\begin{align}
\eta=C_\eta \frac{\e+\p}{\e^{1/4}}, \quad &\text{and} \quad \tp=C_\tau \e^{-1/4},\\
\tilde\eta=\tilde{C}_\eta \frac{\et+\pt}{\et^{1/4}}, \quad &\text{and} \quad \ttp=\tilde{C}_\tau \et^{-1/4}.
\end{align}
In the simulations that follow, we will usually choose
\begin{align}\label{parameters}
\tilde{C}_\eta=\frac{1}{4\pi}, \quad \tilde{C}_\tau=\frac{2-\log2}{2\pi},\quad
C_\eta=10 \tilde{C}_\eta,  \quad \text{and} \quad C_\tau=5 C_\eta.
\end{align} 
motivated by the values obtained in holographic descriptions and in kinetic theories, respectively. This marks the tilded sector as more strongly coupled than the untilded one.

We shall set
\begin{align}\label{parameters2}
 r\equiv-\gamma'/\gamma=2
\end{align}
unless noted otherwise. 
In the inviscid Minkowski case \cite{Kurkela:2018dku}, the condition of UV completeness restricts the value of $r$ to $r>1$. Moreover, it was found in the inviscid case that there is a second order phase transition for $1<r\lessapprox 1.11$, which we discuss in Appendix~\ref{sec:phase}. 
{Note that the coupling $\gamma$ is the only dimensionful scale in our framework.}
Further, we will work with the dimensionless anisotropy parameter, $\x = \phi/(\epsilon + P)$. 
Thus the conservation and the MIS equations read
\begin{align}
\frac{4}{3} \x \left(\frac{\B'}{\B}-\frac{\C'}{\C}\right)+\frac{4}{3} \left(\frac{2
   \B'}{\B}+\frac{\C'}{\C}\right)+\frac{\e'}{\e} &= 0,\\
   \frac{4}{3} \xt \left(\frac{\Bt'}{\Bt}-\frac{\Ct'}{\Ct}\right)+\frac{4}{3} \left(\frac{2
   \Bt'}{\Bt}+\frac{\Ct'}{\Ct}\right)+\frac{\et'}{\et} &= 0,\\
-\frac{\A \x {\e}^{1/4}}{C_{\text{$\tau 
   $}}}-\Big(\frac{4 C_{\text{$\eta $}}}{3  C_{\text{$\tau
   $}}}-\frac{4
   \x{}^2}{3}\Big)\Big(\frac{\B'}{\B}-\frac{\C'}{\C} \Big)-\x'&= 0,\\
-\frac{\At \xt {\et}^{1/4}}{\ctpt}-\Big(\frac{4 \tilde{C}_\eta}{3  \tilde{C}_{\text{$\tau
   $}}}-\frac{4
   \xt{}^2}{3}\Big)\Big(\frac{\Bt'}{\Bt}-\frac{\Ct'}{\Ct} \Big)-\xt'&= 0,
\end{align}
where the prime denotes a derivative with respect to the proper time, e.g. $\varepsilon^\prime=\partial_\tau \varepsilon.$
The full energy momentum tensor \eqref{eq:fullemt} is given by
\begin{align}
    \mathcal{T}^\mn=\text{diag}(\mathcal{E},P_T,P_T,P_L),
\end{align}
where the total energy density, total transverse pressure and longitudinal pressure are
\begin{align}\label{total-energy}
   & \mathcal{E}=-\frac{\A \B^2 \C \e +\At \Bt^2 \Ct \et}{\tau }+\frac{ \gamma}{36 \A \At \C \Ct \tau^2} \Big(\A^2 \At^2 \Big[2 b^2 \tau ^2 (\e -3 \phi )
   \left(\Bt^2 (r-1) \tau ^2 (\et-3 \tilde{\phi})+\Ct^2 r
   (2 \et+3 \tilde{\phi})\right)\nonumber\\
   &+c^2 (2 \e +3 \phi ) \left(2
   \Bt^2 r \tau ^2 (\et-3 \tilde{\phi})+\Ct^2 (2 r-1) (2
   \et+3 \tilde{\phi})\right)\Big]\nonumber\\
   &-6 a^2 \Bt^2 \Ct^2 r
   \et \left(b^2 \tau ^2 (\epsilon -3 \phi )+c^2 (2 \e +3 \phi
   )\right)\nonumber
   \\
   &-6 b^2 c^2 \e  \left(\Bt^2 \left(\At^2 r \tau ^2
   (\et-3 \tilde{\phi})-3 \Ct^2 (r-1) \et \right)+\At^2 \Ct^2 r (2 \et+3 \tilde{\phi})\right)\Big),\\
&P_T=\frac{a b^2 c (2 \e +3 \phi )+\At \Bt^2 \Ct (2 \et+3 \tilde{\phi})}{6 \tau }\nonumber\\
    &+\frac{\gamma }{36 \A\C \At \Ct \tau ^2} \Big[\C\A \At^2 \Big(2 b^2 \tau ^2 (\e -3 \phi )
   \left(\Bt^2 (r-1) \tau ^2 (\et-3 \tilde{\phi})+\Ct^2 r
   (2 \et+3 \tilde{\phi})\right)\nonumber\\
   &+\C^3 (2 \e +3 \phi ) \left(2
   \Bt^2 r \tau ^2 (\et-3 \tilde{\phi})+\Ct^2 (2 r-1) (2
   \et+3 \tilde{\phi})\right)\Big)\nonumber\\
   &-6 b^2 \C^2 \e 
   \left(\Bt^2 \left(\At^2 r \tau ^2 (\et-3 \text{$\phi
   $t})-3 \Ct^2 (r-1) \et\right)+\At^2 \Ct^2 r (2
   \et+3 \tilde{\phi})\right)\nonumber\\
   &-6 \A \C \Bt^2 \Ct^2 r
   \et \left(b^2 \tau ^2 (\e -3 \phi )+c^2 (2 \epsilon +3 \phi
   )\right)\Big],\\
&P_L=P_T-\frac{3 \left(a b^2 c \phi +\At \Bt^2 \Ct \tilde{\phi}\right)}{2
   \tau }.
\end{align}

\subsection{Determining the attractor surface}\label{sec:detatt}

\begin{figure}
\center
\includegraphics[width=.7\columnwidth]{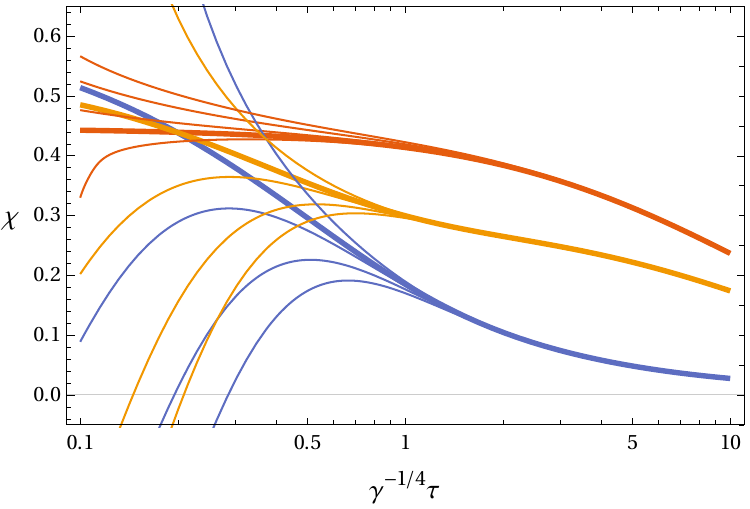}%
\caption{Anisotropies $\chi$ of one particular attractor solution (thick lines) and four neighboring trajectories (thin lines). 
The less viscous (strongly coupled) system is displayed by blue curves,
the more viscous (more weakly coupled) system by red curves, the total system by orange curves. From \cite{Mitra:2020mei}, where more plots are given for this particular solution.
\label{fig:attractor}}
\end{figure}

 As discussed above, the hybrid fluid system has a four-dimensional phase space with the energy densities and non-equilibrium pressures, $(\e, \et, \phi, \tilde{\phi})$, as the dynamical variables. It was shown in \cite{Mitra:2020mei} that the full system has a two-dimensional attractor surface. For any initial condition $(\e, \et, \phi, \tilde{\phi})(\tau_0)$, the system evolves towards a particular curve on the attractor surface. The curves which compose this attractor surface are characterized by dynamical evolutions where the dimensionless anisotropies, $\chi$ and $\tilde{\chi}$ do not diverge as we trace back in time to the moment of collision $\tau=0$. Furthermore,
 \begin{equation}\label{Eq:ChChtET}
     \lim_{\tau\rightarrow0}\chi = \sigma_1, \quad \lim_{\tau\rightarrow0}\tilde\chi = \sigma_2,
 \end{equation}
where
\begin{align}\label{Eq:s1s2def}
    \sigma_1 = \sqrt{\frac{C_{\eta}}{C_{\tau}}}, \quad 
\sigma_2 = \sqrt{\frac{\tilde{C}_\eta}{\tilde{C}_\tau}}.
\end{align}
See \cite{Denicol:2017lxn} for similar analytic behavior of the attractor.
{In Fig.~\ref{fig:attractor} we show the evolution of the anisotropies in the different sectors and the total system for one particular attractor solution
and neighboring trajectories (from \cite{Mitra:2020mei}, where more plots are given for this particular solution).}

In order to determine the attractor curves on the attractor surface
we proceed as follows. We initialize at a time $\tau_1$ and choose our initial energy densities and non-equilibrium pressures, $(\e, \et, \phi, \tilde{\phi})(\tau_1)$. We then solve the coupling equations \eqref{coup1}-\eqref{coup6} at the initial time to determine the metric components at the initial time. Subsequently, we evolve the system backwards in time via \eqref{exp-ward}, \eqref{exp-wardt}, \eqref{exp-mis} and \eqref{exp-mist}, and tune the values of $(\phi,\tilde{\phi})(\tau_1)$ with fixed $(\e, \et)(\tau_1)$ such that $\chi$ and $\tilde{\chi}$ tend to constants for early times following \eqref{Eq:ChChtET}. Having thus found an attractor curve to a satisfying level of precision (in this paper, $10^{-15})$, we then evolve the system forward in time to study the late time approach to hydrodynamics.

With parameters chosen as in \eqref{parameters} and \eqref{parameters2}, we have $\sigma_1 =\sqrt{1/5}$ and $\sigma_2 =\sqrt{1/(2(2 - \ln 2))}$. 
For illustrative purposes, we will discuss one approximate attractor solution, which can be found by choosing {$\tau_1=\gamma^{1/4}$} and
\begin{align}
   \gamma \e(\tau_1) &= \gamma\et(\tau_1) = 0.18, \nonumber \\\gamma\phi_1 (\tau_1) &= 0.10002105 \quad \text{and}\quad \gamma\phi_2 (\tau_1) = 0.0501539602,
\end{align} and we evolve the system backward  in time.
\begin{figure}\centering
\includegraphics[width=0.44\linewidth]{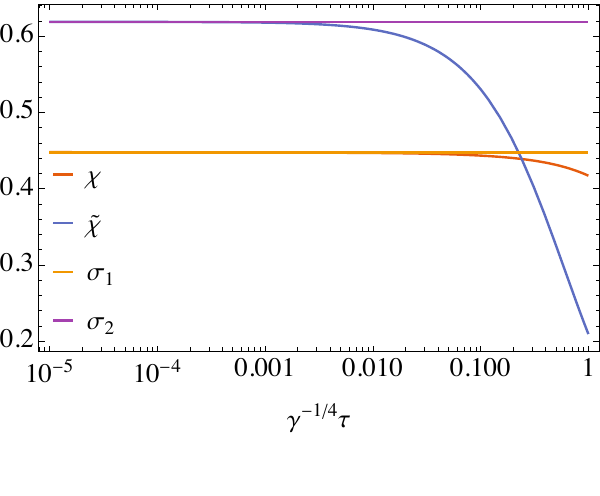}\\
\includegraphics[width=0.45\linewidth]{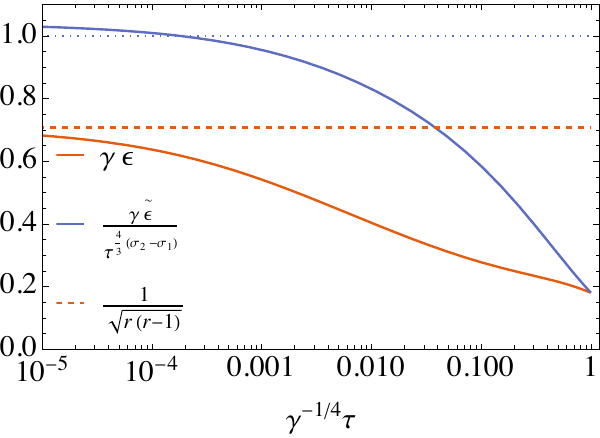}
\caption{{Illustrative behaviour of an attractor curve -- see text in Sec. \ref{sec:detatt} for the choice of parameters. Top: $\x$ and $\xt$ approach $\sigma_1$ and $\sigma_2$ respectively as $\tau\rightarrow0$. 
Bottom: $\e$ and $\et$ satisfy the scaling laws \eqref{Eq:scaling1att1} at early time. We also see that $\gamma \e$ tends towards $\sqrt{(r-1)/r}=1/\sqrt{2}$ (orange dotted line) as claimed in \eqref{Eq:scaling1att1}.} }\label{fig:epsilonplots1}
\end{figure}
We see that in the left panel of Fig.~\ref{fig:epsilonplots1} that with these initial conditions, $\x \rightarrow \sigma_1$ and $\xt\rightarrow\sigma_2$ for early times indicating that we are close to the attractor. Fig.~\ref{fig:epsilonplots1} further shows that the scalings \eqref{Eq:scaling1att1} for $\e$ and $\et$ (see below) come out to be as predicted in the limit $\tau\rightarrow0$.

\section{Mapping the attractor surface at early and late times}\label{sec:early}

\AR{Although in the application to Pb-Pb and $p$-$p$ collisions, only
times $\tau\gtrsim \gamma^{1/4}$ are relevant phenomenologically, we shall begin
by mapping out the attractor surface for arbitrary times.
Even when the early-time behavior of the attractor surface is an insufficient
approximation to a full semi-holographic treatment (let alone actual high-energy collisions), its constituent lines can be usefully parametrized by its parameters at some early reference time $\tau_0\ll \gamma^{1/4}$.
}

\subsection{Early time behavior: 
dominance
of the weakly coupled sector}
The attractor surface has a universal scaling behavior in the limit $\tau \to 0$ which is determined in terms of the two dimensionless viscous parameters $\sigma_1$ and $\sigma_2$ defined in \eqref{Eq:s1s2def}.

Crucially, we need
\begin{equation}\label{Eq:condsn}
    0 <\sigma_1 < 1 \quad {\rm and} \quad \frac{3}{4}\sigma_1 + \frac{1}{4} < \sigma_2 < 1.
\end{equation}
The relations $\sigma_{1,2} <1 $ are imposed by causality \cite{Romatschke:2017ejr}. The two relations above imply that $\sigma_2 > \sigma_1$ as should be the case if the second subsystem is more strongly coupled than the first. The stricter lower bound on $\sigma_2$ is necessary for reasons to be described below. 

We first note that with the above conditions, the anisotropies, $\chi$ and $\tilde{\chi}$, go to $\sigma_{1}$ and $\sigma_{2}$ respectively at $\tau = 0$ on the attractor surface, as we see in Fig.~\ref{fig:epsilonplots1}. Moreover, we find numerically that $\et$ goes to zero while $\e$ goes to a constant, namely $\gamma^{-1} \sqrt{r-1}/\sqrt{r}$. With these numerical inputs, assuming \eqref{Eq:condsn}, and by expanding the conservation, MIS and coupling equations in $\tau$ around $\tau =0$, we readily determine the following universal early time scaling behavior of the energy densities and the anisotropies:
 \begin{align}\label{Eq:scaling1att1}
\gamma \e &\rightarrow \sqrt{\frac{r-1}{r}}, \quad \quad
\gamma\et \rightarrow{\gtwo} \tau ^{\frac{4}{3} (\sigma_2 - \sigma_1)},\nonumber\\
\x &\rightarrow  \sigma_1 + k_1 \tau^{\frac{1}{3}(\sigma_1 +2)}, \quad \quad 
\xt \rightarrow  \sigma_2 + k_2 \tau^{\frac{5}{3}\sigma_2 - \frac{4}{3}\sigma_1 + \frac{2}{3}},
\end{align}
and the following scaling behaviors of the components of the effective metrics:
\begin{align}\label{Eq:scaling1att2}
\A  &\rightarrow  a_{10} \tau ^{\frac{1}{3}(\sigma_1 - 1)}, \quad
\B  \rightarrow  b_{10} \tau ^{\frac{1}{3}(\sigma_1 - 1)}, \quad
\C  \rightarrow  c_{10} \tau ^{\frac{1}{3}(\sigma_1 + 2)}, \nonumber\\
\At  &\rightarrow  a_{20} \tau ^{\frac{4}{3}\sigma_2 - \sigma_1 - \frac{1}{3}}, \quad
\Bt  \rightarrow  b_{20} \tau ^{\frac{1}{3}(\sigma_1 - 1)}, \quad
\Ct  \rightarrow  c_{20} \tau ^{\frac{1}{3}(\sigma_1 + 2)}.
\end{align}

\begin{figure}[ht]\centering{
\includegraphics[width=0.8\linewidth]{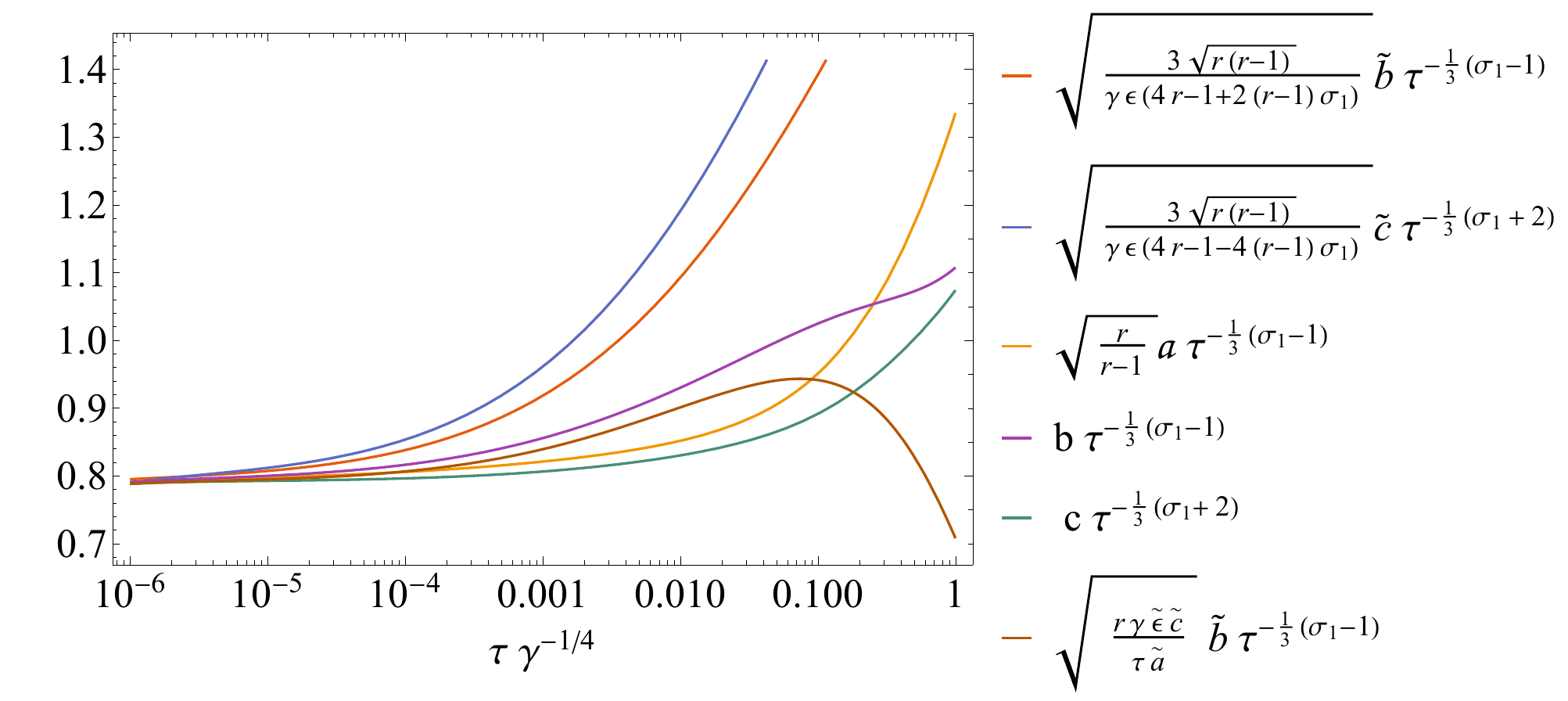}}
\caption{
Check of the identities \eqref{grand-identity} at early times. Similar identities given in \eqref{Eq:grand-identity2} also holds for the disperser surface, and can be verified to a very high accuracy numerically as evident from Fig. \ref{fig:abcplotsdisp}.}\label{fig:abcallplots1}
\end{figure}

Since the attractor surface is two-dimensional, there should be only two independent coefficients of the above early time expansions, which can be chosen to be $k_2$ and $g_2$. The latter determine the remaining coefficients as follows:
\begin{align}
    & b_{10} =  k_2 g_2 ^{-5/4}\gamma^{{1}/{4}} \tilde{C}_\tau \frac{ \sqrt{3}(13\sigma_2 -4\sigma_1+2)\sqrt{r}}
{ \sigma_2 (1 -4r + 2(1-r)\sigma_1)\sqrt{1-4r-4(1-r)\sigma_1}},\nonumber\\
    & a_{10} = \sqrt{\frac{r-1}{r}} b_{10},\nonumber\\
    & c_{10} = b_{10},\nonumber\\
    & a_{20} = - k_2 g_2^{-1/4} \gamma^{{1}/{4}} \tilde{C}_\tau\frac{13\sigma_2 -4\sigma_1+2}{3\sigma_2},\nonumber\\
    & b_{20} =   \sqrt{\frac{4r-1 + 2(r-1)\sigma_1}{3 {r(r-1)}}} b_{10},\nonumber\\
    & c_{20} =  \sqrt{\frac{4r-1 - 4(r-1)\sigma_1}{3 {r(r-1)}}}b_{10}, \nonumber\\
    & k_1 = - C_{\tau}^{-1} \gamma^{-\frac{1}{4}} r^{-5/8}(r-1)^{5/8}\frac{3\sigma_1}{2 +9\sigma_1}b_{10}.
\end{align}
As a result, in the limit $\tau\rightarrow0$, the following five identities should hold 
\begin{align}\label{grand-identity}
&\frac{\B}{\tau^{\frac{1}{3}(\sigma_1 -1)}} =\frac{\C}{\tau^{\frac{1}{3}(\sigma_1 +2)}}= \sqrt{\frac{r}{r-1}}\frac{\A}{\tau^{\frac{1}{3}(\sigma_1 -1)}} =\frac{\Bt}{\tau^{\frac{1}{3}(\sigma_1 -1)}} \sqrt{ r\gamma \et\frac{\Ct}{\tau \At}}\\\nonumber &=\sqrt{\frac{3\sqrt{r(r-1)}}{\gamma\e(4r-1-2(r-1)\x)}} \frac{\Bt}{\tau^{\frac{1}{3}(\sigma_1 -1)}} = \sqrt{\frac{3\sqrt{r(r-1)}}{\gamma\e(4r-1+4(r-1)\x)}} \frac{\Ct}{\tau^{\frac{1}{3}(\sigma_1 +2)}} 
\end{align}
on any curve on the attractor surface. Fig.~\ref{fig:abcallplots1} demonstrates the numerical check of the above identities on an attractor curve.

The stricter lower bound on $\sigma_2$ in Eq. \eqref{Eq:condsn} is necessary because the above scaling relations follow from solving the coupling equations, etc., with the assumption that $\At$ vanishes as $\tau \rightarrow 0$. From the explicit scaling of $\At$ in \eqref{Eq:scaling1att2}, it is clear that the lower bound is then self-consistent. Furthermore, it is also clear from \eqref{Eq:scaling1att1} that $\xt$ also does not diverge as $$\sigma_2 - \frac{4}{5}\sigma_1 + \frac{2}{5} >0 $$ as a consequence of the lower bound on $\sigma_2$ and $\sigma_1 <1$.

If instead of Eq. \eqref{Eq:condsn} we impose 
\begin{equation}\label{Eq:condsn2}
    0 <\sigma_2 < 1 \quad {\rm and} \quad {\rm max}\left(0,\frac{3}{4}\sigma_2 -\frac{1}{2}\right)< \sigma_1 < \sigma_2,
\end{equation}
we obtain that $\At$ diverges as $\tau^{(5/3)\sigma_1-(4/3)\sigma_2 -(1/3)}$ as $\tau\rightarrow 0$ and all other scaling relations discussed above are also altered. Nevertheless, $\x$ and $\xt$ still reaches $\sigma_1$ and $\sigma_2$ as $\tau\rightarrow 0$ on the attractor surface. However, the above condition involves a lower bound on $\sigma_1$ which cannot be motivated otherwise, while \eqref{Eq:condsn} can be readily satisfied with reasonable weak coupling and strong coupling parameters. Therefore, we do not further investigate the above condition.

It is easy to see from the above that the physical energy densities of the two subsystems behave as follows at early time
\begin{equation}\label{eq:E1}
    \mathcal{E}_1 = \e \frac{a b^2 c}{\tau} \sim \tau^{\frac{4}{3}\sigma_1 - \frac{4}{3}},\quad \mathcal{E}_2 = \e \frac{\At \Bt^2 \Ct}{\tau} \sim \tau^{\frac{8}{3}\sigma_2 - \frac{4}{3} \sigma_1 -\frac{4}{3}},
\end{equation}
and therefore
\begin{equation}
  \frac{  \mathcal{E}_2}{\mathcal{E}_1}\sim \tau^{\frac{8}{3}(\sigma_2- \sigma_1)}.
\end{equation}
Since $\sigma_2> \sigma_1$, it follows that the energy density of weakly self-interacting sector is always dominant at sufficiently early times, leading to the universal feature of an energy distribution \textit{on the attractor surface}
\AR{which is reminiscent of} the so-called {bottom-up thermalization} scenario in heavy-ion collisions \cite{Baier:2000sb}.\footnote{We emphasize, however, that this is just a remarkable feature
of the hybrid attractor, and it does not have any physical implications for
the later application to heavy-ion or $p$-$p$ collisions in Sect.~\ref{sec:pheno-pp-pb}, 
where we identify the energy scale
$\gamma^{-1/4}$ with the saturation scale $Q_s$ and have initial
times $\tau Q_s\sim 1$ for the evolution of the system, where initial
data can be arbitrarily far off the attractor surface.}
Also note that $\mathcal{E}_1$ always diverges as $\tau\rightarrow 0$ since $\sigma_1<1$. However, $\mathcal{E}_2$ may or may not diverge in this limit, without contradicting any of the two conditions in \eqref{Eq:condsn}, depending on the choice of parameters. 

While all initial conditions evolve towards specific curves on the attractor surface, they also get repelled from a \textit{disperser surface} in which $\x$ and $\xt$ approach $-\sigma_1$ and $-\sigma_2$ respectively in the limit $\tau \rightarrow 0$. On the disperser surface, the opposite of the energy distribution in
bottom-up thermalization takes place, since both $\mathcal{E}_2$ and $\mathcal{E}_2/\mathcal{E}_1$ diverge as $\tau\rightarrow 0$. In contrast to \eqref{Eq:scaling1att1}, $\gamma\et$  goes to the constant $\sqrt{r-1}/\sqrt{r}$ while $\gamma \e$ vanishes as $\tau\rightarrow 0$ on the disperser surface. For more details, see  Appendix~\ref{sec:double-disp}. One can also find two other such \textit{half-disperser surfaces}, in which $\x$ and $\xt$ approach $\sigma_1$ and $-\sigma_2$ respectively, or $\x$ and $\xt$ approach $-\sigma_1$ and $\sigma_2$ respectively as $\tau\rightarrow 0$. These can be analyzed in a similar manner. All these three surfaces are not relevant for understanding hydrodynamization with generic initial conditions.

\subsection{Initialization and bounds on initial conditions}

We initialize the evolution at $\tau_0 >0$ on the two-dimensional attractor surface. Therefore, the two parameters $k_2$ and $g_2$ determining the early time expansions about $\tau =0$ discussed above are not convenient. Instead we choose $\gamma\e(\tau_0)$ and $\gamma\et(\tau_0)$, or rather $\gamma\e(\tau_0)$ and the ratio $\e(\tau_0)/\et(\tau_0)$ as our parameters. Clearly $\phi(\tau_0)$ and $\tilde{\phi}(\tau_0)$ should be fine-tuned such that $\x$ and $\xt$ reach $\sigma_1$ and $\sigma_2$ as $\tau \rightarrow 0$.

Crucially, $\gamma\e(\tau_0)$ is bounded from above for any choice of $\e(\tau_0)/\et(\tau_0)$, as otherwise we do not obtain 
solutions. However, the 
energy densities and the total energy density can be arbitrarily large. This is similar to the equilibrium solutions studied in \cite{Kurkela:2018dku} where $\e$ and $\et$ were bounded from above, but the total energy density and pressure were unbounded. 

\begin{figure}
    \centering
    \includegraphics[width=0.48\textwidth]{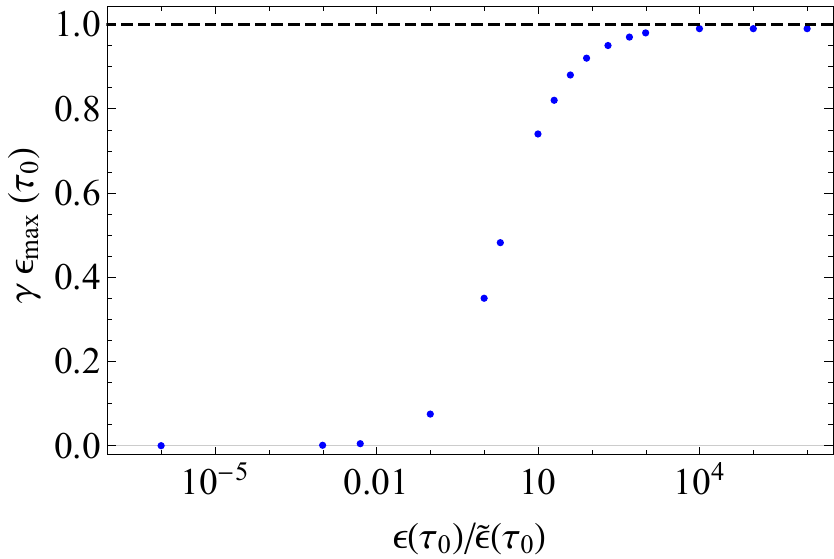}
    \includegraphics[width=0.48\textwidth]{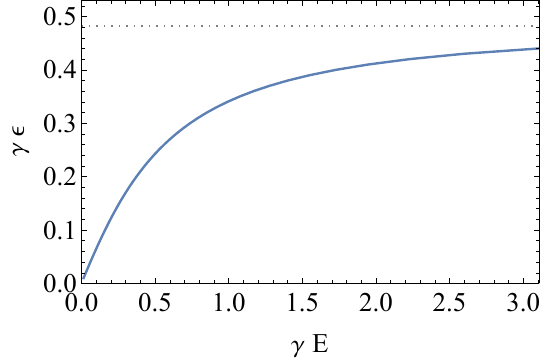}
    \caption{Left: Plot of upper bound on $\gamma\e(\tau_0)$ for different initial ratios $\e(\tau_0)/\et(\tau_0)$. The parameters used are found in \eqref{parameters} and \eqref{parameters2}, and $\tau_0=10^{-3}\gamma^{1/4}$. Right: $\gamma\e(\tau_0)$ vs the total energy density, for the ratio $\varepsilon/\et =2$ at $\tau_0$. The subsystem energy,  $\gamma\e(\tau_0)$, saturates to its upper bound of $\sim 0.482$ for large total energy density, $\gamma E$.  
    } 
    \label{fig:upper-bound-plot}
\end{figure}

As shown on the left panel of Fig. \ref{fig:upper-bound-plot}, the upper bound on $\gamma\e(\tau_0)$ is a function of the ratio $\e(\tau_0)/\et(\tau_0)$ and it goes to $1$ as $\e(\tau_0)/\et(\tau_0)\rightarrow \infty$. As shown on the right panel of Fig. \ref{fig:upper-bound-plot}, the total energy density diverges as $\gamma\e(\tau_0)$ reaches its upper bound for any fixed ratio $\e(\tau_0)/\et(\tau_0)$.

To initialize on the attractor surface and find the upper bound on $\gamma\e(\tau_0)$ at the \ARnew{reference} time $\tau_0=10^{-3}\gamma^{1/4}$ we can proceed as follows. For each $\gamma\e(\tau_0)$ and $\e(\tau_0)/\et(\tau_0)$, we fix 
the values of $\phi$ and $\tilde{\phi}$ by setting $\x =\sigma_1$ and $\xt = \sigma_2$ at $\tau = \tau_0$. Then with chosen precision (approx.\ $10^{-3}$) we determine the upper limit of $\gamma\e(\tau_0)$ such that we obtain 
solutions. We then tune $\phi$ and  $\tilde{\phi}$  
again until we find that $\x =\sigma_1$ and $\xt = \sigma_2$ at $\tau=10^{-6}\gamma^{1/4}$ up to a chosen precision -- this implies that our evolution is closer to the attractor surface. We also recompute the upper bound on $\gamma\e(\tau_0)$ for each $\e(\tau_0)/\et(\tau_0)$. The new $\phi$ and $\tilde\phi$ for each $\gamma\e(\tau_0)$ and $\e(\tau_0)/\et(\tau_0)$, and also the new upper bound on $\gamma\e(\tau_0)$ for each $\e(\tau_0)/\et(\tau_0)$ differ from the previous ones at a subpercent level giving estimate of our errors.

We can systematically correct this error by iterative fine-tuning. At each step of the iteration, we fine tune $\phi$ and  $\tilde{\phi}$ for each $\gamma\e(\tau_0)$ and $\e(\tau_0)/\et(\tau_0)$ such that $\x =\sigma_1$ and $\xt = \sigma_2$ at progressively earlier times. Practically, the very first step of iteration gives a very precise estimate of initial values of $\phi$ and $\tilde{\phi}$ on the attractor surface for each $\gamma\e(\tau_0)$ and $\e(\tau_0)/\et(\tau_0)$, and also the upper bound on $\gamma\e(\tau_0)$ for each $\e(\tau_0)/\et(\tau_0)$ if we initialize at $\tau_0=10^{-3}\gamma^{1/4}$ with our choices of parameters. We also find that the estimates of hydrodynamization times and other results reported below change only sub-percent if we correct for our deviation from the attractor surface systematically in the iterative procedure.

\subsection{Hydrodynamic expansion}
At late time, the full system can be mapped to a hydrodynamic derivative expansion with powers of $\tau^{-2/3}$ counting number of derivatives. The hydrodynamic expansion is determined only by two dimensionless parameters, which could be chosen for instance as
\begin{equation}
    \lim_{\tau\rightarrow\infty}\gamma^{2/3} \e \tau^{4/3} \quad {\rm and}\quad\lim_{\tau\rightarrow\infty}\gamma^{2/3} \et \tau^{4/3}.
\end{equation}
(For reasons to be mentioned below, we will choose a different pair of parameters.) Each pair of these parameters determines a unique curve on the attractor surface. Thus any initial condition in the four-dimensional phase space evolves to a specific curve on the attractor surface labelled by a specific pair of parameters determining the late-time hydrodynamic expansion. On the attractor surface itself, the pair of parameters $k_2$ and $g_2$ determining the early time expansion should relate to the pair of parameters $\gamma\e(\tau_0)$ and $\e(\tau_0)/\et(\tau_0)$ giving the initialization at time $\tau_0$. In turn the latter pair of parameters should match uniquely to a pair of parameters giving the late-time hydrodynamic expansion. We postpone the discussion of the latter matching and turn now to the hydrodynamic expansion itself.

The entire hydrodynamic expansion can be expressed in terms of the perfect fluid components which give the leading $\tau$ dependence of $\e$ and $\et$ at late time as follows
\begin{equation}
\e = \epsilon_{1,pf} + \ldots,\quad \et = \epsilon_{2,pf} + \ldots,\quad \epsilon_{1,pf} = \epsilon_{10}\left(\frac{\tau}{\tau_0}\right)^{-4/3} , \quad \epsilon_{2,pf} = \epsilon_{20}\left(\frac{\tau}{\tau_0}\right)^{-4/3},
\end{equation}
with $\cdots$ denoting viscous corrections and $\tau_0$ is an arbitrary moment of time. Note that at late time the two fluids decouple. The expansion dilutes the energy densities and therefore the mutual coupling via the effective metrics. As a result, the leading order perfect fluid expansions assume the usual forms for both systems. It is clear that the choice of $\tau_0$ above is arbitrary. A different choice $\tildetauzero$ simply rescales $\epsilon_{10}$ and $\epsilon_{20}$ as follows:
\begin{equation}
\epsilon_{10} = \tilde\epsilon_{10}(\tildetauzero/\tau_0)^{4/3}, \quad \epsilon_{20} = \tilde\epsilon_{20}(\tildetauzero/\tau_0)^{4/3}.
\end{equation}

We want to label hydrodynamic curves by variables which are dimensionless and invariant under reparametrizations $\tau_0 \rightarrow \tildetauzero$.
It is straightforward to see that the relevant variables are
\begin{align}\label{alpha}
\alpha := & \gamma(\epsilon_{20}\tau_0^{4/3}) \sqrt{\epsilon_{10} \tau_0^{4/3}} = \gamma \epsilon_{20}\sqrt{\epsilon_{10}} \tau_0^{2}, \\
\beta := & \gamma(\epsilon_{10}\tau_0^{4/3}) \sqrt{\epsilon_{20} \tau_0^{4/3}} = \gamma \epsilon_{10}\sqrt{\epsilon_{20}} \tau_0^{2}. \label{beta}
\end{align}
These two parameters uniquely label each attractor (hydrodynamic) curve as shown below. It is also useful to define
\begin{equation}
\mathfrak{r}:= (\epsilon_{20}/\epsilon_{10})^{1/4} =\sqrt{\alpha/\beta} = \lim_{\tau\rightarrow\infty}{(\tilde{\e}/\e)^{1/4}}.
\end{equation}
For the sake of convenience, we define dimensionless time:
\begin{equation}
\omega_1 := \epsilon_{1,pf}^{1/4}\tau, \quad \omega_2 := \epsilon_{2,pf}^{1/4}\tau.
\end{equation}
Note however that $\omega_1$ and $\omega_2$ are not independent time variables. In fact
\begin{equation}
\omega_2 = \mathfrak{r}\, \omega_1.
\end{equation}
The hydrodynamic expansion in $\tau$, obtained by solving all coupling equations, the two MIS equations and the two conservation equations, can be readily stated in terms of these variables as follows
\begin{align}
\e^{1/4}\tau  &= \omega_1 - \frac{2}{3} \cet  + \frac{3 \alpha-4 \cet \ctp }{18} \frac{1}{\omega_1} + \mathcal{O}(\omega_1^{-2}),  \label{invert}\\
\x &= \frac{4\cet }{3\omega_1} +  \frac{8\cet \ctp }{9\omega_1^2} +  \mathcal{O}(\omega_1^{-3}) ,\\
\et^{1/4}\tau  &= \omega_2 - \frac{2}{3} \cett  + \frac{3 \beta-4 \cett \ctpt }{18} \frac{1}{\omega_2} + \mathcal{O}(\omega_2^{-2}), \\
\xt &= \frac{4\cett }{3\omega_2} +  \frac{8\cett \ctpt }{9\omega_2^2} +  \mathcal{O}(\omega_2^{-3}),
\end{align}
The above equations show all terms up to second order in derivatives. Defining 
\begin{equation}
w_1:= \e^{1/4}\tau, \quad w_2:= \et^{1/4}\tau 
\end{equation}
we can invert \eqref{invert}
\begin{equation}
\omega_1 = w_1 + \frac{2}{3} \cet  - \frac{3 \alpha-4 \cet \ctp }{18} \frac{1}{w_1} + \mathcal{O}(\omega_1^{-2}).
\end{equation}
In terms of $w_1$ we can readily express the hydrodynamic expansion as a phase space curve
\begin{align}
\x &= \frac{4\cet }{3w_1} +  \frac{8\cet \ctp }{9w_1^2} +  \mathcal{O}(w_1^{-3}) ,\\
w_2  
&=\sqrt{ \frac{\alpha}{\beta}}w_1 - \frac{2}{3} \left(\cett - \sqrt{ \frac{\alpha}{\beta}} \cet \right)\\&+\left(\sqrt{ \frac{\beta}{\alpha}}\frac{3\beta-4 \cett \ctpt -24 \cett^2}{18} - \sqrt{ \frac{\alpha}{\beta}}\frac{3 \alpha- \cet \ctp -24 \cet ^2}{18} \right)\frac{1}{ w_1} + \mathcal{O}(w_1^{-2}), \nonumber \\
\xt
&=\sqrt{ \frac{\beta}{\alpha}} \frac{4\cett }{3 w_1} +  \frac{\beta}{\alpha}\frac{8\cett \left(\cett +\cet \sqrt{ \frac{\beta}{\alpha}}+\ctpt \right)}{9 w_1^2} +  \mathcal{O}(w_1^{-3}).
\end{align}
With $w_1$, $w_2$, $\x$ and $\xt$ as the phase space variables, the phase space curves representing the hydrodynamic expansion (or the attractor) are then given by the above expressions, $\x(w_1)$, $w_2(w_1)$ and $\xt(w_1)$. As manifest from above, each curve is labelled by two dimensionless, time-reparametrisation invariant parameters, $\alpha$  and $\beta$.  
 
The hydrodynamic late-time expansion of the components of the effective metrics are:
 \begin{align}
  \A &= 1 - \frac{1}{2}\alpha\frac{1}{\omega_1^2} + \frac{4}{3}\cett \sqrt{\alpha\beta}\frac{1}{\omega_1^3}+ \mathcal{O}(\omega_1^{-4}), \\
 \B &= 1 + \frac{1}{6}\alpha\frac{1}{\omega_1^2} + \mathcal{O}(\omega_1^{-4}),\nonumber\\ 
 \C &= 1 + \frac{1}{6}\alpha\frac{1}{\omega_1^2} -\frac{4}{3}\cett \sqrt{\alpha\beta}\frac{1}{\omega_1^3}+ \mathcal{O}(w_1^{-4}),\\
 \At &= 1 - \frac{1}{2}\frac{\beta^2}{\alpha}\frac{1}{\omega_1^2}+ \frac{4}{3}\cet \frac{\beta^2}{\alpha}\frac{1}{\omega_1^3} + \mathcal{O}(\omega_1^{-4}), \\
 \Bt &= 1 + \frac{1}{6}\frac{\beta^2}{\alpha}\frac{1}{\omega_1^2} +\mathcal{O}(\omega_1^{-4}),\nonumber\\ 
\Ct &= 1 + \frac{1}{6}\frac{\beta^2}{\alpha}\frac{1}{\omega_1^2}-\frac{4}{3}\cet \frac{\beta^2}{\alpha}\frac{1}{\omega_1^3}  + \mathcal{O}(\omega_1^{-4}).
 \end{align}
Remarkably, up to $\mathcal{O}(w_1^{-4})$ the volume factors are:
\begin{equation}
\A \B^2 \C = 1 + \mathcal{O}(w_1^{-4}), \quad \At \Bt^2 \Ct= 1 + \mathcal{O}(w_1^{-4}).
\end{equation}

We can readily compute the full system energy density which behaves as
\begin{equation}\label{Eq:wtotal}
\mathcal{E}^{1/4}\tau = \omega -\frac{2}{3}\frac{C_{\eta_1}\epsilon_{10}^{3/4}+C_{\eta_2}\epsilon_{20}^{3/4}}{(\epsilon_{10}+\epsilon_{20} )^{3/4}}+ \mathcal{O}(\omega^{-1}),
\end{equation}
where
\begin{equation}
\omega := \mathcal{E}_{pf}^{1/4}\tau = (\epsilon_{1,pf}+\epsilon_{2,pf} )^{1/4} \tau.
\end{equation}
We find that the full energy-momentum tensor can be mapped to the standard hydrodynamic Bjorken flow expansion (in the usual Milne metric) of a single fluid, but with effective equation of state and effective transport coefficients dependent on the parameters $\alpha$ and $\beta$.

We can explicitly map the full system to a \textit{single} expanding fluid at late time and compare to the above discussion. For a single fluid, the evolution of the energy density in the Milne background metric takes the generic form (up to first order)
\begin{equation}\label{Bjorken-gen-eqn}
\mathcal{E}' = -\frac{1}{\tau}\left( \mathcal{E} + \mathcal{P}(\mathcal{E})\right) +\frac{1}{\tau^2}\left(\zeta(\mathcal{E}) + \frac{2}{3} \eta(\mathcal{E})\right),
\end{equation}
where $\zeta$ and $\eta$ are the bulk and shear viscosity, respectively.
Furthermore, the trace of the (first-order) hydrodynamic energy-momentum tensor takes the form
\begin{equation}\label{Bjorken-trT}
T^\mu_{\,\mu} = 3 \mathcal{P}(\mathcal{E}) -\mathcal{E}  - \frac{3}{\tau} \zeta(\mathcal{E}).
\end{equation}
We can assume that
\begin{align}\label{fullP-form}
\mathcal{P}(\mathcal{E}) &= \frac{\mathcal{E}}{3} + k \mathcal{E}^2 + \cdots\\
\eta(\mathcal{E}) &= \kappa \mathcal{E}^{3/4} + \cdots, \quad \zeta(\mathcal{E}) = \tildekappa \mathcal{E}^{3/4} + \cdots.
\end{align}
with $k$,  $ \kappa $ and $\tildekappa$ constants. Solving \eqref{Bjorken-gen-eqn} with inputs from \eqref{fullP-form} in the large proper time expansion, we obtain 
\begin{align}\label{Eq:EtotalSingle}
\mathcal{E} &= \mathcal{E}_0 \left(\frac{\tau_0}{\tau}\right)^{4/3} - \frac{\tau_0}{2\tau^2} (2 \kappa + 3 \tilde{\kappa})\mathcal{E}_0^{3/4}
\nonumber\\ & +\frac{3}{32 \tau^{8/3}}\left( \mathcal{E}_0^{1/2}\tau_0^{2/3}(4\kappa^2 + 12 \kappa\tildekappa+9\tildekappa^2) - 8\mathcal{E}_0^{2}\tau_0^{8/3}k\right) + \cdots,
\end{align}
and also 
\eqref{Bjorken-trT} 
\begin{equation}\label{Tr-T-1}
{\rm Tr}(T) = - \frac{3}{\tau^2} \mathcal{E}_0^{3/4} \tau_0 \tildekappa + \frac{3}{8\tau^{8/3}}\left( \mathcal{E}_0^{1/2}\tau_0^{2/3}(6\kappa\tildekappa+9\tildekappa^2) + 8\mathcal{E}_0^{2}\tau_0^{8/3}k\right)+ \cdots.
\end{equation}

We can determine the constants $k$, $ \kappa $ and $\tilde{\kappa}$ simply by matching the above expansions to those obtained by solving our system of equations. We readily see that these constants depend on the specific curve on the attractor surface. Comparing Eq.\ \eqref{Eq:EtotalSingle} with Eq. \eqref{Eq:wtotal}, we explicitly find that the effective viscosity of the full system is
\begin{equation}\label{eff-shear}
C_{\eta, eff} = \frac{\cet \epsilon_{10}^{3/4}+\cett\epsilon_{20}^{3/4}}{(\epsilon_{10}+\epsilon_{20} )^{3/4}} = \frac{\cet\beta^{3/4}+\cett\alpha^{3/4}}{(\alpha^2+\beta^2 )^{3/4}}
\end{equation}
which depends manifestly on which curve on the attractor surface the full system evolves to at late time.In fact $ \kappa $ can be obtained from \eqref{eff-shear}:
\begin{equation}\label{kappa}
\kappa = \frac{4}{3}C_{\eta, eff} 
\end{equation}
Similarly, the hydrodynamic expansion of our hybrid system yields that
\begin{equation}\label{Tr-T-2}
{\rm Tr}(T) = - \frac{8\tau_0^{8/3}}{3\tau^{8/3}}\gamma \epsilon_{10}\epsilon_{20}+ \cdots.
\end{equation}
We can immediately conclude that
\begin{itemize}
    \item $\mathcal{E}_0 = \epsilon_{10}+ \epsilon_{20}.$
\item $k = - \frac{1}{3} \gamma \frac{\epsilon_{10}\epsilon_{20}}{\mathcal{E}_0^2} = - \frac{1}{3} \gamma \frac{\epsilon_{10}\epsilon_{20}}{(\epsilon_{10}+ \epsilon_{20})^2} = - \frac{1}{3} \gamma \frac{\alpha^2\beta^2}{(\alpha^2+ \beta^2)^2},$
\item and $\tildekappa = 0,$
\end{itemize}
i.e. the effective bulk viscosity vanishes since the $\tau^{-2}$ term vanishes in \eqref{Tr-T-2}.

Since $k$ and $\kappa$ both depend on $\alpha$ and $\beta$, it's clear that the effective equation of state and the transport coefficients of the single fluid description of the full system depends on the specific curve on the attractor surface.

\begin{figure}
\centering
\begin{minipage}{\linewidth}
\includegraphics[width=0.5\linewidth]{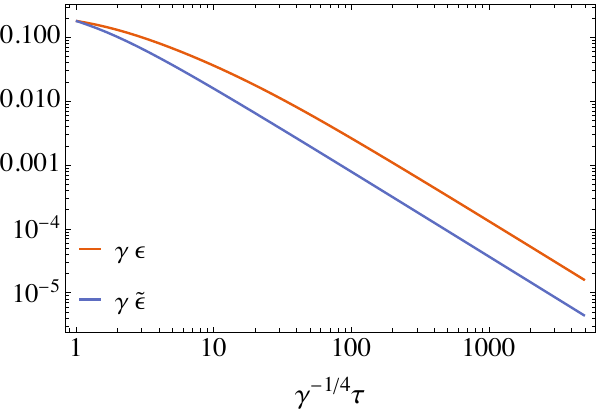}
\includegraphics[width=0.5\linewidth]{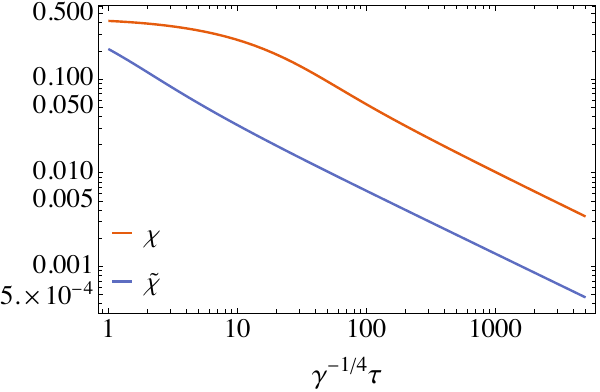}
\end{minipage}
\caption{\AS{The hydrodynamic tails for the attractor with $(\e,\et, {\phi},{\tilde{\phi}})(\tau_0)$ given by $(0.18, 0.18, 0.1000210481, 0.05015396018505)$  {in units where $\gamma=1$}.  Left: The energy densities of both sectors attain the slope of $-4/3$ showing the onset of perfect fluid flow. Right: The anisotropies of both sectors attain the slope of $-2/3$ showing the onset of first order hydrodynamics.
} }\label{fig:epsilonplotsfull2}
\end{figure}

In order to interpolate to hydrodynamics, it is useful to do a {doubly logarithmic} plot for $\x$, $\xt$, $\e$ and $\et$. 
Perfect hydrodynamics begins when the logarithmic plots of $\e(\tau)$ and $\et(\tau)$ become parallel to each other with slope $-4/3$, i.e. when they both reach the perfect fluid hydrodynamic tail $1/\tau^{4/3}$, which is evident in Fig. \ref{fig:epsilonplotsfull2} for the first-order hydrodynamic tails of $\e$ and $\et$ for the first and second attractors, respectively. 

At late time
\begin{equation}
\ln \e \approx - (4/3) \ln \tau + \ln (\epsilon_{10}\tau_0^{4/3}),\quad \ln \et \approx - (4/3) \ln \tau + \ln (\epsilon_{20}\tau_0^{4/3}).
\end{equation}
Therefore the intercepts of the energy densities give us the time-reparametrisation invariant $\epsilon_{10}\tau_0^{4/3}$ and $\epsilon_{10}\tau_0^{4/3}$ with the slopes fitting well to $-4/3$. We obtain for the attractor
\begin{equation}
\epsilon_{10}\tau_0^{4/3} \approx 1.346, \, \,\epsilon_{20}\tau_0^{4/3} \approx 0.370, \,\, \alpha = 0.424,\,\,\beta = 0.801.
\end{equation}
First-order hydrodynamics begins when the logarithmic plots of $\x(\tau)$ and $\xt(\tau)$ become parallel to each other with slope $-2$, i.e. when they both reach the hydrodynamic tail $1/\tau^2$ given by first-order hydrodynamics (note in the perfect fluid limit $\chi$ vanishes). This is shown in Fig.~\ref{fig:epsilonplotsfull2} for the first-order hydrodynamic tails of $\x$ and $\xt$ for the first and second attractors, respectively. 

We can also match using leading order hydrodynamic expansion of {$\chi$ ($\tilde{\chi}$)}:
\begin{align}
\ln {\chi} \approx - 2 \ln \tau + \ln\frac{16\cet }{9} +3/4\ln (\epsilon_{10}\tau_0^{4/3}),\nonumber\\ \ln {\tilde{\chi}} \approx - 2 \ln \tau + \ln \frac{16\cett }{9}  +3/4\ln (\epsilon_{20}\tau_0^{4/3}).
\end{align}
As a consistency check, we see that the intercepts reproduce $\epsilon_{10}\tau_0^{4/3}$ and $\epsilon_{20}\tau_0^{4/3}$ to a high accuracy.

\ASnew{In Appendix~\ref{app:borel} we illustrate how to understand the hydrodynamic gradient expansion and see how its Borel resummation can describe the attractor surface in terms of the parameters $\alpha$ and $\beta$, defined in \eqref{alpha} and \eqref{beta}.}

\subsection{Late time approach to attractor: fluctuation analysis}\label{sec:late}\label{sec:fluc}

The approach to the attractor at late time can be simply understood by perturbing around the late-time hydrodynamic expansion computed above. 

The leading order linear fluctuations of energy densities, anisotropies and effective metric variables about the hydrodynamic expansion, i.e. the attractor solution at late time are defined as follows:
\begin{align}
\e(\tc) &={\epsilon_{10}} {\tc}^{-4/3} +  {\tc}^{-4/3} \left( {\epsilon_{11}}\tc^{-2 /3} + {\epsilon_{12}}\tc^{-4/3} \right) + \mathcal{O}\left(\tc^{-10/3}\right)+ \de (\tc),\\
\phi(\tc) &= f_{10} + \left( f_{11} \tc^{-2 /3} + f_{12} \tc^{-4/3} + f_{13} \tc^{-2} + + f_{14} \tc^{-8/3} \right)+ \mathcal{O}\left(\tc^{-10/3}\right) + \dph(\tc) ,\\
a(\tc) &= 1+ \tc^{-2/3} \left( A_{11} \tc^{-2 /3} + A_{12} \tc^{-4 /3}  \right)+ \mathcal{O}\left(\tc^{-10/3}\right) + \delta a (\tc) ,\\
b(\tc) &= 1 + \tc^{-2/3} \left( B_{11} \tc^{-2 /3} + B_{12} \tc^{-4 /3}  \right)+ \mathcal{O}\left(\tc^{-10/3}\right) + \delta b (\tc),\\
c(\tc) &= \tau_0 \tc + {\tc}^{1/3} \left( C_{11}  \tc^{-2 /3}  +  C_{12}  \tc^{-4/3} \right) + \mathcal{O}\left(\tc^{-10/3}\right)+ \delta c(\tc),
\end{align}
and similarly for the tilded sector. Above $\tc:=\tau/\tau_0$ and $\tde (\tc)$, $\dph(\tc)$, etc. denote the fluctuations about the hydrodynamic expansion. We have truncated to the first few terms of the hydrodynamic expansions which can be found explicitly in Appendix~\ref{app:grad-exp}, because only these are sufficient to obtain the leading order behavior of the fluctuations. The fluctuations have two independent non-hydrodynamic modes capturing the missing information of the initial conditions in the four-dimensional phase space, and two hydrodynamic modes which span the two-dimensional attractor surface. Since the non-hydrodynamic modes, which do not admit a power series expansion $\tc^{-2/3}$, tell us how a generic evolution relaxes towards the attractor surface, we will focus on finding these explicitly here.

One can eliminate the fluctuations in six effective metric variables in terms of other four fluctuations ($\de(\tc)$, $\tde(\tc)$,  $\dph(\tc)$, $\dpht(\tc)$) by using six metric coupling equations. This leaves us with four first-order coupled ordinary differential equations in four variables, which are given in Appendix~\ref{app:grad-exp}. One can  further solve these equations until one is left with a pair of decoupled fourth-order differential equations for the anisotropy fluctuations.
Depending on the relative values of parameters, the fluctuations, $\dph(\tc)$ and $\dpht(\tc)$, have different leading order forms, which we detail below.

In the case when $\ctp \: {\epsilon_{20}}^{1/4} \neq \ctpt \: {\epsilon_{10}}^{1/4}$,
the two independent non-hydrodynamic solutions (at leading order) are:
\begin{align}\label{sol-phi1}
\dph(\tc) &=\coef \; e^{-\frac{3 {\epsilon_{10}}^{1/4} \; \tau_0}{2 \ctp}  \tc^{2/3}}  \tc^{-\frac{4}{3} + 
\frac{2 \cet}{3 \ctp}} + \tcoef \; e^{-\frac{3 {\epsilon_{20}}^{1/4} \; \tau_0}{2 \ctpt}  \tc^{2/3}}  \tc^{  
-\frac{10}{3} + 
\frac{2 \cett}{3 \ctpt}},\\
\label{sol-phi2}
\dpht(\tc) &= \coefp \; e^{-\frac{3 {\epsilon_{10}}^{1/4} \; \tau_0}{2 \ctp}  \tc^{2/3}} \tc^{  
-\frac{10}{3} + 
\frac{2 \cet}{3 \ctp}}   + \tcoefp \; e^{-\frac{3{\epsilon_{20}}^{1/4} \; \tau_0}{2 C_{\tau_2}}  \tc^{2/3}}  \tc^{-\frac{4}{3} + 
\frac{2 \cett}{3 \ctpt}}
\end{align}
where $\coef, \tcoef, \coefp, \tcoefp$ are constants, of which only two are independent: 
\begin{eqnarray}
\label{cp1-c1}
\coef &=& \coefp \frac{3 \cet \left( \ctpt \; {\epsilon_{10}}^{1/4} - \ctp \; {\epsilon_{20}}^{1/4} \right)}{2 \cett \ctp^2 {\epsilon_{20}}  \gamma}\tau_0 ,\nonumber\\
\label{cp2-c2}
\tcoefp &=& \tcoef \frac{3 \cett \left( - \ctpt \; {\epsilon_{10}}^{1/4} + \ctp \; {\epsilon_{20}}^{1/4} \right)}{2 \cet \ctpt^2  {\epsilon_{10}} \gamma}\tau_0.
\end{eqnarray}

If $\ctp \: {\epsilon_{20}}^{1/4} > \ctpt \: {\epsilon_{10}}^{1/4}$, i.e. if 
\begin{equation}\label{Eq:poss1}
    \mathfrak{r}=\sqrt{\alpha/\beta} > \ctpt/\ctp
\end{equation}
then 
\begin{equation}
\exp{\left(-\frac{3 {\epsilon_{10}}^{1/4} \; \tau_0}{2 \ctp}  \tc^{2/3}\right)} =\exp{\left(-\frac{3 {\epsilon_{10}}^{1/4} \; \tau_0^{1/3}}{2 \ctp}  \tau^{2/3}\right)} = \exp{\left(-\frac{3}{2 \ctp}\left(\frac{\beta}{\sqrt{\alpha\gamma}}\right)^{1/3}\tau^{2/3}\right)}
\end{equation} 
will be the dominant exponential and give the relaxation towards the attractor. The powers of $\tc$ accompanying this exponential in \eqref{sol-phi1} and \eqref{sol-phi2} imply that $\dpht (\tc)$ will decay faster towards the attractor surface than $\dph (\tc)$ in this case. This implies that the strongly coupled sector will hydrodynamize faster than the weakly coupled one.
If $\ctp \: {\epsilon_{20}}^{1/4} < \ctpt \: {\epsilon_{10}}^{1/4}$, i.e. if 
\begin{equation}\label{Eq:poss2}
   \mathfrak{r}=\sqrt{\alpha/\beta}< \ctpt/\ctp,
\end{equation}
then 
\begin{equation}
\exp{\left(-\frac{3 {\epsilon_{20}}^{1/4} \; \tau_0}{2 \ctpt}  \tc^{2/3}\right)} =\exp{\left(-\frac{3 {\epsilon_{20}}^{1/4} \; \tau_0^{1/3}}{2 \ctpt}  \tau^{2/3}\right)} = \exp{\left(-\frac{3}{2 \ctpt}\left(\frac{\alpha}{\sqrt{\beta\gamma}}\right)^{1/3}\tau^{2/3}\right)}
\end{equation} 
is the dominant mode. By an analogous argument, $\dph (\tc)$ will decay faster than $\dpht (\tc)$ implying that the weakly coupled sector will hydrodynamize faster than the strongly coupled one.

Full numerical solutions do confirm that indeed when \eqref{Eq:poss1} is realized, the strongly coupled sector hydrodynamizes faster and the reverse occurs when \eqref{Eq:poss2} holds. However, realizing \eqref{Eq:poss2} requires choosing large ratios of $\et(\tau_0)/\e(\tau_0)$. This is also unnatural because this requires extreme fine-tuning at early time, because of the bottom-up behavior on the attractor surface in which $\e$ goes to a constant while $\et$ vanishes as $\tau\rightarrow 0$. 

In the case when $
\ctp \; {\epsilon_{20}}^{1/4} = \ctpt \; {\epsilon_{10}}^{1/4}$, i.e. 
\begin{equation}
    \mathfrak{r}=\sqrt{\alpha/\beta} = \ctpt/\ctp ,
\end{equation}
the two independent solutions to leading order are: 
\begin{align}\label{sol-phi1-case2}
\dph(\tc) &= e^{-\frac{3 {\epsilon_{10}}^{1/4} \tau_0}{2 \ctp}  \tc^{2/3}} \left(   \coef  \; \tc^{-\frac{4}{3} + \frac{2 \cet}{3 \ctp}}   +  \tcoef \; \tc^{-\frac{8}{3}  + \frac{2 \cett}{3 \ctpt} }  \right),\\
\label{sol-phi2-case2}
\dpht(\tc) &= e^{-\frac{3 {\epsilon_{10}}^{1/4} \tau_0}{2 \ctp}  \tc^{2/3}} \left(   \coefp  \;  \tc^{-\frac{8}{3} + \frac{2 \cet}{3 \ctp}}   + \tcoefp \;  \tc^{- \frac{4}{3}  + \frac{2 \cett}{3 \ctpt} }  \right),
\end{align}
where $\coef $, $ \tcoef$, $\coefp$, and $\tcoefp$ are arbitrary constants only two of which are independent: 
\begin{align}
\coef = - \frac{\cet \ctp^2 \left( - \cett \ctp + \cet \ctpt - 2 \ctp \ctpt  \right)}{\cett \ctpt^4  {\epsilon_{10}}  \gamma} \coefp ,\nonumber\\
\tcoefp = - \frac{\cett  \left( \cett \ctp - \cet \ctpt  - 2 \ctp \ctpt  \right)}{\cet  \ctpt ^4  {\epsilon_{10}}  \gamma}\tcoef.
\end{align}

\subsection{Matching with late times
}
Now we are prepared to study how parameters setting initial conditions on the Bjorken flow attractor surface map to $\alpha$ and $\beta$, which are defined in \eqref{alpha} and \eqref{beta}, and which determine the hydrodynamic expansion at late time. In particular, we find that they scale as a power of the total energy density on the attractor surface. For $r=2$, the scaling of $\alpha$ and $\beta$ is found to be
\begin{align}
    \alpha&\sim (\gamma \mathcal{E}(\tau = \gamma^{1/4}))^{3/2},\\
    \beta&\sim (\gamma \mathcal{E}(\tau = \gamma^{1/4}))^{3/2},
    \end{align}
where $ \mathcal{E}$ is the total initial energy density. Using the following definitions of the total entropy density and the individual subsystem entropy densities, {which are valid when both subsystems reach the perfect fluid limit:}
\begin{align}
    S_{tot} &=s  +\tilde{s},\label{Eq:Seq}\\
    s&=\frac{4}{3 \tau} \e^{3/4} \B^2 \C,\label{Eq:seq}\\
    \tilde{s}&=\frac{4}{3 \tau}  \et^{3/4} \Bt^2 \Ct,\label{Eq:steq}
\end{align}
we can readily see from the hydrodynamic expansions discussed above that the individual system entropy densities behave at late time as
\begin{align}
    \lim_{\tau\rightarrow\infty}\tau s = \beta/\sqrt{\alpha}, \quad \lim_{\tau\rightarrow\infty}\tau \tilde{s} = \alpha/\sqrt{\beta},
\end{align}
which means that the total entropy scales like
\begin{align} \lim_{\tau\rightarrow\infty}    \tau S_{\rm tot} \equiv (   \tau S_{\rm tot} )_{\rm hydro}&\sim (\gamma \mathcal{E}(\tau = \gamma^{1/4}))^{3/4}.
\end{align}
We can compare this to the pocket formula in \cite{Giacalone:2019ldn} giving the final state particle multiplicity as a power of the total initial energy density for single conformal fluid attractors via $(S\tau)_{\rm hydro}\sim (e\tau)_0^{2/3}$ to see that the interactions due to the effective metric coupling in our hybrid model increases the scaling exponent from $2/3$ to $3/4$ \AMnew{when the total energy density is evaluated at time $\gamma^{1/4}$}. The pocket formula of \cite{Giacalone:2019ldn} states that the total particle multiplicity per unit rapidity\footnote{Here, the rapidity, $\eta$, should not to be confused with the shear viscosity.} follows 
\begin{align}
    \frac{dN}{d\eta} \approx  (S\tau)_{\rm hydro}.
\end{align}
We would need more microscopic inputs to relate our results to particle multiplicities, and therefore it is beyond the scope of the present work. Interestingly, in the case of one fluid in Bjorken flow, the attractor solution has been shown to provide a maximum for  {dilepton} production (while the repulsor/disperser represents a minimum) \cite{Naik:2021yph}. {It would be interesting to do a similar study in our context, however we would then need to replace the perturbative MIS sector with a kinetic theory.}

\begin{figure}
\center
\includegraphics[width=0.6\linewidth]{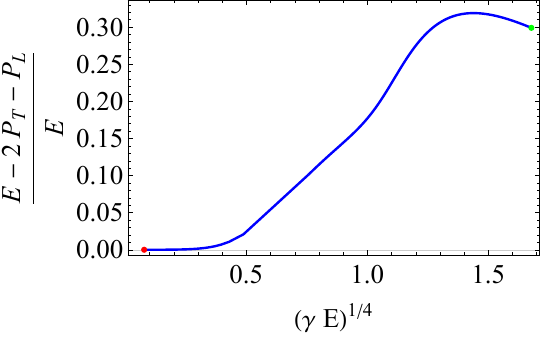}
\caption{The trace of the full energy momentum tensor as a function of the total energy density, which acts as a proxy for the effective temperature. This plot is for an initial energy density of the hard sector of $\gamma\varepsilon=0.8$, $\e /\et=40$. The green dot denotes the value at $\tau_0=10^{-3}\gamma^{1/4}$ and the red point is at $\tau=10^{+3}\gamma^{1/4}$.}\label{fig:trace}
\end{figure}

{We can also study the trace of the full energy momentum tensor, which encodes the interaction energy. For late times, we find that the trace falls like the typical energy density, $$\frac{\vert{\rm Tr}\, T\vert}{\mathcal{E}}=\frac{\mathcal{E}-P_L - 2P_T}{\mathcal{E}}\sim \tau^{-4/3},$$ as follows from \eqref{Eq:wtotal} and \eqref{Tr-T-2}.  An illustrative example for the full time-evolution of the interaction energy is shown in Fig.~\ref{fig:trace}. We find that the interaction energy is non-vanishing only in the intermediate time-scales when both systems hydrodynamize, however it vanishes both at early and late times.}

\section{Hydrodynamization on the attractor surface}\label{sec:hydrodynamization}
Figure~\ref{fig:epsilonplotsfull2} 
shows the existence of hydrodynamic tails of the attractor at a sufficiently late time, giving evidence of hydrodynamization in the hybrid attractor. This preliminary observation motivates a  systematic study of the hydrodynamization times of the two sectors on the attractor surface.

\subsection{Hydrodynamization time}\label{sec:hydro-time}
\begin{figure}[H]
\centering
\includegraphics[scale=0.8]{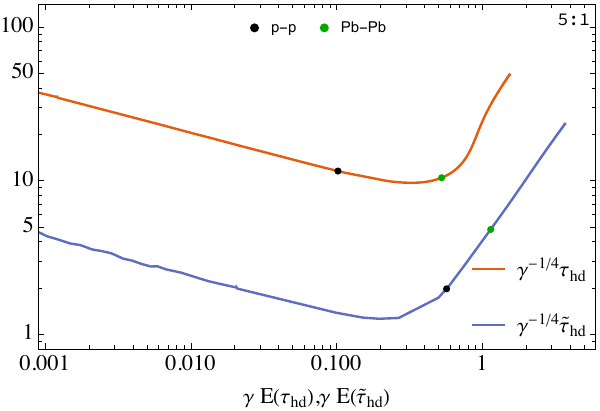}
\caption{The figure shows a doubly logarithmic plot of the hydrodynamization time for the hard (in red) and the soft sector (in blue) against the total energy density at the respective $ \tau_{hd}$ and $ \tilde{\tau}_{hd}$ for ratio $\epsilon(\tau_0):\tilde\epsilon(\tau_0)=5:1$. The hydrodynamization time of the hard sector is longer than the soft sector. The black and the green dots corresponds to the hydrodynamization times for initial conditions which are physically realizable in the $p$-$p$ and Pb-Pb collision.} \label{fig:ratio_5:1}
\end{figure}

 Most commonly, hydrodynamization time is defined as the time when both the longitudinal and transverse pressure can be described by the first-order hydrodynamics up to a certain accuracy \cite{Attems:2017zam}. For the hybrid hydrodynamic attractor, we define the hydrodynamization time of  \AM{the hard \AR{(more weakly coupled)} sector}  \AM{via}
\begin{eqnarray}
\frac{|\Delta P_L|}{P} := \frac{|\phi-\phi_{1st}|}{P}< 0.1, \hspace{0.4cm}\text{for} \hspace{0.4cm} \tau>\tau_{hd}, 
\end{eqnarray}
and similarly for the strongly coupled (tilde) sector,
so that $ \tau_{hd}$ and $ \tilde{\tau}_{hd}$ denote the hydrodynamization times of the \AR{hard} and \AR{soft} sectors, respectively. {This implies that the ratio of the departure of the anisotropic pressures ($\phi$ and $\tilde\phi$ respectively) from that given by first order hydrodynamics ($\phi_{1st}$ and $\tilde\phi_{1st}$ respectively) to the respective isotropic parts of the pressures for both sectors is less than $10$ percent after the corresponding hydrodynamization times.} In general, the hydrodynamization time of perturbative sector $ \tau_{hd}$ is larger than that of the holographic sector $ \tilde{\tau}_{hd}$ as shown in Fig \ref{fig:ratio_5:1}, where the blue curve corresponds to the \AR{soft} sector and the red one to the \AR{hard} sector.

Further, if we analyze the hydrodynamization time of the two components of the hybrid attractor by classifying the total energy density at the hydrodynamization time into three regions, one can find the following generic features of them,

First, when the total energy density at the hydrodynamization time of each sector lies within the region,
$$ 0.0001 \gamma^{-1} \lessapprox \mathcal{E}(\tau_{hd}),  \mathcal{E}(\tilde{\tau}_{hd}) \AR{\ll} \gamma^{-1}    $$ 
(negative slope in the Fig \ref{fig:ratio_5:1}),
we have found that the hydrodynamization time of both the sectors follows conformal behaviours i.e.
\begin{equation}
\tau_{hd} \approx p \times ( 4 \pi C_\eta) \times \mathcal{E}(\tau_{hd})^{-1/4}     
\end{equation}  
and 
\begin{equation} \tilde{\tau}_{hd} \approx \tilde{p} \times ( 4 \pi \tilde{C}_\eta) \times \mathcal{E}(\tilde{\tau}_{hd})^{-1/4} 
\end{equation}  
where $ p $ and $ \tilde{p}$ are constants. \AMnew{We note from Fig. \ref{fig:ratio_5:1} that the range of the energy density at hydrodynamization time in the conformal window is slightly larger in the weakly self-interacting subsector.}

\begin{figure}[ht]
\centering
\includegraphics[scale=0.9]{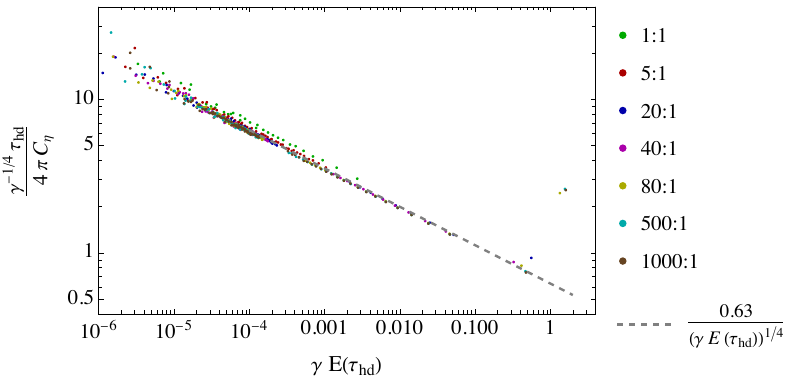}
\caption{The doubly logarithmic plot of the scaled hydrodynamization time of the weakly coupled (hard) sector vs the total energy density at the  ${\tau}_{hd}$ shows universality and indistinguishability from other conformal attractors, {when the hard sector is dominant at a reference time (here $\tau_0 = 0.001 \gamma^{1/4}$). The various color labels refer to different ratios of energy densities of the hard to the strongly coupled (soft) sector at the initial reference time. We note that the hydrodynamization time of the dominant hard sector is independent of these ratios.} }\label{fig:hardconformality}
\end{figure}

\begin{figure}[ht]
\centering
\includegraphics[scale=0.9]{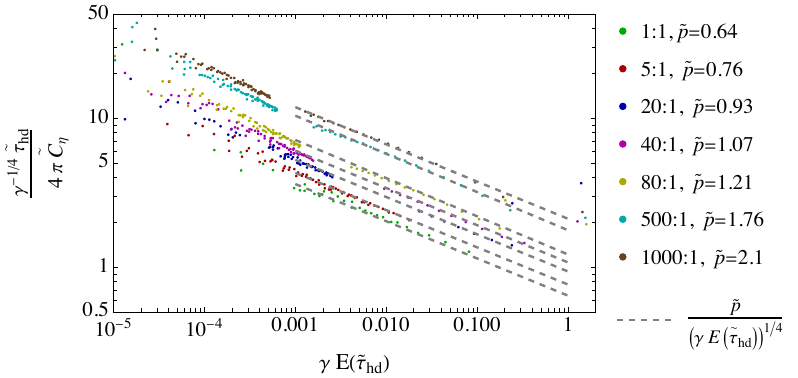} 
\caption{The figure shows the doubly logarithmic plot of the scaled hydrodynamization time of the strongly coupled (soft) sector against the total energy density at the respective $ \tilde{\tau}_{hd}$, {when the weakly coupled (hard) sector is dominant at an initial reference time (here $\tau_0 = 0.001 \gamma^{1/4}$). The various color labels refer to different ratios of energy densities of the hard to the soft sector at the initial reference time.} Values of $ \tilde{p} $ for ratios from $ 1:1$ to $ 1000:1$ of the initial energy densities of the subsectors increases from \textbf{\textcolor{Green}{0.64}}, \textbf{\textcolor{BrickRed}{0.76}}, \textbf{\textcolor{Blue}{0.93}}, \textbf{ \textcolor{RedViolet}{1.07}}, \textbf{\textcolor{SpringGreen}{1.21}}, \textbf{ \textcolor{cyan}{1.76}} to \textbf{\textcolor{Sepia}{2.1}}. These different values of $ \tilde{p}$ shows the dependency of $ \tilde{\tau}_{hd}$ on the ratio 
\AR{$\epsilon(\tau_0):\tilde\epsilon(\tau_0)$}. 
}\label{fig:softconformality}
\end{figure}

The value of $p$,  $\tilde{p}$ and the characteristic of the hydrodynamization time in this conformal window is determined based on which sector dominates at early time. For instance, with dominance of the perturbative sector at reference time $\tau_0= 10^{-3}\gamma^{1/4}$,
we find
\begin{equation}
p \approx 0.63 .
\end{equation}  
This is the same value {(within numerical accuracy) that is} 
obtained in the case of conformal attractors. We find that the hydrodynamization time of the perturbative sector in this conformal region is \textit{universal} as shown in Fig \ref{fig:hardconformality} and the same as in conformal attractors studied earlier in the literature. However for the holographic sector as in Fig \ref{fig:softconformality}, the value of the $ \tilde{p}$ increases from  $ 0.64 $ to $ 2.1 $ as the ratio of the energy densities of the individual sectors at some early time $ \tau_0 = 0.001\gamma^{1/4} $ increases from $ 1:1 $ to $ 1000:1 $. Thus the hydrodynamization time of the holographic sector shows sensitivity towards the initial ratio of the energy densities of the subsectors. Hence one finds that even in the conformal window the hydrodynamization time of the soft sector can be substantially larger than the conformal attractors.

\begin{figure}[ht]
\centering
\includegraphics[scale=0.9]{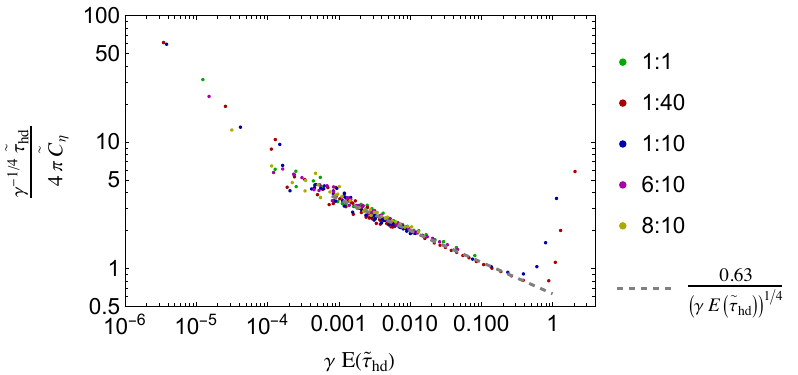} 
\caption{The figure shows the doubly logarithmic plot of the scaled hydrodynamization time of the strongly coupled (soft) sector against the total energy density at the respective $ \tilde{\tau}_{hd}$ {when the soft sector is dominant at the reference time $\tau_0 = 0.001 \gamma^{1/4}.$ The various color labels refer to different ratios of energy densities of the hard to the soft sector at the initial reference time. We note that the hydrodynamization time of the dominant soft sector is independent of these ratios. The plot shows the universal behaviour of the hydrodynamization time of the dominant sector in the conformal window, as the value of $ \tilde{p} $ is $ 0.63$, and is the same as in Fig.~\ref{fig:hardconformality} corresponding to the cases where the hard sector was dominant at initial reference time.} }\label{fig:ratiol1softconformality}
\end{figure}

\begin{figure}[ht]
\centering
\includegraphics[scale=0.9]{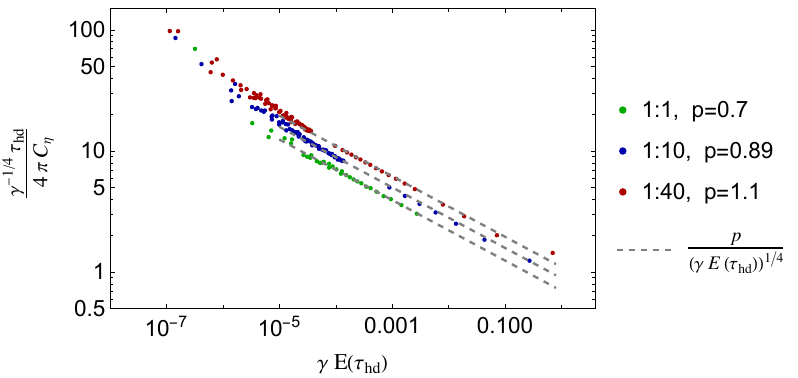}
\caption{The doubly logarithmic plot of the scaled hydrodynamization time of the weakly coupled (hard) sector vs the total energy density at the  ${\tau}_{hd}$ {when the soft sector is dominant at the reference time $\tau_0 = 0.001 \gamma^{1/4}.$ The various color labels refer to different ratios of energy densities of the hard to the soft sector at the initial reference time. We note that the hydrodynamization time of the subdominant hard sector depends  on the ratio of the energy densities. Comparing with Fig.~\ref{fig:softconformality}, where the hard sector was dominant at initial reference time, we observe that in both of these cases, the more dominant one sector is, the more delayed is the hydrodynamization of the sub-dominant sector.} }\label{fig:ratiol1hardconformality}
\end{figure}

{However,} with initial conditions where the energy density of the holographic sector is dominating on the attractor surface, the roles of conformal window in the subsector reverses i.e., $ \tilde{p}$ assumes the conformal value $ 0.63 $, thus showing \textit{universality} as in Fig \ref{fig:ratiol1softconformality} and coinciding with the usual conformal attractors. Meanwhile from Fig \ref{fig:ratiol1hardconformality}, we find that the perturbative sector shows dependence on the initial energy densities of the subsectors and the value of the constant $ p $ increases with the increase in the initial energy density of the perturbative sector. This reversal in the behaviour of hydrodynamization time in the conformal window can be attributed to the fact that the sub-dominant sector is forced to share its energy with the dominating sector even before the hydrodynamization, while the dominating sector is dynamically driven to act universally.

Next, when the energy densities at hydrodynamization times of both the sectors are close to $ \gamma^{-1} $ i.e., 
$$ \mathcal{E}(\tau_{hd}), \mathcal{E}(\tilde{\tau}_{hd}) \sim \gamma^{-1}  $$
the hydrodynamization times on the attractor surface takes minimum values as shown in Fig \ref{fig:ratio_5:1}.

When $ \mathcal{E}(\tau_{hd}) $ and $  \mathcal{E}(\tilde{\tau}_{hd})$ are very large, the hydrodynamization times increase rapidly and do not show any scaling behaviour like that in conformal window. This corresponds to the positive slope region in Fig \ref{fig:ratio_5:1}. 

\subsection{Rough phenomenological match to $p$-$p$ and Pb-Pb collisions }\label{sec:pheno-pp-pb}

In order to contemplate possible phenomenological implications of these features, we interpret attractor curves
with lower total energy densities at $\tau\sim \gamma^{1/4}$
as corresponding to small systems (high-multiplicity $p$-$p$ or $A$-$p$ collisions) as opposed to $A$-$A$ heavy-ion collisions (Pb-Pb or Au-Au). To relate the two,
we identify the energy scale of the intersystem coupling $\gamma^{-1/4}$ with the saturation scale $Q_s^{p\mbox{-}p}$ of $p$-$p$ collisions, and we assume that $\mathcal{E}(\tau)\propto Q_s^4$ at $\tau\sim Q_s^{-1}$
with large $A$-$A$ systems having a larger $Q_s$.
In the IPsat and bCGC models studied in \cite{Kowalski:2007rw}
$Q_{s,A}$ at large nucleon number $A$, impact parameter $b=0$, and
Feynman $x$ parameter around 0.001 was found to be given roughly by
\begin{equation}\label{Eq:match-pp}
    Q_{s,A}^2\sim 0.4 (x/0.001)^{-0.3} A^{1/3} Q_{s,p}^2.
\end{equation}
For Pb-Pb collisions, where $A^{1/3}\approx 6$, we take simply
\begin{equation}\label{Eq:match-PbPb}
Q_{s,Pb}^2 = 2.5\,  Q_{s,p}^2
\end{equation}
as a typical value
and get that at $ \tau \sim Q_{s,Pb}^{-1}$ the total physical energy density is $\mathcal{E}_\mathrm{Pb\mbox{-}Pb}(\tau \sim 0.63) \sim 6.25\,
\mathcal{E}_{p\mbox{-}p}(\tau \sim 1)$ in units where $\gamma=1$.

The Tables \ref{pp-table} and \ref{pb-table} show the hydrodynamization times \AR{of a selection of attractor solutions} of the accordingly defined $p$-$p$ and Pb-Pb collisions for various ratios of 
energy densities of the two subsectors \ARnew{at the conveniently chosen early reference time $\tau_0=10^{-3}\gamma^{1/4}$, which we use to parametrize the attractor surface}. 

\begin{table}[H] 
\centering
 \begin{tabular}{|c| c| c| c| c|c|c|c|c| } 
 \hline
$\varepsilon (\tau_0)/ \tilde{\varepsilon} (\tau_0)$ & $\varepsilon (\tau_0 = 10^{-3}) $ & $\tau_{hd}$ & $\tilde{\tau}_{hd}$ & $w_{hd}$/10 & $\tilde{w}_{hd}$ & $\mathcal{E}(\tau_{hd})$ & $\mathcal{E}(\tilde{\tau}_{hd})$ \\ [0.5ex] 
 \hline\hline
 0.1 & 0.070341996 &  20.86 & 1.328 & 0.6016 & 1.195 & 0.02747 & 0.7321 \\ 
0.2 & 0.1278994995 &  17.49 & 1.557 & 0.6007 & 1.314 & 0.03887 & 0.6260 \\ 
 1 & 0.354767 &  13.32 & 1.952 & 0.6055 & 1.432 & 0.07180 & 0.5345 \\ 
 3 & 0.556843 &  11.92 & 2.016 & 0.6092 & 1.344 & 0.09415 & 0.5492 \\
 4 & 0.603962 &  11.65 & 2.004 & 0.6103    &  1.300   & 0.09995 & 0.5620 \\
 5 & 0.63547095 &  11.50 & 1.969 & 0.6109 & 1.250 & 0.1026 & 0.5710 \\
 6 & 0.657225 & 11.36  & 1.938 & 0.6109 & 1.208 & 0.1054 &0.5821 \\
 7 & 0.672626 &  11.23  & 1.918 & 0.6114 & 1.177 & 0.1087 & 0.5947 \\
 8 & 0.6839 &  11.15  & 1.887 & 0.6116  & 1.142 & 0.1103 & 0.6040 \\
 10 & 0.69876 & 11.04 & 1.825 & 0.6113 & 1.076  & 0.1119 & 0.6196 \\[1ex] 
 \hline
 \end{tabular}
 \caption{Hydrodynamization times for $p$-$p$ initial conditions \ARnew{parametrized by $\varepsilon (\tau_0)/ \tilde{\varepsilon} (\tau_0)$ at a reference time $\tau_0=10^{-3}\ll \tau_i\sim 1$} \AR{(in units where $\gamma=1$)}.
{Here $w= \mathcal{E}_1^{1/4}\tau$, where $\mathcal{E}_1$ is defined in \eqref{eq:E1}, $w_{hd}=w(\tau_{hd})$, and similarly for the second subsystem. \AM{Note that for $\varepsilon (\tau_0)/ \tilde{\varepsilon} (\tau_0)< 1$, the hydrodynamization times for the hard (soft) sector are typically larger (smaller).}}}\label{pp-table}
\end{table}

\begin{table}[H]
\centering
 \begin{tabular}{|c| c| c| c| c| c|c|c|c|} 
 \hline
$\varepsilon (\tau_0)/ \tilde{\varepsilon} (\tau_0)$ & $\varepsilon (\tau_0 = 10^{-3}) $ & $\tau_{hd}$ & $\tilde{\tau}_{hd}$ & $w_{hd}$/10 & $\tilde{w}_{hd}$ & $\mathcal{E}(\tau_{hd})$ & $\mathcal{E}(\tilde{\tau}_{hd})$\\ [0.5ex] 
 \hline\hline
 0.1 & 0.06911777 & 12.80 & 3.152 & 0.6165  & 3.043  & 0.2101 & 1.069  \\
 0.2 & 0.126617865 & 11.73 & 3.690  & 0.6586 & 3.333  & 0.2712  & 0.9724  \\
 0.5 & 0.2423754 & 11.17 & 4.332 & 0.7285  & 3.586  & 0.3520 & 0.9494  \\
 1 & 0.355793523 & 10.98   & 4.671   & 0.7763  & 3.627 & 0.4092  & 0.9817 \\ 
 2 & 0.4830497244 &   10.79   & 4.850  &  0.8117  & 3.532  & 0.4623 &1.038\\
 3 & 0.555141806 & 10.64  & 4.870 & 0.8248 & 3.413  & 0.4918 &1.080 \\
 4 & 0.6002567 &  10.51 & 4.842 & 0.8304 & 3.300 & 0.5122  &1.115  \\
 5 & 0.62995522 &  10.40  & 4.795  & 0.8326 & 3.196 & 0.5276 &1.146 \\
 8 & 0.674906  & 10.12  & 4.617  & 0.8315 & 2.935  & 0.5587 &1.222  \\
 10 & 0.688444  & 9.978 & 4.493  & 0.8286  & 2.792 & 0.5726 & 1.266  \\[1ex]
 \hline
 \end{tabular}
\caption{Hydrodynamization times for Pb-Pb initial conditions \AR{(in units where $\gamma=1$)}. 
}\label{pb-table}
\end{table}

\begin{figure}[H]
\centering
\includegraphics[scale=0.72]{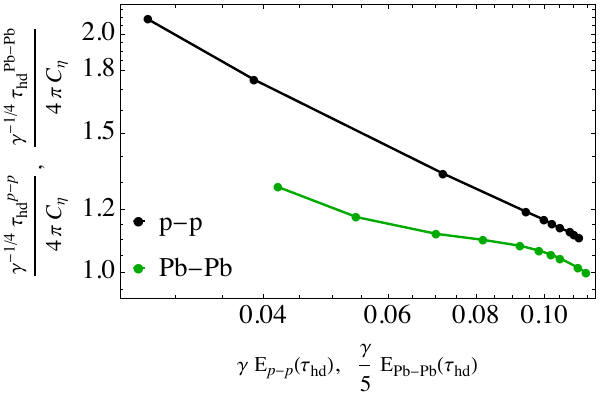}\quad
\includegraphics[scale=0.72]{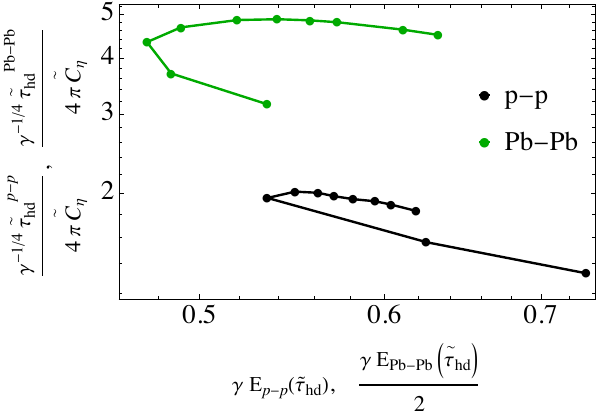}
\caption{The figures show a doubly logarithmic plot of the scaled hydrodynamization time for the perturbative sector (in the left) and holographic sector (in the right) against the total energy density at the respective $ \tau_{hd}$ and $ \tilde{\tau}_{hd}$. The black and the green dots are the hydrodynamization times for initial conditions which are physically realizable in the $p$-$p$ and Pb-Pb collision as listed in Tables \ref{pp-table} and \ref{pb-table}. The total energy density at hydrodynamization time for Pb-Pb collisions is scaled by a factor of $ 1/5$ and $ 1/2$ to fit in the same plot with $p$-$p$ collision. 
}\label{fig:phenopp-pb}
\end{figure} 

From these tables one finds that in our $p$-$p$ collisions $ \mathcal{E}(\tau_{hd}) $ and $ \mathcal{E}(\tilde{\tau}_{hd}) $ are \AMnew{smaller, while} the hydrodynamization time of the perturbative sector is larger and that of the holographic sector is smaller compared to the Pb-Pb collisions. This behaviour is {reflected} in Fig \ref{fig:phenopp-pb}, where the $ \mathcal{E}(\tau_{hd})$ and  $ \mathcal{E}(\tilde{\tau}_{hd})$ of the Pb-Pb collision is scaled by the factor of $ 1/5$ and $1/2 $ to fit in the plot with $p$-$p$ collision. Also one observes that this phenomenologically realizable hydrodynamization time of the perturbative sector lies in the conformal region (\AMnew{\textit{maximal} value of $\gamma\mathcal{E}(\tau_{hd})$ is about $0.1$ for $p$-$p$ and $0.6$ for Pb-Pb collisions}) while that of the non-perturbative sector lies in the non-conformal region for both $p$-$p$ and Pb-Pb collisions (\AMnew{\textit{minimal} value of $\gamma\mathcal{E}(\tilde\tau_{hd})$ is about $0.5$ for $p$-$p$ and $0.9$ for Pb-Pb collisions}). These features of the hydrodynamization time of perturbative and holographic sector for the initial conditions of $p$-$p$ and Pb-Pb collisions are captured in the Fig \ref{fig:ratio_5:1} \AMnew{as mentioned below (note also from this figure that the conformal windows for the two sectors are slightly different from each other):}
\begin{itemize}
    \item In case of the perturbative sector for both $p$-$p$ and Pb-Pb collisions, the total energy densities at the hydrodynamization time, $ \mathcal{E}(\tau_{hd}) $, lies approximately within the conformal region denoted by the black and the green dots in the figure.
    In addition, since the $ \mathcal{E}(\tau_{hd}) $ is minimal in $p$-$p$ collisions, the hydrodynamization time of the perturbative sector $ \tau_{hd} $ is sufficiently larger which is evident from the scaling behaviour of $ \tau_{hd}$ in the conformal window.
    
    \item For the non-perturbative sector, $ \mathcal{E}(\tilde{\tau}_{hd}) > \gamma^{-1} $ and lies outside the conformal region for both $p$-$p$ and Pb-Pb collisions as shown in the figure. Further due to the small total energy density, $ \mathcal{E}(\tau_{hd}) $, in $p$-$p$ collisions, the hydrodynamization time $ \tilde{\tau}_{hd}$ of the soft sector of $p$-$p$ collision is smaller than that of Pb-Pb collisions, as indicated from the growth of $ \tilde{\tau}_{hd}$ with $\mathcal{E}( \tilde{\tau}_{hd})$.
\end{itemize}
\AM{Note from the Tables \ref{pp-table} and \ref{pb-table} that for smaller ratios of the hard to soft energy densities at \ARnew{reference} time  ($\varepsilon (\tau_0)/ \tilde{\varepsilon} (\tau_0)\lesssim 1$), the hydrodynamization times for hard/soft sectors are typically larger/smaller, in both $p$-$p$ and Pb-Pb collisions.} For details on the evolution of energy densities in these scenarios, see Appendix \ref{sec:thermalization}.

\subsection{Relevance of results for general initial conditions}
\label{sec:phys-ini}

The above study of hydrodynamization, in the phenomenological context of small and large system collisions, has been restricted to the hydrodynamic attractor surface. However, the attractor surface is not necessarily relevant at $\tau \sim Q_s^{-1}$, where we use conditions such as \eqref{Eq:match-PbPb} to match the evolution with a more microscopic description, such as the glasma effective theory, via the total energy density. A natural question then is whether our results for hydrodynamization in large vs small system collisions remain valid for generic evolutions that may be far away from the attractor surface at \ARnew{a glasma matching time} $\taum \sim Q_s^{-1}$.
\begin{figure}[ht]
\subfigure[Pb-Pb  Hard]{\includegraphics[height=4.2cm]{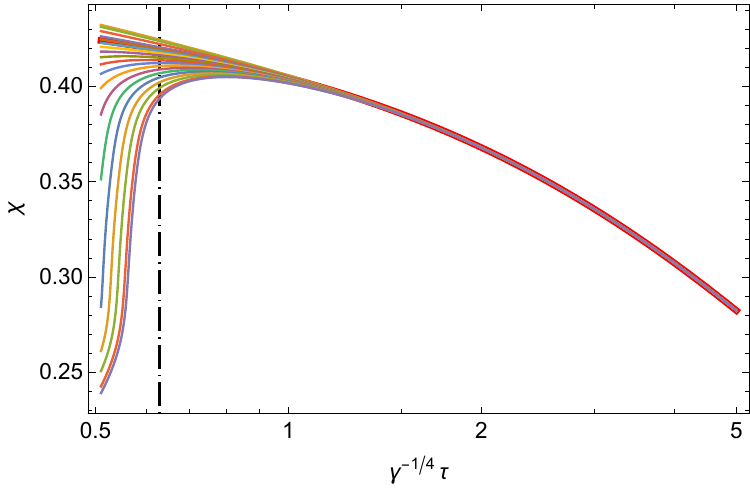} }
\subfigure[p-p Hard]{\includegraphics[height=4.2cm]{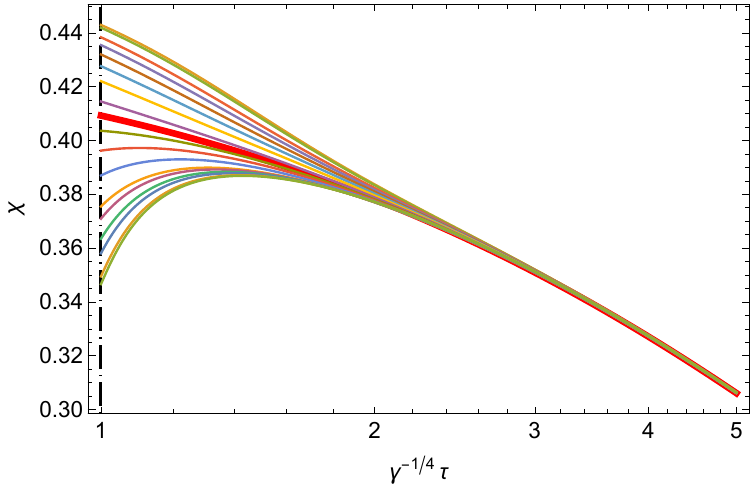}}
\vfill
\hspace*{2mm}\subfigure[Pb-Pb Soft]{\includegraphics[height=4.2cm]{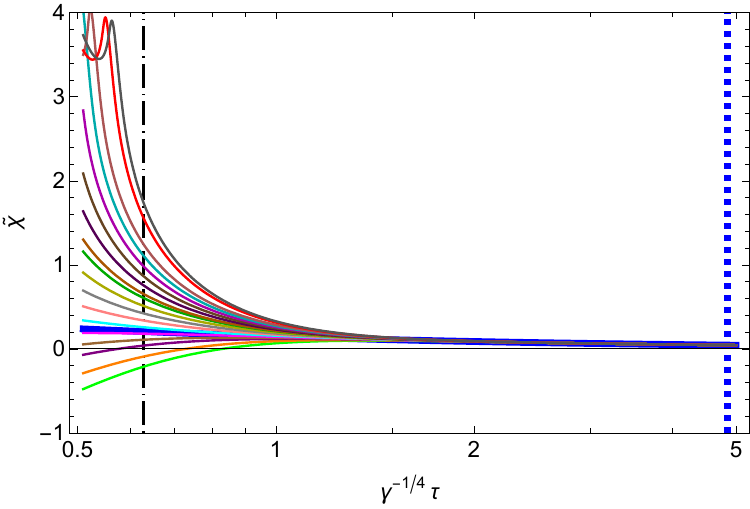}}
\hspace*{2mm}\subfigure[p-p Soft]{\includegraphics[height=4.2cm]{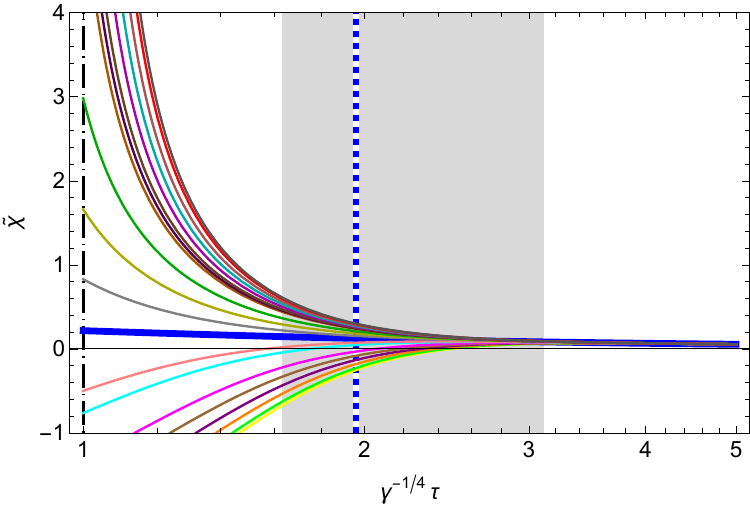}}
\caption{Left: Generic evolutions of the dimensionless anisotropy which deviate from the attractor surface in Pb-Pb collisions for both hard sector (top) and soft sector (bottom). In all cases, the glasma matching condition $\mathcal{E}_\mathrm{Pb\mbox{-}Pb}(\taum \sim 0.63) \sim 6.25$ (in units where $\gamma =1$) is satisfied (see Sec. \ref{sec:pheno-pp-pb}) at the glasma matching time $\taum \sim 0.63 = Q_{s,Pb}^{-1}$ which is indicated as a black \AR{dash-dotted} vertical line. All evolutions converge to the corresponding attractor curve which is marked in bold red (for hard) and bold blue (for soft), and this occurs much \AR{earlier than} hydrodynamization in both cases. The hydrodynamization time for the soft sector is marked as a vertical blue dotted line, while that for the hard sector is \AR{beyond the} range of the plot.
Right: 
The same for $p$-$p$ collisions, where the glasma matching condition $\mathcal{E}_{p\mbox{-}p}(\taum \sim 1) \sim 1$ is satisfied (see Sec. \ref{sec:pheno-pp-pb}) at the glasma matching time $\taum \sim 1 = Q_{s,p}^{-1}$ indicated as a black \AR{dash-dotted} vertical line. All evolutions converge to the corresponding attractor curve which is marked in bold red (for hard) and bold blue (for soft). In \AR{the} case of the hard sector, the convergence to the attractor occurs \ARnew{also much} before hydrodynamization time, \ARnew{but in} the soft sector \ARnew{of $p$-$p$ collisions}, the convergence to the attractor is simultaneous with hydrodynamization. The hydrodynamization times for various evolutions are spread over a range marked as a gray column which includes the blue dotted vertical time corresponding to the attractor \AR{(the corresponding range in the Pb-Pb soft case is negligibly small)}. Note that \AR{the} hydrodynamization time for the soft sector in Pb-Pb collisions is larger than the maximum value for that of the soft sector in $p$-$p$ collisions.}
\label{fig:off-attractor}
\end{figure}

We have explored this issue by studying evolutions in the context of both $p$-$p$ and Pb-Pb type collisions where the anisotropies of both hard and soft sectors are far from the attractor values around $\taum\sim Q_s^{-1}$, retaining the constraints on the total energy densities used in Sec.~\ref{sec:pheno-pp-pb}, which are $\mathcal{E}_\mathrm{Pb\mbox{-}Pb}(\taum \sim 0.63) \sim 6.25$ and $\mathcal{E}_{p\mbox{-}p}(\taum \sim 1) \sim 1$ respectively (in units $\gamma =1$), in order that we could match with glasma models at the time where the latter just cease to be applicable and need to be replaced by kinetic-like descriptions. (Recall that we have identified $\gamma$ with $Q_s^{-4}$ for $p$-$p$ collisions.) Numerically, this has been implemented by initializing with values different from known attractor solutions both before and after $\taum\sim Q_s^{-1}$, and evolving forwards/backwards in time. While it is hard to explore the full four-dimensional phase space, initializing with different anisotropies (i.e. $\phi$ and $\tilde{\phi}$) away from the attractor surface typically produce maximal deviations from the attractor surface around $\taum\sim Q_s^{-1}$.

\AM{As shown in Fig.\ \ref{fig:off-attractor}, in the Pb-Pb collision scenario, the convergence to the attractor happens much earlier than hydrodynamization.\footnote{Because of the nonlinearity of the dynamics, one cannot give a precise initial rate of approach to the attractor as this depends strongly on the specific solution. In the example shown in Fig.\ \ref{fig:off-attractor}, the rate $\Gamma_\mathrm{init}=|d\ln(\chi-\chi_\mathrm{attractor})/d\tau|$ at initial times $\gamma^{-1/4}\taum=1$
and 0.63, respectively, for $p$-$p$ and Pb-Pb, is, however, always much larger than $\tau_{hd}^{-1}$, for both, the hard and the soft sectors. Close to the attractor, $\Gamma_\mathrm{init}/\tau_{hd}^{-1}\sim 60$ for the hard sector in both $p$-$p$ and Pb-Pb, and $\tilde\Gamma_\mathrm{init}/\tilde\tau_{hd}^{-1}\sim 20$ (30) for the soft sector in $p$-$p$ (Pb-Pb). Further away from the attractor, this ratio varies between 40 and 130 (20 and 50) for the hard (soft) sector of Pb-Pb, and between 10 and 70 (2 and 300) for the hard (soft) sector of $p$-$p$.} The hydrodynamization times for the hard/soft sectors for Pb-Pb collision type scenarios are identical to that of the corresponding attractor curves to which the evolution converges. The same is also true for the hard sector in p-p collision type scenario as shown in Fig.\ \ref{fig:off-attractor}. However, for the soft sector in $p$-$p$ collisions, as shown in Fig.\ \ref{fig:off-attractor}, the convergence to the attractor is simultaneous with hydrodynamization, and therefore the hydrodynamization times in off-attractor evolutions do not coincide with that of the corresponding attractor curve, and can be larger than the latter by \AR{typically up to 50} percent \AR{(gray column in Fig.\ \ref{fig:off-attractor}(d))}. 

Nevertheless, we note that even the largest hydrodynamization time for the off-attractor evolution in p-p type collision type scenario is smaller than the hydrodynamization time in Pb-Pb scenario. The latter is determined by the attractor. Therefore, we conclude that the hydrodynamization of the soft sector in $p$-$p$ type collisions occurs earlier than that in Pb-Pb type collisions in generic evolutions even if they deviate significantly from the attractor surface at $\tau \sim \gamma^{1/4}$.}

Thus the insights gleaned from the study of the attractor surface should apply also for more generic phenomenologically relevant initial conditions. \AM{Particularly, the facilitation of hydrodynamization of the soft sector due to smaller energy densities in small system collisions compared to that in large system collisions (since both the hydrodynamization times lie outside of the conformal window) should be valid for generic phenomenologically relevant initial conditions.}

\section{Discussion}

The main physical insights which we have gained from the Bjorken flow attractor of our hybrid fluid model is how hydrodynamization \AR{may work} in an asymptotically free gauge theory with both weakly interacting and strongly interacting degrees of freedom. In our model \AR{we represent those by two fluid components with viscosities differing by one order of magnitude, coupled
in accordance with the semi-holographic framework developed in \cite{Faulkner:2010tq,Mukhopadhyay:2013dqa,Iancu:2014ava,Mukhopadhyay:2015smb,Banerjee:2017ozx,Kurkela:2018dku,Ecker:2018ucc}. This model}
has only one dimensionful energy scale, namely $\gamma^{-1/4}$, which sets the inter-system coupling. We \AR{have studied the hydrodynamic attractor of this system and we} have shown analytically that at very early time, {$\tau\ll\gamma^{1/4}$}, the energy density \AR{in the weakly coupled sector} always dominates over that of the strongly coupled sector with an exponent which \AR{is} determined by the transport coefficients and the relaxation times of the two sectors. 
\AR{This feature, which is reminiscent of the bottom-up thermalization scenario of Ref.~\cite{Baier:2000sb}, is, however, only relevant physically when the
system is initialized correspondingly early; when in the context of heavy-ion collisions $\gamma^{-1/4}$ is identified with the saturation scale $Q_s$, no such hierarchy arises at $\tau\sim Q_s^{-1}$, where we make contact with glasma descriptions, although the weakly coupled sector
is typically the dominant one.} 

\AR{The hydrodynamic attractor of our hybrid fluid model is given by a (phase-space) surface of attractor lines that can be characterized by the ratio of the energy densities in the two sectors
at some arbitrary early reference time together with the overall energy, and we have determined the resulting
hydrodynamization times in the two sectors for a wide range of these parameters.} 

When the total energy densities at the respective hydrodynamization times ($\mathcal{E}(\tau_{hd})$, $\mathcal{{E}}(\tilde{\tau}_{hd})$) is less than $\gamma^{-1}$, both sectors hydrodynamize as in a conformal hydrodynamic attractor ($\tau_{hd} \propto \mathcal{E}(\tau_{hd})^{-1/4}$ and $\tilde{\tau}_{hd} \propto \mathcal{E}(\tilde{\tau}_{hd})^{-1/4}$) up to a pre-factor $p$. Remarkably, the dominant sector (which is the perturbative sector unless we fine-tune the initial conditions) always hydrodynamize in a universal way, with the pre-factor $p$ \AR{being the} same as in a decoupled conformal hydrodynamic attractor. The inter-system coupling does not affect $p$ for the dominant sector, however the pre-factor for the sub-dominant (typically strongly interacting) sector increases with the ratio of energy densities at the initialization time -- {the higher} the dominance of the dominant sector, {the} more delayed is the hydrodynamization of the sub-dominant sector. When the energy densities at the respective hydrodynamization times increases beyond $\gamma^{-1}$,  the hydrodynamization times of both sectors increase rapidly after reaching their minimum values.

Another relevant insight of our model is that although the full system behaves as a single fluid at late time, its effective equation of state and effective transport coefficients depend on which curve on the attractor surface that the system converges to, and is thus process dependent. Although for the equation of state the process-dependence is sub-leading, the effective shear viscosity depends on how the full energy is finally shared between the two subsectors.

We also have gained new insights {with potential relevance} for heavy-ion collision experiments. Setting initial conditions motivated by saturation scenarios, we see that the total energy density at the hydrodynamization time of the strongly coupled sector $\mathcal{{E}}(\tilde{\tau}_{hd})$ lies outside the conformal window, while that corresponding to the time of hydrodynamization of the perturbative sector, namely $\mathcal{{E}}({\tau}_{hd})$, lies within the conformal window. \AM{It follows from our general results that while the perturbative sector hydrodynamizes later, the non-perturbative strongly interacting sector hydrodynamizes \textit{earlier} in $p$-$p$ collisions compared to Pb-Pb collisions on the attractor surface. This is because the hydrodynamization times for the soft sector in both cases lie outside of the conformal window, and the total energy density is smaller in the case of $p$-$p$ collisions than in Pb-Pb collisions at the respective hydrodynamization times.} Thus we get a new perspective on how collective flow develops in small system collisions. 

Furthermore, we have shown that these results remain valid for generic phenomenologically relevant initial conditions with energy densities matched to the glasma models at time scales of order the saturation scale. 
\AM{In Pb-Pb collisions, the off-attractor evolutions for both sectors converge to the attractor surface much earlier than hydrodynamization, while in $p$-$p$ collisions this happens only for the hard sector. In these cases, the attractor surface determines the hydrodynamization times. However, the approach to the attractor is simultaneous with hydrodynamization for the soft sector in $p$-$p$ collsions, and the hydrodynamization times show \AR{typically} up to \AR{50} percent departure from the corresponding attractor value. Nevertheless, our conclusion that the hydrodynamization time of the soft sector in $p$-$p$ collisions is smaller than in Pb-Pb collisions remains true even for evolutions which deviate significantly from the attractor surface at the time scales when we match with glasma models.}

\ARnew{We therefore conclude that the hybrid attractor of our two component system with a more weakly and a more strongly interacting subsector is relevant for a range of phenomenologically relevant initial conditions well before the onset of hydrodynamics.
The hybrid fluid model thus provides a potentially useful model for describing regimes with different importance of weakly and strongly coupled physics that will be interesting to confront with experiment.}
\AR{It would be interesting \ARnew{in particular} to explore whether the different off-equilibrium evolution of the soft, more strongly interacting IR sector in small systems could lead to characteristic differences between $p$-$p$ and Pb-Pb collisions.}
In the future, we hope to study a fuller semi-holographic set-up with a kinetic theory coupled to a five-dimensional geometry governed by Einstein's equations. This is necessary for {confronting our approach} with experimental data, also for the study of dilepton production \cite{Naik:2021yph,Coquet:2021lca} and hadronization \cite{Giacalone:2019ldn} on the attractor surface. 

Another interesting direction of study is to understand how stochastic statistical and quantum fluctuations, which are suppressed in the large $N$ limit, affect the hydrodynamic attractor. This needs to be understood separately for both weakly interacting and strongly interacting theories. In our hybrid set-up, these fluctuations are needed for equi-partition of energy between the perturbative and holographic sectors and thus for complete thermalization (we will discuss this more in an upcoming work). In the context of the attractor, this may lead the Bjorken attractor surface to collapse to a single curve where the ratio of physical subsystem temperatures goes to one asymptotically, over a long time-scale. Such fluctuations could be enhanced if the system passes near a critical point,\footnote{See \cite{Grossi:2021gqi} for studies of effects of stochastic critical fluctuations on hydrodynamics, and \cite{Kovtun:2012rj} for a review on stochastic statistical fluctuations in hydrodynamics. See also \cite{Mitra:2020hbj,Rodgers:2022fuv} for studies of superfluid Bjorken flow.} and therefore such a study could be interesting for
{the search of a critical endpoint in heavy-ion collisions}.

\acknowledgments

The authors would like to thank Michal Heller, Edmond Iancu and Jean-Yves Ollitrault for useful discussions. AM acknowledges support from the Ramanujan Fellowship and ECR award of the Department of Science
and Technology of India and also the New Faculty Seed Grant of IIT Madras. AS is supported by the Austrian
Science Fund (FWF), project no. J4406.

\appendix

\section{Late time gradient expansion}\label{app:grad-exp}
The coefficients present in the late time expansion of our variables are given by (note that the subscripts 1 and 2 denote the untilded and tilded sectors, respectively)
\begin{align}
 {\epsilon_{11}}&= -\frac{8}{3} \cet {\epsilon_{10}}^{3/4} \tau_0^{-1},   \; \;   {\epsilon_{12}}=-\frac{2}{9} {\epsilon_{10}}^{1/2} \tau_0^{-2} \left( -12 \cet^2 + 4 \cet \ctp + 3 {\epsilon_{10}}^{1/2} {\epsilon_{20}} \tau_0^2 \gamma  \right),\\
{\epsilon_{21}}&= -\frac{8}{3} \cet {\epsilon_{20}}^{3/4} \tau_0^{-1},   \; \;   {\epsilon_{22}}=-\frac{2}{9} {\epsilon_{20}}^{1/2}  \tau_0^{-2} \left( -12 \cett^2 + 4 \cett \ctpt + 3 {\epsilon_{10}}^{1/2} {\epsilon_{20}} \tau_0^2 \gamma  \right),\\
 f_{10} &= 0,\; f_{11} = 0, \; f_{12} = 0, \; f_{13}=\frac{16}{9}\cet {\epsilon_{20}}^{3/4} \tau_0^{-1} , \; f_{14}=-\frac{32}{27}\cet {\epsilon_{10}}^{1/2}  \tau_0^{-2} \left( 3 \cet - \ctp \right),\\
f_{20} &= 0,\; f_{21} = 0, \; f_{12} = 0, \; f_{23}=\frac{16}{9}\cett {\epsilon_{10}}^{3/4} \tau_0^{-2}, \; f_{24}=-\frac{32}{27}\cett {\epsilon_{20}}^{1/2} \tau_0^{-2}  \left( 3 \cett - \ctpt \right),\\
 A_{11}&= -\frac{1}{2}{\epsilon_{20}}  \gamma , \; A_{12}= \frac{4}{3} \cett {\epsilon_{20}}^{3/4} \tau_0^{-1} \gamma ,\\
A_{21}&= -\frac{1}{2}{\epsilon_{10}}  \gamma , \; A_{22}= \frac{4}{3} \cet {\epsilon_{10}}^{3/4} \tau_0^{-1} \gamma ,\\
 B_{11}&= \frac{1}{6} {\epsilon_{20}}  \gamma,  \;  B_{12}= 0,\\
B_{21}&= \frac{1}{6} {\epsilon_{10}} \gamma,  \;  B_{22}= 0,\\
C_{11}&=\frac{1}{6} {\epsilon_{20}}  \gamma \tau_0,  \;   C_{12}=- \frac{4}{3} \cett {\epsilon_{20}}^{3/4}   \gamma,\\
C_{21}&=\frac{1}{6} {\epsilon_{10}}  \gamma \tau_0 ,  \;   C_{22}=- \frac{4}{3} \cet {\epsilon_{10}}^{3/4}  \gamma.
\end{align}

The equations of the fluctuations are given by
\begin{align}
      \dph'(\tc) &= \dph(\tc) \left( -\frac{{\epsilon_{10}}^{1/4}  \tau_0 }{\ctp } \frac{1}{\tc^{1/3}} +\frac{2   \left( \cet -2  \ctp   \right)}{3 \ctp }\frac{1}{\tc}   + \frac{2 \left( \cet  \ctp + 3  \gamma {\epsilon_{10}}^{1/4} {\epsilon_{20}}  \tau_0^2 \right)  }{9  \ctp  {\epsilon_{10}}^{1/4} \tau_0 }\frac{1}{\tc^{5/3}} \right) \nonumber\\
      & + \de(\tc) \left( \frac{4 \cet }{3 \ctp  }\frac{1}{\tc} -\frac{8 \cet   }{27 {\epsilon_{10}}^{1/4} \tau_0} \frac{1}{\tc^{5/3}}   \right) \label{fluc-eqns-first-order-1}
   \\
   \de'(\tc)&=\tde (\tc) \left(   \frac{2 \gamma  {\epsilon_{10}}  }{9} \frac{1}{\tc^{7/3}}-\frac{8 \gamma   \cet  {\epsilon_{10}}^{3/4} }{9 \tau_0} \frac{1}{\tc^{3}}   \right)+ \dpht(\tc) \left( -\frac{2 \gamma {\epsilon_{10}} }{3} \frac{1}{\tc^{7/3}}  + \frac{8\gamma 
   \cet  {\epsilon_{10}}^{3/4} }{3 \tau_0 }\frac{1}{\tc^{3}} \right) \nonumber
   \\&  +\de(\tc) \left(-\frac{4 }{3 \tc }  +\frac{8 \gamma
    \e_{20} }{9} \frac{1}{\tc^{7/3}}-\frac{32 \gamma  \cett {\epsilon_{20}}^{3/4} }{9 \tau_0 } \frac{1}{\tc^{3}}\right)  +  \dph(\tc) \left(\frac{8 \gamma  \cett {\epsilon_{20}}^{3/4} }{3 \tau_0 } \frac{1}{\tc^3} + \frac{1}{ \tc} \right)
    \label{fluc-eqns-first-order-2}
    \end{align}
    \begin{align}
        \dpht'(\tc)&= \dpht(\tc) \left(  -\frac{ {\epsilon_{20}}^{1/4}  \tau_0 }{ \ctpt }\frac{1}{\tc^{1/3}} +\frac{2
   (\cett-2  \ctpt) }{3 \ctpt  } \frac{1}{\tc } + \frac{2 \left( \cett  \ctpt + 3  \gamma \e_{20}^{1/4} {\epsilon_{10}}  \tau_0^2 \right)}{9  \ctpt  {\epsilon_{20}}^{1/4} \tau_0} \tc^{1/3}\right) \nonumber
   \\& +\tde(\tc) \left(\frac{4 \cett }{3 \ctpt  }\tc-\frac{8 \cett   }{27 \e_{20}^{1/4} \tau_0 } \tc^{1/3} \right)\label{fluc-eqns-first-order-3}\\
        \tde'(\tc)&= \de(\tc) \left(\frac{2 \gamma  {\epsilon_{20}}  }{9} \frac{1}{\tc^{7/3}}-\frac{8 \gamma   \cett  {\epsilon_{20}}^{3/4} }{9 \tau_0 } \frac{1}{\tc^3}\right)+\dph(\tc) \left( -\frac{2 \gamma {\epsilon_{20}} }{3} \frac{1}{\tc^{7/3}}  + \frac{8\gamma 
   \cet  {\epsilon_{20}}^{3/4} }{3 \tau_0 }\frac{1}{\tc^{3}}\right)\nonumber
   \\&  +\tde(\tc) \left( -\frac{4 }{3  \tc } +\frac{8 \gamma
    {\epsilon_{10}} }{9} \frac{1}{\tc^{7/3}}-\frac{32 \gamma  \cet \e_{10}^{3/4} }{9 \tau_0 } \frac{1}{\tc^3}  \right)  + \dpht(\tc)\left(\frac{8 \gamma  \cet {\epsilon_{10}}^{3/4} }{3 \tau_0} \frac{1}{\tc^3} + \frac{1}{ \tc}\right)\label{fluc-eqns-first-order-4}
\end{align}

\section{Phase transition}\label{sec:phase}

As was explored in the inviscid case in \cite{Kurkela:2018dku}, the coupling equations as presented exhibit a phase transition depending on the value of $r$.
Thermal equilibrium is best explored in the Minkowski background. For the purposes of this section
\begin{align}
    \varepsilon=3P, \quad P=n T^4_1, \quad \text{and} \quad \mathcal{T}=\A T_1=\At T_2,
\end{align}
where the last relation is the requirement of thermal equilibrium, which means that the physical system temperature, $\mathcal{T}$, can be expressed in terms of the individual subsystem temperature via 
\begin{align}
\sqrt{-\eta_{00}}\mathcal{T}=\sqrt{-g_{00}} T_1=\sqrt{-\tilde{g}_{00}} T_2. 
\end{align}

Further, we introduce the light cone velocity
\begin{align}
    v=\A/\B, \quad \text{and} \quad \tilde{v}=\At/\Bt,
\end{align}
which represents the effective causal light cone of each individual subsystem. Interactions between the two systems change the effective light cone velocity. Note that $c>v,\tilde{v}>0$, where $c=1$ is the speed of light.

We first outline some results in the case of identical subsystems. It is instructive to understand the phase transition by considering the inviscid Minkowski equations, for details see \cite{Kurkela:2018dku}. 

For identical inviscid conformal  subsystems in thermal equilibrium, we have
that the Minkowski coupling equations can be solved until there is one left, which is given by \cite{Kurkela:2018dku}
\begin{align}\label{mink-coup}
    \gamma n_1 \mathcal{T}^4=\frac{v^5(1-v^2)(3+v^2)}{(3+v^4-3r(1-v^2)^2)^2}.
\end{align}
Requiring that $0<\mathcal{T}<\infty$ and remembering that $0<v<1,$ we see that $r>1$. Analytically, a phase transition was found in \cite{Kurkela:2018dku} (see Appendix D for more details) when the coupling takes the value 
\begin{align}
r_c=\frac{1}{540}\left(195+43\sqrt{15}+\sqrt{30(4082-557\sqrt{15})}\right)\approx 1.114.
\end{align}

We can explore this phase transition in the Bjorken case of identical subsystems by mapping from Bjorken to Minkowski. The identical subsystem equations for the Bjorken case are given by \eqref{exp-ward} and \eqref{coup4}-\eqref{coup6}, after identifying the tilded quantities with the untilded ones, i.e. $\At \rightarrow \A$.

\begin{figure}[h]
\includegraphics[width=0.5\linewidth]{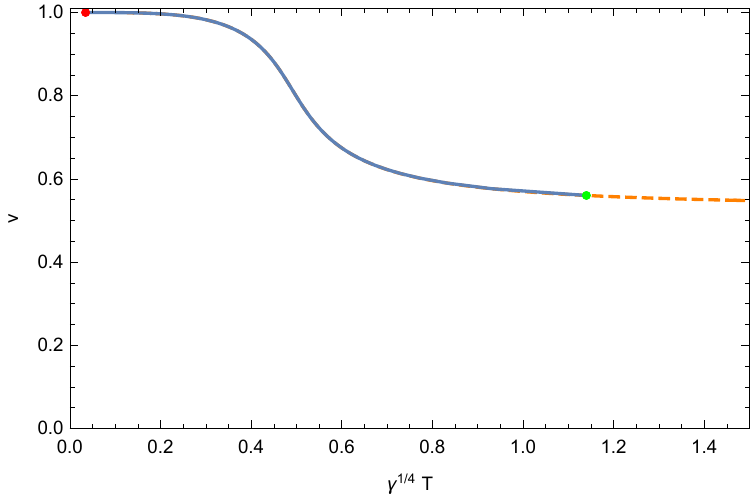}
\includegraphics[width=0.5\linewidth]{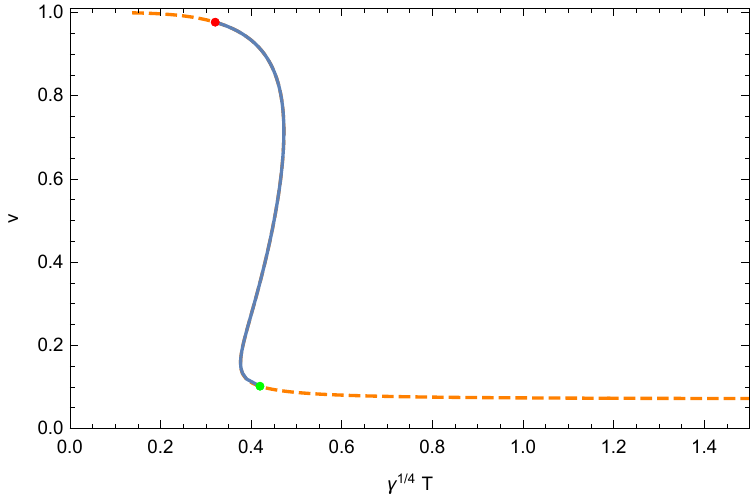}
\caption{Light cone velocity as a function of temperature for the inviscid Bjorken and Minkowski cases of identical subsystems for two different values of $r$. The orange dashed curve is the result of working directly in the Minkowski background \eqref{mink-coup}, as studied in \cite{Kurkela:2018dku}. The blue line denotes the Bjorken evolution with increasing $\tau$ going from the green to red dot (right to left on the plot). Left: in the case $r=2$, the green dot is for $\gamma\varepsilon(\tau_0=0.0001)=0.348$, while the red dot denotes $\tau=20,$ the time we choose to stop the simulation. Right: below the phase transition, for $r=1.01$ the Bjorken system is initialized at the green dot with $\gamma\varepsilon(\tau_0=0.01)=0.1$.}\label{fig:inviscid-vel}
\end{figure}

First, as a sanity check, we do the matching for the inviscid case. We show an illustrative example with $r=2$ in the left panel of Fig.~\ref{fig:inviscid-vel}. In the Minkowski case, we solve \eqref{mink-coup} directly and arrive at $v=v(\mathcal{T})$. In the Bjorken case, we parametrically plot the light cone velocity $v(\tau)$ as a function of temperature $\mathcal{T}=\A(\tau)(\varepsilon(\tau)/3)^{1/4}$. Early time is represented by the green dot, with late time coinciding with the red dot. The combined choice of initial energy density and initialization time sets the value of the green point, which in the case of the plot we take $\gamma\varepsilon(\tau_0=0.0001)=0.348$. 

Likewise, in the right panel of Fig.~\ref{fig:inviscid-vel} we take a value of the coupling below the critical value of $r$, namely $r=1.01<r_c$. In this case, we see the presence of the phase transition via the multivaluedness of the light cone velocity as a function of $\mathcal{T}$. Of course, the light cone velocity can only take one value at a time, which means there is a discontinuous jump around the critical temperature. As expected, we see that the there is complete agreement between the two sets of equations. 

Next, we consider adding viscous corrections. We now have to contend with the evolution of $\phi$ given by \eqref{exp-mis} with values of the transport coefficients $\eta=4C_\eta\varepsilon(\tau)^{3/4}/3$ and $\tau_\pi=C_{\tau_\pi}\varepsilon^{-1/4}$ with $C_\eta=1/4\pi$ and $C_{\tau_\pi}=(2-\log{2})/2\pi.$ The matching with the inviscid Minkowski case will not be perfect since we are no longer in thermal equilibrium, but it will serve as a guide. We find that there is little difference to the above parametric plots for late times.

\begin{figure}[h]
\includegraphics[width=0.49\linewidth]{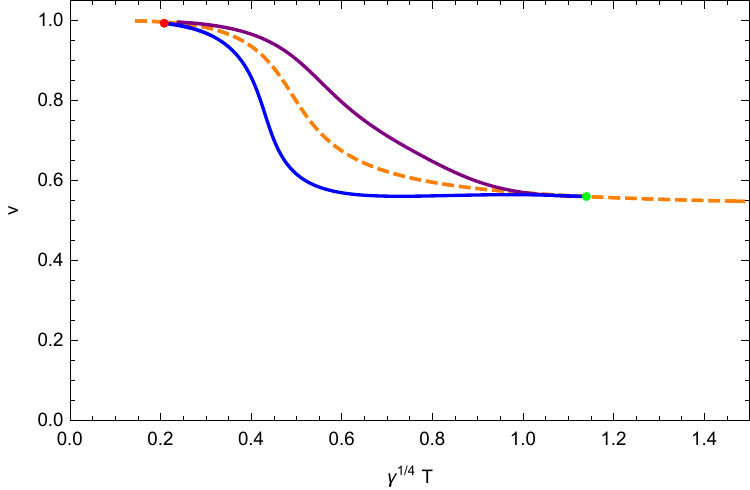}
\includegraphics[width=0.49\linewidth]{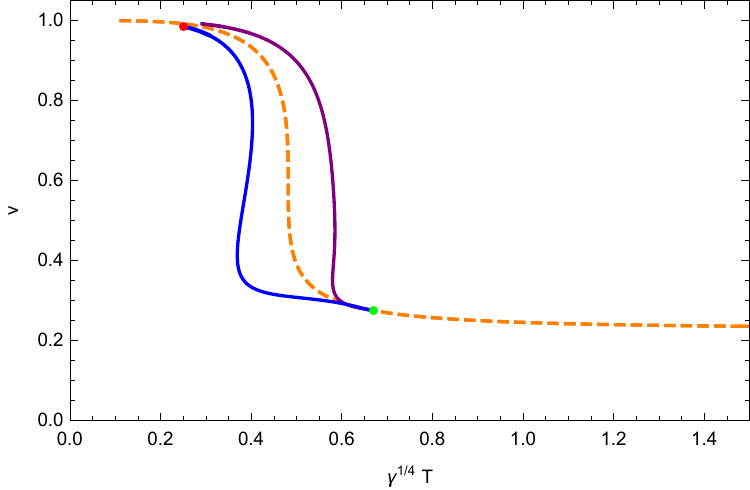}
\centering{\includegraphics[width=0.49\linewidth]{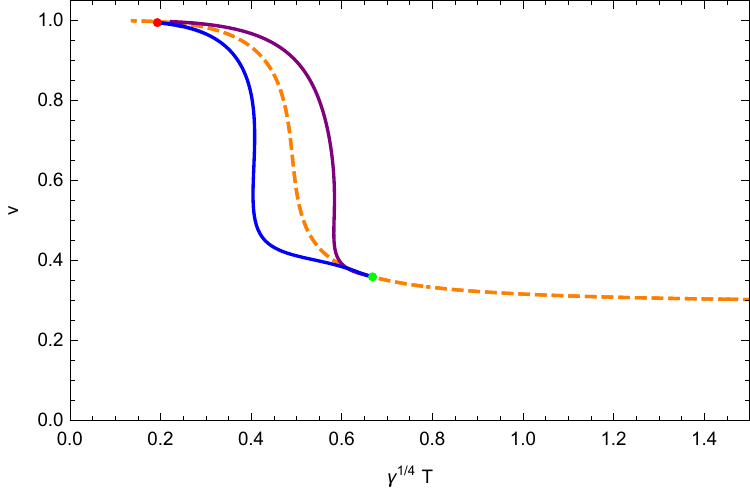}}
\caption{Light cone velocity as a function of temperature for the viscous Bjorken and Minkowski cases for a variety of $r$. The orange dashed curve is the inviscid case in the Minkowski background as shown in Fig.~\ref{fig:inviscid-vel}. As before, the evolution in the $\tau$ coordinate follows from the green to red dot (right to left). The upper solid purple line denotes the more viscous system, while the lower solid blue line is the weakly coupled system. The initial conditions are identical for both systems for each $r$ and with initially no dissipation $\phi(\tau_0)=\tilde{\phi}(\tau_0)=0$ at $\tau_0=1$, i.e. top: $r=2$ and $\gamma\varepsilon=\gamma \et=0.348$, middle: $r=r_c$ with  $\gamma\varepsilon=\gamma \et=0.248$, and finally bottom: $r=1.2$ with $\gamma\varepsilon=\gamma \et=0.3$,  while the transport coefficients are given by \eqref{parameters}. }\label{fig:uneq-vel}
\end{figure}

Now we go to the full case considered in this paper: unequal subsystems. To study this numerically, we consider equal initial conditions for both subsystems. As we see in Fig.~\ref{fig:uneq-vel}, due to the choice of the initial conditions, the both subsystems have initially the same light cone velocities. Then for intermediate times, the individual subsystems have a markedly different evolution at intermediate times before the light cone velocities approach the vacuum value of 1. Interestingly, we can determine situations where the light cone velocity in one system is multivalued, while in the other it is not. Of course, it is worthwhile to repeat that we are not in thermal equilibrium, which means we are not exactly mapping to the physical temperature $\mathcal{T}$, but instead to an effective temperature.

\section{The disperser surface}\label{sec:double-disp}
On the disperser surface, we have the following scaling behaviors near $\tau = 0$:
 \begin{align}
 \gamma \et &\rightarrow \sqrt{\frac{r-1}{r}}, 
\gamma \e \rightarrow {g_1} \tau ^{\frac{4}{3} (\sigma_2 - \sigma_1)},\\
\x &\rightarrow  -\sigma_1 + k_1 \tau^{\frac{1}{3}(2 -\sigma_2)},
\xt \rightarrow  -\sigma_2 + k_2 \tau^{-\frac{5}{3}\sigma_1 +\frac{4}{3}\sigma_2 + \frac{2}{3}},\\
\A   &\rightarrow  a_{10} \tau ^{\sigma_2-\frac{4}{3}\sigma_1  - \frac{1}{3}},
\B  \rightarrow  b_{10} \tau ^{-\frac{1}{3}(\sigma_2 + 1)},
\C  \rightarrow  c_{10} \tau ^{\frac{1}{3}(2-\sigma_2 )},\\
\At  &\rightarrow  a_{20} \tau ^{-\frac{1}{3}(\sigma_2 + 1)},
\Bt  \rightarrow  b_{20} \tau ^{-\frac{1}{3}(\sigma_2+ 1)},
\Ct  \rightarrow  c_{20} \tau ^{\frac{1}{3}(2-\sigma_2)}.
\end{align}
We note that $\x$ and $\xt$ go to $-\sigma_1$ and $-\sigma_2$ respectively, $\gamma\e$ vanishes while $\gamma\et$ goes to $\sqrt{r-1}/\sqrt{r}$  as $\tau\rightarrow 0$. The above scalings follow on the disperser surface if instead of \eqref{Eq:condsn}, we require that $(4/3)\sigma_1< \sigma_2 < 1$ with $0<\sigma_1<1$ which is satisfied by our choices of parameters.\footnote{More specifically, we get these scalings if $\A\e/\At \rightarrow 0$ as $\tau\rightarrow0$.}

The five identities analogous to those in Eq. \eqref{grand-identity} which should hold at $\tau = 0$ are:
\begin{align}\label{Eq:grand-identity2}
&\frac{\Bt}{\tau^{-\frac{1}{3}(\sigma_2 +1)}} =\frac{\Ct}{\tau^{\frac{1}{3}(2-\sigma_2)}}= \sqrt{\frac{r}{r-1}}\frac{\At}{\tau^{-\frac{1}{3}(\sigma_2 +1)}}\\\nonumber &=\sqrt{\frac{3\sqrt{r(r-1)}}{\gamma\et(4r-1-2(r-1)\sigma_2)}} \frac{\B}{\tau^{-\frac{1}{3}(\sigma_2 +1)}} = \sqrt{\frac{3\sqrt{r(r-1)}}{\gamma\et(4r-1+4(r-1)\sigma_2)}} \frac{\C}{\tau^{\frac{1}{3}(2-\sigma_2)}} \\\nonumber & =\frac{\B}{\tau^{-\frac{1}{3}(\sigma_2 +1)}} \sqrt{ r\gamma \e\frac{\C}{\tau a_1}}
\end{align}
We also find that the other six coefficients of the early time expansion can be determined in terms of $k_1$ and $g_1$ in consistency with the fact that the disperser surface is two-dimensional.

\begin{figure}
\includegraphics[width=0.5\linewidth]{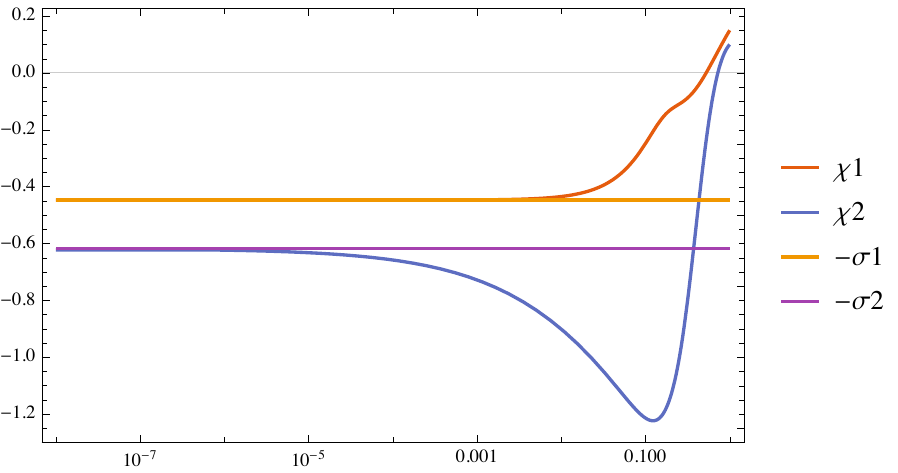}\label{fig:chiplotsdisp}
\includegraphics[width=0.5\linewidth]{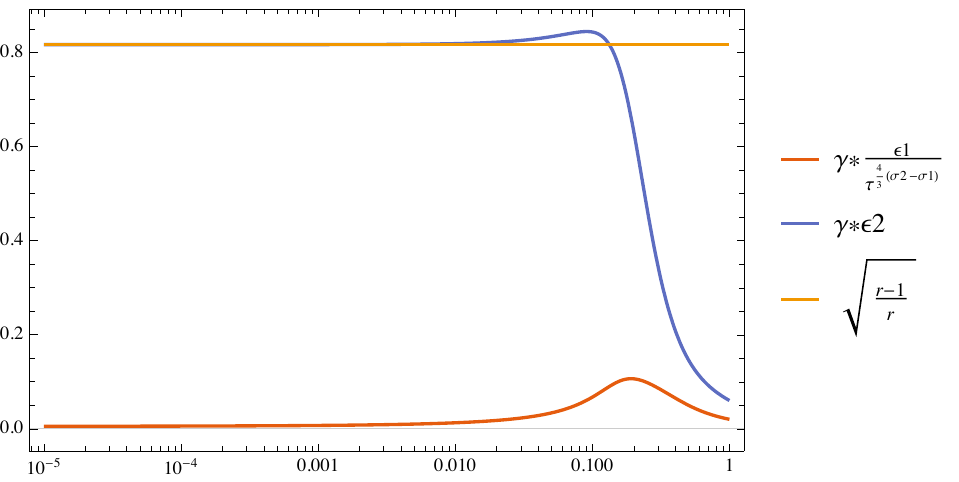}\label{fig:epsilonplotsdisp}
\caption{A disperser solution is studied with initial conditions $\e(\tau_0) = 0.1$, $\et(\tau_0) = 0.3$, $\phi_1(\tau_0) = 0.02$ and $\phi_2(\tau_0) = 0.02$ at $\tau_0 = 1$ with all parameters same as in the case of our attractor solutions except $r =3$. Left panel: In the limit, $\tau \rightarrow0$, $\x$ and $\xt$ go to $-\sigma_1$ and $-\sigma_2$ respectively. Right panel: The opposite of bottom-up happens: $\gamma \et$ goes to a constant $\sqrt{(r-1)/r}$ and $\gamma \epsilon$ goes to zero as $\tau^{\frac{4}{3}(\sigma_2-\sigma_1)}$.}\label{fig:dispsoln}
\end{figure}

\begin{figure}
 \includegraphics[width=\linewidth]{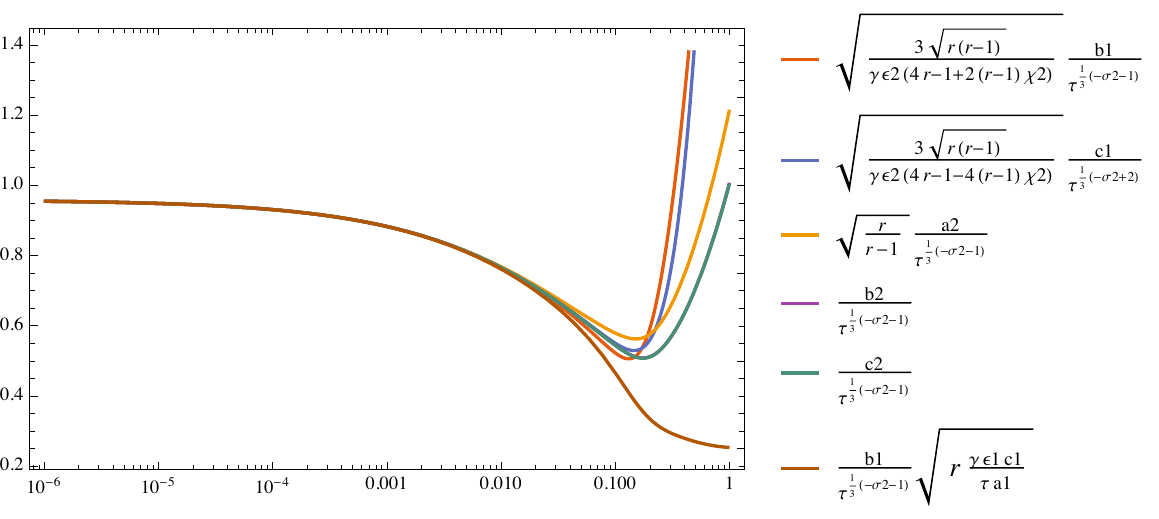}
\caption{A check of the grand identity \eqref{Eq:grand-identity2} for the disperser solution in Fig. \ref{fig:dispsoln}.}\label{fig:abcplotsdisp}
\end{figure}

In Fig. \ref{fig:dispsoln}, the early time behaviors of $\x$, $\xt$, $\e$ and $\et$ in a disperser solution have been shown, and in Fig. \ref{fig:abcplotsdisp}, we have shown the numerical verification of \eqref{Eq:grand-identity2} to a very high accuracy.

\section{Borel resummation of the hydrodynamic expansion}\label{app:borel}

The full system at late time behaves as a single fluid with the transport coefficient being determined by which attractor curve on the 2D attractor surface the system goes to as discussed above.  The curves on the attractor surface are parameterized by the dimensionless parameters, $\alpha$ and $\beta$ (defined in \eqref{alpha} and \eqref{beta}) which govern the late time hydrodynamic expansion and thus the effective transport coefficients of the full system. 
Here we aim to understand the hydrodynamic gradient expansion and see how its Borel resummation can describe the attractor surface in terms of these two parameters.

\begin{figure}[ht]
\includegraphics[width=0.5\linewidth]{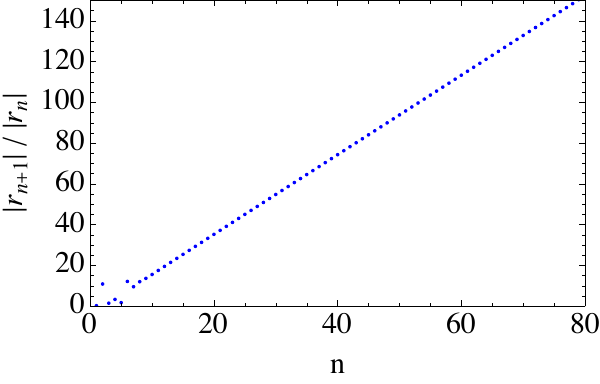}
\includegraphics[width=0.5\linewidth]{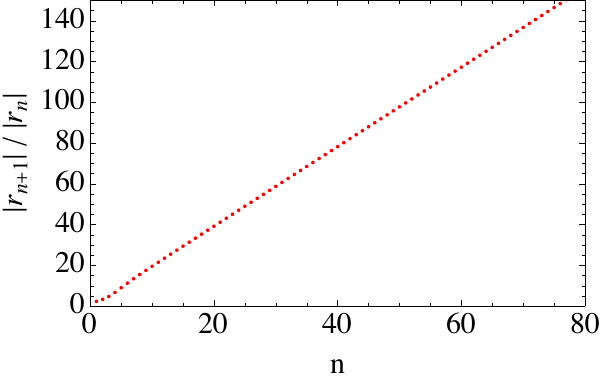} \\
\includegraphics[width=0.5\linewidth]{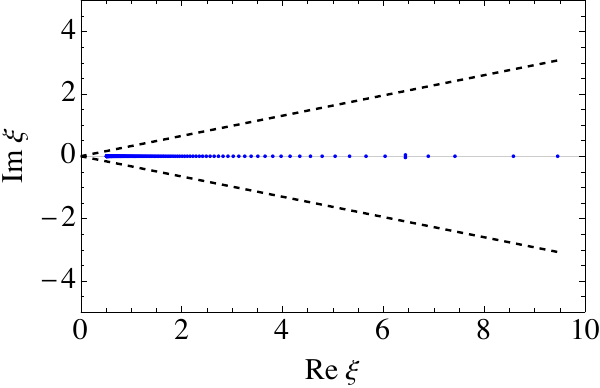}
\includegraphics[width=0.5\linewidth]{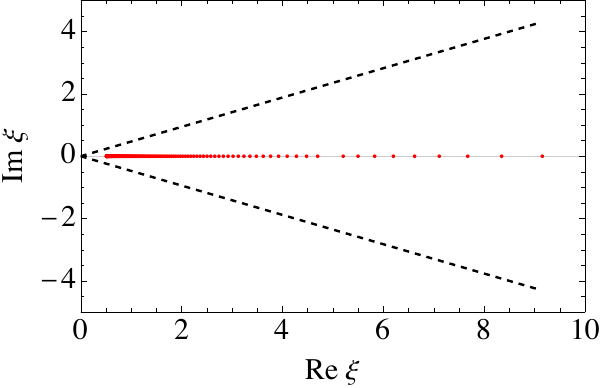}
\caption{The top figures shows the ratio of the absolute values of the $n+1$-th to $n$-th series coefficients of the hydrodynamic expansion of $\chi$ and $\tilde\chi$ as function of $n$, where the left panel is for the strong system and the right panel is for the weak system. The linear growth of this ratio implies factorial growth of $r_n$ for both weak and strong systems. The bottom figures show the pole structure of both the subsystems on the Borel plane.  The poles of the strong sector and the weak sector are shown in blue (on left) and red (on right) respectively. It starts from $  \xi_{1,0} = \xi_{2,0} =\frac{ 3 } {2 C_{\tau}} \Big(\frac{\beta}{ \sqrt{\alpha \gamma}} \Big)^{1/3} $, for $ \alpha = 1.169 $ and $ \beta = 2.747 $. The black dotted lines shows the contour of integration at an angle $ \pm \theta$.}\label{fig:slope_pade}
\end{figure}

We will expand the anisotropies, at late time in powers of inverse proper time, $x=\tau^{-2/3},$
\begin{eqnarray} \label{chis}
\x = \frac{4 \cet }{3 } \Big(\frac{\sqrt{\alpha \gamma} }{\beta }\Big)^{1/3}  x + \frac{8 }{9} \Big(\frac{\alpha \gamma }{\beta^2 }\Big)^{1/3} (\cet^2 + \cet \ctp) x^{2} +\mathcal{O}(x^3)\, , \nonumber  \\
\xt = \frac{4 \cett }{3}\Big(\frac{\sqrt{\beta \gamma} }{\alpha}\Big)^{1/3} x + \frac{8}{9} \Big(\frac{\beta \gamma }{\alpha^2 }\Big)^{1/3} (\cett^2 + \cett \ctpt) x^{2} +\mathcal{O}(x^3)\, .
\end{eqnarray}
Let's use the notation $\xi_{i}$ with $i=1,2$ and $\chi_1 = \chi$ and $\chi_2 = \tilde\chi$. At arbitrarily high order the coefficients of these two series expansions $r_{i,n}$ for $\chi_{i}$ show factorial growth following
\begin{equation*}
    \frac{r_{i,n}}{r_{i,n+1}} = \frac{\xi_{i,0}}{n+1} \left(1+\frac{\gamma_i +1}{n+1}+\mathcal{O}(n^{-2})\right)
\end{equation*}
indicating the series is asymptotic in nature with zero radius of convergence, which is seen in subsystems in the top panels of Fig.~\ref{fig:slope_pade}. The parameters $\xi_{i,0}$ and $\gamma_i$ for the weak and strong systems depend on $\alpha$ and $\beta$.

To promote the divergent series to a well-defined function the series needs to have finite radius of convergence. This is successfully done by adopting the standard technique of Borel resummation \cite{Aniceto:2011nu}.  
The Borel transform of the series is given by
\begin{eqnarray} \label{gborel}
 \chi_{B,i} (\xi)= \sum_{n=0}^{\infty} \frac{r_{i,n}}{n!} {\xi}^n  
 \end{eqnarray}
 $ r_{i,n}$ are the coefficients from the series (\ref{chis}). This Borel transformed function has a finite radius of convergence about the origin  in the complex $\xi $ plane (i.e. the Borel plane) and is related to the original series via Laplace transform.  However the function possesses poles in the complex plane as a consequence of the finite radius of convergence.  Hence to perform the integral one has to  do analytic continuation of the function $ \chi_{B,i}({\xi})$.  Since we only have access to a finite number of terms, we will need to approximate the integrand, which we do via Pad{\'e} approximant
 \begin{equation}
 \chi_{B,i}({\xi}) \approx  P_{N,M,i}({\xi}) =  \frac{\sum_{i=0}^{N} n_i {\xi}^i }{1+ \sum_{j=1}^{M} m_j {\xi}^j}
 \end{equation}
where the coefficients of the Pad{\'e} approximant are fixed to agree with series coefficients (\ref{gborel}). The Pad{\'e} series can go up to arbitrary orders with a constraint $N + M = k$,  where $ k $ is the order up to which Borel transformed anisotropy is truncated.  
Most commonly, one considers the symmetric case with $ N=M= k/2$.  In our case we have taken the order $ k$ to be as high as $ 500$ with precision set to thousand digits. 

 The singularities of the function $  \chi_{B,i}({\xi})$  appear as concentrated poles of the Pad\'e-approximant in the Borel plane and {they depend} on the dimensionless parameter $ \alpha$ and $ \beta$.  In Fig.~\ref{fig:slope_pade}, we show the pole structure of both sectors in the Borel plane.
The poles start from a point nearest to the origin, namely $\xi_{1,0}$ and $\xi_{2,0}$ for $\chi_{B}(\xi)$ and $\tilde\chi_{B}(\xi)$ respectively, and accumulate along the real axis giving an image of branch cut as in the case of decoupled MIS \cite{Heller:2015dha}.   The starting value of the branch cut is proportional to the inverse of the slope of the coefficients shown in Fig.~\ref{fig:slope_pade} and can be identified with the non-hydrodynamic mode that decays exponentially at late-times. From the analysis in Sec.~\ref{sec:fluc} we readily see that $ \xi_{1,0} = \xi_{2,0} =\frac{ 3 } {2 C_{\tau}} \Big(\frac{\beta}{ \sqrt{\alpha \gamma}} \Big)^{1/3}= 3\epsilon_{10}^{1/4}\tau_0^{1/3} /(2 C_{\tau})=\xi_0 $ when the condition \eqref{Eq:poss1} is satisfied, and $ \xi_{1,0} = \xi_{2,0} =\frac{ 3 } {2 C_{\tau}} \Big(\frac{\alpha}{ \sqrt{\beta \gamma}} \Big)^{1/3}= 3\epsilon_{20}^{1/4}\tau_0^{1/3} /(2 C_{\tau})=\xi_0 $ when the condition \eqref{Eq:poss2} holds.

With the analytic continuation of the Borel transformed function, we can now perform the inverse integration \cite{Aniceto:2011nu,Heller:2015dha,Casalderrey-Solana:2017zyh}
\begin{eqnarray} \label{Eq:inverse_Borel}
\chi_i (x) = \frac{1}{x}  \int_\mathcal{C} e^{-\xi/x} P_{N,M,i} ({\xi}) d \xi
\end{eqnarray}
where $ \mathcal{C}$ is the contour from $ 0$ to $ \infty$.  However the presence of singularity in $ \chi_{B,i}(\xi)$ introduces complex ambiguities  implying that the integration depends on the choice of contour $\mathcal{C}$. The choice of the contour actually determine the Stokes coefficients appearing in the trans-series that encode the full details of the initial conditions. Appropriate choices of these Stokes coefficients (contour of integration) correspond to evolution on the attractor surface. In our case, the contours of integration can be chosen to be the straight lines given by two different angles for $\chi$ and $\tilde\chi$, respectively, as shown in Fig. \ref{fig:slope_pade}. We have two Stokes coefficients corresponding to the two exponential decay modes discussed in Sec. \ref{sec:fluc} and these give the two missing data (other than the parameters $\alpha$ and $\beta$ appearing in the hydrodynamic expansion) which specify the initial conditions fully in the four-dimensional phase space.

\section{Different initial ratios of subsector energy densities}\label{sec:thermalization}

\begin{figure}
\centering
\includegraphics[width=0.4\linewidth]{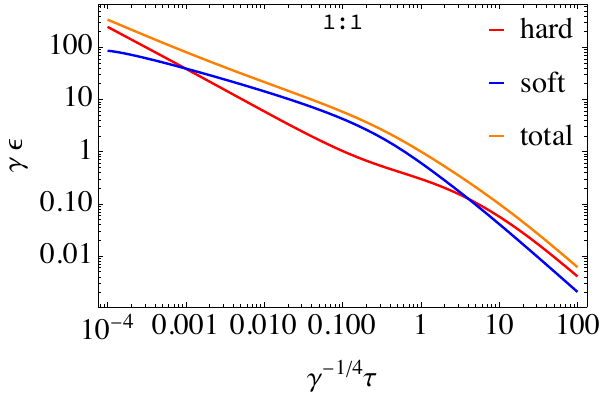}
\includegraphics[width=0.4\linewidth]{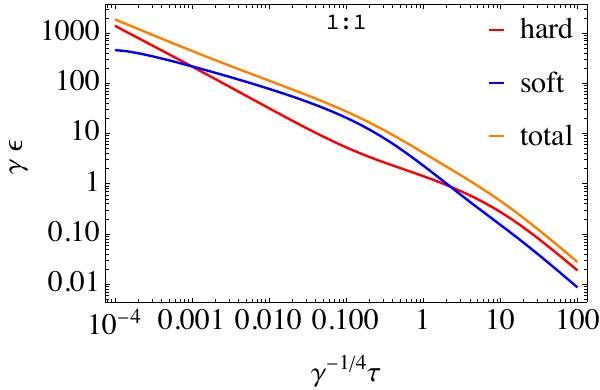}
\includegraphics[width=0.4\linewidth]{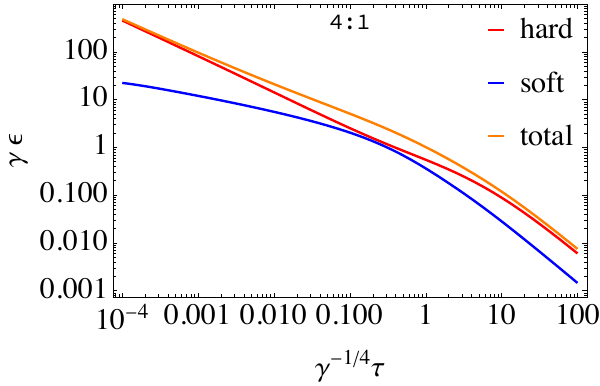}
\includegraphics[width=0.4\linewidth]{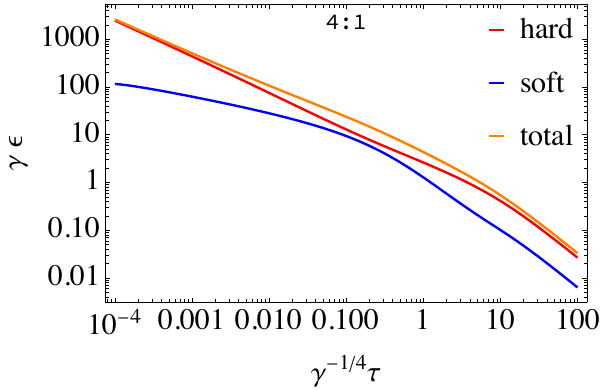}
\includegraphics[width=0.4\linewidth]{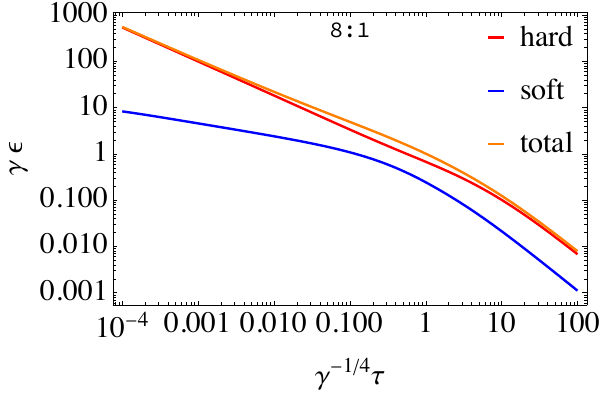} 
\includegraphics[width=0.4\linewidth]{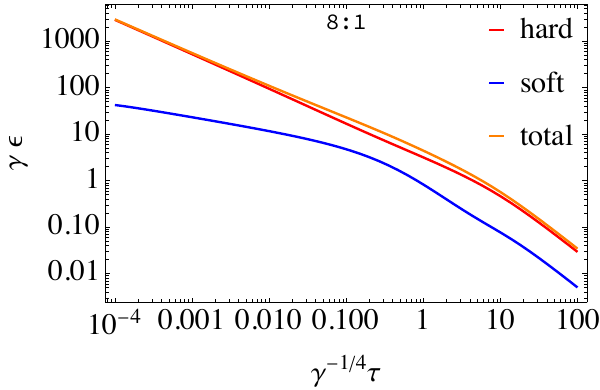} 
\caption{Energy densities for $p$-$p$ collisions (left column) and Pb-Pb (right column) for a variety of initial ratios $\e/\et$. For both $p$-$p$ and Pb-Pb collisions, as the ratio of the initial energy densities of the two sectors increases, the energy of the hard sector dominates for longer time.
} \label{fig:p-p_bottom}\label{fig:Pb-Pb_bottom}
\end{figure}

While at sufficiently early times all attractor solutions have
a dominant weakly coupled component, different initial conditions
lead to different ratios of energy densities at some initial reference time,
chosen here as
$\tau_0=0.001\gamma^{1/4}$. In figure \ref{fig:p-p_bottom} we display the evolution of the energy densities
in the subsectors and the total energy density for different initial
ratios, comparing the two sets of initial conditions we tentatively
associated with $p$-$p$ and Pb-Pb collisions in sect.\ \ref{sec:pheno-pp-pb}.

For a ratio $\mathcal{E}_1:\mathcal{E}_2$ less than 4:1, the energy density in the holographic sector overtakes the one in the perturbative sector for a stretch of time, whereas in the opposite case it never gains dominance.  At sufficiently large time, the ``perturbative'' sector becomes again dominant in each case, but although this is reminiscent of the transition to a weakly coupled hadron gas after hadronization, we consider our model as a toy of heavy-ion collisions only
in the deconfined early stages.


\bibliography{main}

\providecommand{\href}[2]{#2}\begingroup\raggedright\begin{thebibliography}{10}

\bibitem{Chesler:2009cy}
P.~M. Chesler and L.~G. Yaffe, \emph{{Boost invariant flow, black hole
  formation, and far-from-equilibrium dynamics in N = 4 supersymmetric
  Yang-Mills theory}},
  \href{https://doi.org/10.1103/PhysRevD.82.026006}{\emph{Phys. Rev. D}
  {\bfseries 82} (2010) 026006},
  [\href{https://arxiv.org/abs/0906.4426}{{\ttfamily 0906.4426}}].

\bibitem{Chesler:2010bi}
P.~M. Chesler and L.~G. Yaffe, \emph{{Holography and colliding gravitational
  shock waves in asymptotically AdS$_5$ spacetime}},
  \href{https://doi.org/10.1103/PhysRevLett.106.021601}{\emph{Phys. Rev. Lett.}
  {\bfseries 106} (2011) 021601},
  [\href{https://arxiv.org/abs/1011.3562}{{\ttfamily 1011.3562}}].

\bibitem{Heller:2011ju}
M.~P. Heller, R.~A. Janik and P.~Witaszczyk, \emph{{The characteristics of
  thermalization of boost-invariant plasma from holography}},
  \href{https://doi.org/10.1103/PhysRevLett.108.201602}{\emph{Phys. Rev. Lett.}
  {\bfseries 108} (2012) 201602},
  [\href{https://arxiv.org/abs/1103.3452}{{\ttfamily 1103.3452}}].

\bibitem{Chesler:2015bba}
P.~M. Chesler, \emph{{Colliding shock waves and hydrodynamics in small
  systems}}, \href{https://doi.org/10.1103/PhysRevLett.115.241602}{\emph{Phys.
  Rev. Lett.} {\bfseries 115} (2015) 241602},
  [\href{https://arxiv.org/abs/1506.02209}{{\ttfamily 1506.02209}}].

\bibitem{Attems:2016tby}
M.~Attems, J.~Casalderrey-Solana, D.~Mateos, D.~Santos-Oliv{\'a}n, C.~F.
  Sopuerta, M.~Triana et~al., \emph{{Holographic Collisions in Non-conformal
  Theories}}, \href{https://doi.org/10.1007/JHEP01(2017)026}{\emph{JHEP}
  {\bfseries 01} (2017) 026},
  [\href{https://arxiv.org/abs/1604.06439}{{\ttfamily 1604.06439}}].

\bibitem{Attems:2017zam}
M.~Attems, J.~Casalderrey-Solana, D.~Mateos, D.~Santos-Oliv\'an, C.~F.
  Sopuerta, M.~Triana et~al., \emph{{Paths to equilibrium in non-conformal
  collisions}}, \href{https://doi.org/10.1007/JHEP06(2017)154}{\emph{JHEP}
  {\bfseries 06} (2017) 154},
  [\href{https://arxiv.org/abs/1703.09681}{{\ttfamily 1703.09681}}].

\bibitem{Romatschke:2017vte}
P.~Romatschke, \emph{{Relativistic Fluid Dynamics Far From Local Equilibrium}},
  \href{https://doi.org/10.1103/PhysRevLett.120.012301}{\emph{Phys. Rev. Lett.}
  {\bfseries 120} (2018) 012301},
  [\href{https://arxiv.org/abs/1704.08699}{{\ttfamily 1704.08699}}].

\bibitem{Florkowski:2017olj}
W.~Florkowski, M.~P. Heller and M.~Spalinski, \emph{{New theories of
  relativistic hydrodynamics in the LHC era}},
  \href{https://doi.org/10.1088/1361-6633/aaa091}{\emph{Rept. Prog. Phys.}
  {\bfseries 81} (2018) 046001},
  [\href{https://arxiv.org/abs/1707.02282}{{\ttfamily 1707.02282}}].

\bibitem{Keegan:2015avk}
L.~Keegan, A.~Kurkela, P.~Romatschke, W.~van~der Schee and Y.~Zhu, \emph{{Weak
  and strong coupling equilibration in nonabelian gauge theories}},
  \href{https://doi.org/10.1007/JHEP04(2016)031}{\emph{JHEP} {\bfseries 04}
  (2016) 031}, [\href{https://arxiv.org/abs/1512.05347}{{\ttfamily
  1512.05347}}].

\bibitem{Heinz:2013th}
U.~Heinz and R.~Snellings, \emph{{Collective flow and viscosity in relativistic
  heavy-ion collisions}},
  \href{https://doi.org/10.1146/annurev-nucl-102212-170540}{\emph{Ann. Rev.
  Nucl. Part. Sci.} {\bfseries 63} (2013) 123--151},
  [\href{https://arxiv.org/abs/1301.2826}{{\ttfamily 1301.2826}}].

\bibitem{Heller:2013fn}
M.~P. Heller, R.~A. Janik and P.~Witaszczyk, \emph{{Hydrodynamic Gradient
  Expansion in Gauge Theory Plasmas}},
  \href{https://doi.org/10.1103/PhysRevLett.110.211602}{\emph{Phys. Rev. Lett.}
  {\bfseries 110} (2013) 211602},
  [\href{https://arxiv.org/abs/1302.0697}{{\ttfamily 1302.0697}}].

\bibitem{Heller:2015dha}
M.~P. Heller and M.~Spalinski, \emph{{Hydrodynamics Beyond the Gradient
  Expansion: Resurgence and Resummation}},
  \href{https://doi.org/10.1103/PhysRevLett.115.072501}{\emph{Phys. Rev. Lett.}
  {\bfseries 115} (2015) 072501},
  [\href{https://arxiv.org/abs/1503.07514}{{\ttfamily 1503.07514}}].

\bibitem{Heller:2016gbp}
M.~P. Heller, \emph{{Holography, Hydrodynamization and Heavy-Ion Collisions}},
  \href{https://doi.org/10.5506/APhysPolB.47.2581}{\emph{Acta Phys. Polon. B}
  {\bfseries 47} (2016) 2581},
  [\href{https://arxiv.org/abs/1610.02023}{{\ttfamily 1610.02023}}].

\bibitem{Romatschke:2017acs}
P.~Romatschke, \emph{{Relativistic Hydrodynamic Attractors with Broken
  Symmetries: Non-Conformal and Non-Homogeneous}},
  \href{https://doi.org/10.1007/JHEP12(2017)079}{\emph{JHEP} {\bfseries 12}
  (2017) 079}, [\href{https://arxiv.org/abs/1710.03234}{{\ttfamily
  1710.03234}}].

\bibitem{Denicol:2017lxn}
G.~S. Denicol and J.~Noronha, \emph{{Analytical attractor and the divergence of
  the slow-roll expansion in relativistic hydrodynamics}},
  \href{https://doi.org/10.1103/PhysRevD.97.056021}{\emph{Phys. Rev. D}
  {\bfseries 97} (2018) 056021},
  [\href{https://arxiv.org/abs/1711.01657}{{\ttfamily 1711.01657}}].

\bibitem{Casalderrey-Solana:2017zyh}
J.~Casalderrey-Solana, N.~I. Gushterov and B.~Meiring, \emph{{Resurgence and
  Hydrodynamic Attractors in Gauss-Bonnet Holography}},
  \href{https://doi.org/10.1007/JHEP04(2018)042}{\emph{JHEP} {\bfseries 04}
  (2018) 042}, [\href{https://arxiv.org/abs/1712.02772}{{\ttfamily
  1712.02772}}].

\bibitem{Denicol:2018pak}
G.~S. Denicol and J.~Noronha, \emph{{Hydrodynamic attractor and the fate of
  perturbative expansions in Gubser flow}},
  \href{https://doi.org/10.1103/PhysRevD.99.116004}{\emph{Phys. Rev. D}
  {\bfseries 99} (2019) 116004},
  [\href{https://arxiv.org/abs/1804.04771}{{\ttfamily 1804.04771}}].

\bibitem{Kurkela:2019set}
A.~Kurkela, W.~van~der Schee, U.~A. Wiedemann and B.~Wu, \emph{{Early- and
  Late-Time Behavior of Attractors in Heavy-Ion Collisions}},
  \href{https://doi.org/10.1103/PhysRevLett.124.102301}{\emph{Phys. Rev. Lett.}
  {\bfseries 124} (2020) 102301},
  [\href{https://arxiv.org/abs/1907.08101}{{\ttfamily 1907.08101}}].

\bibitem{Heller:2020anv}
M.~P. Heller, R.~Jefferson, M.~Spali\'nski and V.~Svensson, \emph{{Hydrodynamic
  Attractors in Phase Space}},
  \href{https://doi.org/10.1103/PhysRevLett.125.132301}{\emph{Phys. Rev. Lett.}
  {\bfseries 125} (2020) 132301},
  [\href{https://arxiv.org/abs/2003.07368}{{\ttfamily 2003.07368}}].

\bibitem{Almaalol:2022ijz}
D.~Almaalol, K.~Boguslavski, A.~Kurkela and M.~Strickland,
  \emph{{Non-equilibrium attractor in high-temperature QCD plasmas}},
  \href{https://arxiv.org/abs/2208.00513}{{\ttfamily 2208.00513}}.

\bibitem{Jaiswal:2022mdk}
S.~Jaiswal, S.~Pal, C.~Chattopadhyay, L.~Du and U.~Heinz,
  \emph{{Far-from-equilibrium attractor in non-conformal plasmas}},
  \href{https://arxiv.org/abs/2208.00744}{{\ttfamily 2208.00744}}.

\bibitem{Soloviev:2021lhs}
A.~Soloviev, \emph{{Hydrodynamic attractors in heavy ion collisions: a
  review}}, \href{https://doi.org/10.1140/epjc/s10052-022-10282-4}{\emph{Eur.
  Phys. J. C} {\bfseries 82} (2022) 319},
  [\href{https://arxiv.org/abs/2109.15081}{{\ttfamily 2109.15081}}].

\bibitem{Berges:2020fwq}
J.~Berges, M.~P. Heller, A.~Mazeliauskas and R.~Venugopalan, \emph{{QCD
  thermalization: Ab initio approaches and interdisciplinary connections}},
  \href{https://doi.org/10.1103/RevModPhys.93.035003}{\emph{Rev. Mod. Phys.}
  {\bfseries 93} (2021) 035003},
  [\href{https://arxiv.org/abs/2005.12299}{{\ttfamily 2005.12299}}].

\bibitem{Busza:2018rrf}
W.~Busza, K.~Rajagopal and W.~van~der Schee, \emph{{Heavy Ion Collisions: The
  Big Picture, and the Big Questions}},
  \href{https://doi.org/10.1146/annurev-nucl-101917-020852}{\emph{Ann. Rev.
  Nucl. Part. Sci.} {\bfseries 68} (2018) 339--376},
  [\href{https://arxiv.org/abs/1802.04801}{{\ttfamily 1802.04801}}].

\bibitem{Gelis:2010nm}
F.~Gelis, E.~Iancu, J.~Jalilian-Marian and R.~Venugopalan, \emph{{The Color
  Glass Condensate}},
  \href{https://doi.org/10.1146/annurev.nucl.010909.083629}{\emph{Ann. Rev.
  Nucl. Part. Sci.} {\bfseries 60} (2010) 463--489},
  [\href{https://arxiv.org/abs/1002.0333}{{\ttfamily 1002.0333}}].

\bibitem{Arnold:2002zm}
P.~B. Arnold, G.~D. Moore and L.~G. Yaffe, \emph{{Effective kinetic theory for
  high temperature gauge theories}},
  \href{https://doi.org/10.1088/1126-6708/2003/01/030}{\emph{JHEP} {\bfseries
  01} (2003) 030}, [\href{https://arxiv.org/abs/hep-ph/0209353}{{\ttfamily
  hep-ph/0209353}}].

\bibitem{Kurkela:2015qoa}
A.~Kurkela and Y.~Zhu, \emph{{Isotropization and hydrodynamization in weakly
  coupled heavy-ion collisions}},
  \href{https://doi.org/10.1103/PhysRevLett.115.182301}{\emph{Phys. Rev. Lett.}
  {\bfseries 115} (2015) 182301},
  [\href{https://arxiv.org/abs/1506.06647}{{\ttfamily 1506.06647}}].

\bibitem{Faulkner:2010tq}
T.~Faulkner and J.~Polchinski, \emph{{Semi-Holographic Fermi Liquids}},
  \href{https://doi.org/10.1007/JHEP06(2011)012}{\emph{JHEP} {\bfseries 06}
  (2011) 012}, [\href{https://arxiv.org/abs/1001.5049}{{\ttfamily 1001.5049}}].

\bibitem{Mukhopadhyay:2013dqa}
A.~Mukhopadhyay and G.~Policastro, \emph{{Phenomenological Characterization of
  Semiholographic Non-Fermi Liquids}},
  \href{https://doi.org/10.1103/PhysRevLett.111.221602}{\emph{Phys. Rev. Lett.}
  {\bfseries 111} (2013) 221602},
  [\href{https://arxiv.org/abs/1306.3941}{{\ttfamily 1306.3941}}].

\bibitem{Iancu:2014ava}
E.~Iancu and A.~Mukhopadhyay, \emph{{A semi-holographic model for heavy-ion
  collisions}}, \href{https://doi.org/10.1007/JHEP06(2015)003}{\emph{JHEP}
  {\bfseries 06} (2015) 003},
  [\href{https://arxiv.org/abs/1410.6448}{{\ttfamily 1410.6448}}].

\bibitem{Mukhopadhyay:2015smb}
A.~Mukhopadhyay, F.~Preis, A.~Rebhan and S.~A. Stricker, \emph{{Semi-Holography
  for Heavy Ion Collisions: Self-Consistency and First Numerical Tests}},
  \href{https://doi.org/10.1007/JHEP05(2016)141}{\emph{JHEP} {\bfseries 05}
  (2016) 141}, [\href{https://arxiv.org/abs/1512.06445}{{\ttfamily
  1512.06445}}].

\bibitem{Banerjee:2017ozx}
S.~Banerjee, N.~Gaddam and A.~Mukhopadhyay, \emph{{Illustrated study of the
  semiholographic nonperturbative framework}},
  \href{https://doi.org/10.1103/PhysRevD.95.066017}{\emph{Phys. Rev. D}
  {\bfseries 95} (2017) 066017},
  [\href{https://arxiv.org/abs/1701.01229}{{\ttfamily 1701.01229}}].

\bibitem{Kurkela:2018dku}
A.~Kurkela, A.~Mukhopadhyay, F.~Preis, A.~Rebhan and A.~Soloviev, \emph{{Hybrid
  Fluid Models from Mutual Effective Metric Couplings}},
  \href{https://doi.org/10.1007/JHEP08(2018)054}{\emph{JHEP} {\bfseries 08}
  (2018) 054}, [\href{https://arxiv.org/abs/1805.05213}{{\ttfamily
  1805.05213}}].

\bibitem{Ecker:2018ucc}
C.~Ecker, A.~Mukhopadhyay, F.~Preis, A.~Rebhan and A.~Soloviev, \emph{{Time
  evolution of a toy semiholographic glasma}},
  \href{https://doi.org/10.1007/JHEP08(2018)074}{\emph{JHEP} {\bfseries 08}
  (2018) 074}, [\href{https://arxiv.org/abs/1806.01850}{{\ttfamily
  1806.01850}}].

\bibitem{Muller:1967zza}
I.~M{\"u}ller, \emph{{Zum Paradoxon der W\"armeleitungstheorie}},
  \href{https://doi.org/10.1007/BF01326412}{\emph{Z. Phys.} {\bfseries 198}
  (1967) 329--344}.

\bibitem{Israel:1979wp}
W.~Israel and J.~Stewart, \emph{{Transient relativistic thermodynamics and
  kinetic theory}},
  \href{https://doi.org/10.1016/0003-4916(79)90130-1}{\emph{Annals Phys.}
  {\bfseries 118} (1979) 341--372}.

\bibitem{Mitra:2020mei}
T.~Mitra, S.~Mondkar, A.~Mukhopadhyay, A.~Rebhan and A.~Soloviev,
  \emph{{Hydrodynamic attractor of a hybrid viscous fluid in Bjorken flow}},
  \href{https://doi.org/10.1103/PhysRevResearch.2.043320}{\emph{Phys. Rev.
  Res.} {\bfseries 2} (2020) 043320},
  [\href{https://arxiv.org/abs/2006.09383}{{\ttfamily 2006.09383}}].

\bibitem{Baier:2000sb}
R.~Baier, A.~H. Mueller, D.~Schiff and D.~T. Son, \emph{{'Bottom up'
  thermalization in heavy ion collisions}},
  \href{https://doi.org/10.1016/S0370-2693(01)00191-5}{\emph{Phys. Lett. B}
  {\bfseries 502} (2001) 51--58},
  [\href{https://arxiv.org/abs/hep-ph/0009237}{{\ttfamily hep-ph/0009237}}].

\bibitem{Loizides:2016tew}
C.~Loizides, \emph{{Experimental overview on small collision systems at the
  LHC}}, \href{https://doi.org/10.1016/j.nuclphysa.2016.04.022}{\emph{Nucl.
  Phys. A} {\bfseries 956} (2016) 200--207},
  [\href{https://arxiv.org/abs/1602.09138}{{\ttfamily 1602.09138}}].

\bibitem{Schlichting:2016sqo}
S.~Schlichting and P.~Tribedy, \emph{{Collectivity in Small Collision Systems:
  An Initial-State Perspective}},
  \href{https://doi.org/10.1155/2016/8460349}{\emph{Adv. High Energy Phys.}
  {\bfseries 2016} (2016) 8460349},
  [\href{https://arxiv.org/abs/1611.00329}{{\ttfamily 1611.00329}}].

\bibitem{Kowalski:2007rw}
H.~Kowalski, T.~Lappi and R.~Venugopalan, \emph{{Nuclear enhancement of
  universal dynamics of high parton densities}},
  \href{https://doi.org/10.1103/PhysRevLett.100.022303}{\emph{Phys. Rev. Lett.}
  {\bfseries 100} (2008) 022303},
  [\href{https://arxiv.org/abs/0705.3047}{{\ttfamily 0705.3047}}].

\bibitem{Mondkar:2021qsf}
S.~Mondkar, A.~Mukhopadhyay, A.~Rebhan and A.~Soloviev, \emph{{Quasinormal
  modes of a semi-holographic black brane and thermalization}},
  \href{https://doi.org/10.1007/JHEP11(2021)080}{\emph{JHEP} {\bfseries 11}
  (2021) 080}, [\href{https://arxiv.org/abs/2108.02788}{{\ttfamily
  2108.02788}}].

\bibitem{Baier:2007ix}
R.~Baier, P.~Romatschke, D.~T. Son, A.~O. Starinets and M.~A. Stephanov,
  \emph{{Relativistic viscous hydrodynamics, conformal invariance, and
  holography}},
  \href{https://doi.org/10.1088/1126-6708/2008/04/100}{\emph{JHEP} {\bfseries
  04} (2008) 100}, [\href{https://arxiv.org/abs/0712.2451}{{\ttfamily
  0712.2451}}].

\bibitem{Noronha:2011fi}
J.~Noronha and G.~S. Denicol, \emph{{Transient Fluid Dynamics of the
  Quark-Gluon Plasma According to AdS/CFT}},
  \href{https://arxiv.org/abs/1104.2415}{{\ttfamily 1104.2415}}.

\bibitem{Heller:2014wfa}
M.~P. Heller, R.~A. Janik, M.~Spali\'nski and P.~Witaszczyk, \emph{{Coupling
  hydrodynamics to nonequilibrium degrees of freedom in strongly interacting
  quark-gluon plasma}},
  \href{https://doi.org/10.1103/PhysRevLett.113.261601}{\emph{Phys. Rev. Lett.}
  {\bfseries 113} (2014) 261601},
  [\href{https://arxiv.org/abs/1409.5087}{{\ttfamily 1409.5087}}].

\bibitem{Romatschke:2017ejr}
P.~Romatschke and U.~Romatschke, \emph{{Relativistic Fluid Dynamics In and Out
  of Equilibrium}}.
\newblock Cambridge Monographs on Mathematical Physics. Cambridge University
  Press, 5, 2019,
  \href{https://doi.org/10.1017/9781108651998}{10.1017/9781108651998}.

\bibitem{Giacalone:2019ldn}
G.~Giacalone, A.~Mazeliauskas and S.~Schlichting, \emph{{Hydrodynamic
  attractors, initial state energy and particle production in relativistic
  nuclear collisions}},
  \href{https://doi.org/10.1103/PhysRevLett.123.262301}{\emph{Phys. Rev. Lett.}
  {\bfseries 123} (2019) 262301},
  [\href{https://arxiv.org/abs/1908.02866}{{\ttfamily 1908.02866}}].

\bibitem{Naik:2021yph}
L.~J. Naik, S.~Jaiswal, K.~Sreelakshmi, A.~Jaiswal and V.~Sreekanth,
  \emph{{Hydrodynamical attractor and thermal particle production in heavy-ion
  collision}},  \href{https://arxiv.org/abs/2107.08791}{{\ttfamily
  2107.08791}}.

\bibitem{Coquet:2021lca}
M.~Coquet, X.~Du, J.-Y. Ollitrault, S.~Schlichting and M.~Winn,
  \emph{{Intermediate mass dileptons as pre-equilibrium probes in heavy ion
  collisions}},
  \href{https://doi.org/10.1016/j.physletb.2021.136626}{\emph{Phys. Lett. B}
  {\bfseries 821} (2021) 136626},
  [\href{https://arxiv.org/abs/2104.07622}{{\ttfamily 2104.07622}}].

\bibitem{Grossi:2021gqi}
E.~Grossi, A.~Soloviev, D.~Teaney and F.~Yan, \emph{{Soft pions and transport
  near the chiral critical point}},
  \href{https://doi.org/10.1103/PhysRevD.104.034025}{\emph{Phys. Rev. D}
  {\bfseries 104} (2021) 034025},
  [\href{https://arxiv.org/abs/2101.10847}{{\ttfamily 2101.10847}}].

\bibitem{Kovtun:2012rj}
P.~Kovtun, \emph{{Lectures on hydrodynamic fluctuations in relativistic
  theories}}, \href{https://doi.org/10.1088/1751-8113/45/47/473001}{\emph{J.
  Phys. A} {\bfseries 45} (2012) 473001},
  [\href{https://arxiv.org/abs/1205.5040}{{\ttfamily 1205.5040}}].

\bibitem{Mitra:2020hbj}
T.~Mitra, A.~Mukhopadhyay and A.~Soloviev, \emph{{Hydrodynamic attractor and
  novel fixed points in superfluid Bjorken flow}},
  \href{https://doi.org/10.1103/PhysRevD.103.076014}{\emph{Phys. Rev. D}
  {\bfseries 103} (2021) 076014},
  [\href{https://arxiv.org/abs/2012.15644}{{\ttfamily 2012.15644}}].

\bibitem{Rodgers:2022fuv}
R.~Rodgers and J.~G. Subils, \emph{{Boost-invariant superfluid flows}},
  \href{https://doi.org/10.1007/JHEP09(2022)205}{\emph{JHEP} {\bfseries 09}
  (2022) 205}, [\href{https://arxiv.org/abs/2207.02903}{{\ttfamily
  2207.02903}}].

\bibitem{Aniceto:2011nu}
I.~Aniceto, R.~Schiappa and M.~Vonk, \emph{{The Resurgence of Instantons in
  String Theory}},
  \href{https://doi.org/10.4310/CNTP.2012.v6.n2.a3}{\emph{Commun. Num. Theor.
  Phys.} {\bfseries 6} (2012) 339--496},
  [\href{https://arxiv.org/abs/1106.5922}{{\ttfamily 1106.5922}}].

\end{thebibliography}\endgroup
\bibliographystyle{JHEP}

\end{document}